\documentclass[11pt,a4paper]{article}
\usepackage{alpha}
\usepackage{lineno,hyperref}
\usepackage{booktabs}
\usepackage{amsmath}
\usepackage{amssymb}
\usepackage{wasysym}
\usepackage[flushleft]{threeparttable}
\usepackage{lscape}
\usepackage{rotating}

\def\xSF{\chi{\rm SF}}

\def\rmd{{\rm d}}

\def\rmO{{\rm O}}

\def\proof{\noindent{\sl Proof:}\kern0.6em}

\def\frac#1#2{\hbox{$#1\over#2$}}
\def\dual{\mathstrut^*\kern-0.1em}

\def\lvec#1{\setbox0=\hbox{$#1$}
    \setbox1=\hbox{$\scriptstyle\leftarrow$}
    #1\kern-\wd0\smash{
    \raise\ht0\hbox{$\raise1pt\hbox{$\scriptstyle\leftarrow$}$}}
    \kern-\wd1\kern\wd0}
\def\rvec#1{\setbox0=\hbox{$#1$}
    \setbox1=\hbox{$\scriptstyle\rightarrow$}
    #1\kern-\wd0\smash{
    \raise\ht0\hbox{$\raise1pt\hbox{$\scriptstyle\rightarrow$}$}}
    \kern-\wd1\kern\wd0}
\def\slash#1{\setbox0=\hbox{$#1$}\setbox1=\hbox{$\kern1pt/$}
    #1\kern-\wd0\kern1pt/\kern-\wd1\kern\wd0}

\def\nabstar#1{{\nabla\kern0.5pt\smash{\raise 4.5pt\hbox{$\ast$}}
               \kern-5.5pt_{#1}}}

\def\drvstar#1{{\partial\kern0.5pt\smash{\raise 4.5pt\hbox{$\ast$}}
               \kern-6.0pt_{#1}}}

\def\ldrvstar#1{{\lvec{\,\partial}\kern-0.5pt\smash{\raise 4.5pt\hbox{$\ast$}}
               \kern-5.0pt_{#1}}}

\def\fm{{\rm fm}}
\def\MSbar{\overline{\rm MS\kern-0.5pt}\kern0.5pt}
\def\Nf{{N_{\rm f}}}

\def\psibar{\overline{\psi}}

\def\zetabar{\bar{\zeta}}
\def\zetaprime{\zeta\kern1pt'}
\def\zetabarprime{\zetabar\kern1pt'}

\def\diracstar#1#2{
    \setbox0=\hbox{$\gamma$}\setbox1=\hbox{$\gamma_{#1}$}
    \gamma_{#1}\kern-\wd1\kern\wd0
    \smash{\raise4.5pt\hbox{$\scriptstyle#2$}}}

\def\SUthree{{\rm SU(3)}}

\def\Obs{{\mathcal O}}

\def\Ds{D_{\rm s}}
\def\DsdagDs{\Ds{\Ds}^{\kern-1pt\dagger}}

\def\avg#1{{\kern1.0pt\overline{\kern-1.0pt#1\kern-1.0pt}\kern1.0pt}}

\begin{document}

\title{High precision renormalization of the flavour non-singlet 
       Noether currents in lattice QCD with Wilson quarks}

\author[address1]{Mattia Dalla Brida}

\author[address2]{Tomasz Korzec}
\author[address3]{Stefan Sint}
\author[address4]{Pol Vilaseca}

\address[address1]{Dipartimento di Fisica, Universit\`a di Milano-Bicocca
		    and INFN, sezione di Milano-Bicocca, \\ Piazza della Scienza 3,
		    I-20126 Milano, Italy}

\address[address2]{Department of Physics, Bergische Universit{\"a}t Wuppertal, 
		   Gau{\ss}str. 20, 42119 Wuppertal, Germany}
\address[address3]{School of Mathematics, Trinity College Dublin, Dublin 2, Ireland}
\address[address4]{Dipartimento di Fisica, Universit\`a La Sapienza di Roma 
		   and INFN, sezione di Roma, \\ Piazzale A. Moro 2, I-00185, Roma, Italy}

\begin{abstract}

We determine the non-perturbatively renormalized axial current for O($a$) improved lattice QCD
with Wilson quarks. Our strategy is based on the  chirally rotated Schr\"odinger functional 
and can be generalized to other finite (ratios of) renormalization constants which are traditionally
obtained by imposing continuum chiral Ward identities as normalization conditions.
Compared to the latter we achieve an error reduction by up to one order of magnitude.
Our results have already enabled the setting of the scale for the $\Nf=2+1$ CLS ensembles~\cite{Bruno:2016plf} 
and are thus an essential ingredient for the recent $\alpha_s$ determination by the ALPHA collaboration~\cite{Bruno:2017gxd}.
In this paper we shortly review the strategy and present our results for both $\Nf=2$ and $\Nf=3$ 
lattice QCD, where we match the $\beta$-values of the CLS gauge configurations. 
In addition to the axial current renormalization, we also present precise results for 
the renormalized local vector current. 
\end{abstract}


\maketitle

\section{Introduction}

Lattice regularizations with Wilson type fermions~\cite{Wilson:1974sk}
are widely used in current lattice QCD simulations~\cite{Lin:2008pr,Aoki:2009ix,
	Bietenholz:2010jr,Baron:2010bv,Fritzsch:2012wq,Borsanyi:2014jba,Bruno:2014jqa}.
The ultra-locality of the action enables numerical efficiency and thus
access to a wide range of lattice spacings and spatial volumes. Furthermore,
Wilson fermions maintain the full flavour symmetry of the continuum action, 
as well as the discrete symmetries such as parity, charge conjugation and time reversal. 
Unitarity is either realized exactly, or, in the case
of Symanzik-improved actions, approximately up to cutoff effects which 
vanish in the continuum limit.

The price to pay for these advantages consists in the explicit breaking of
all chiral symmetries by the Wilson term in the action. Well-known consequences
include the additive renormalization of quark masses, the mixing under renormalization
of composite operators in different chiral multiplets and discretization effects
linear in $a$, the lattice spacing. Furthermore, the Noether currents
of chiral symmetry are no longer protected against renormalization. 

The matrix elements of the axial Noether currents 
between pion or kaon states and the vacuum, parametrized by the decay constants $f_{\pi,K}$, e.g.
\begin{equation}
     \langle 0 \vert A_\mu^{ud}(0) \vert \pi^-, {\bf p}\rangle = i p_\mu f_\pi,\qquad 
     A_\mu^{ud}(x) = \psibar_{u}(x) \gamma_\mu\gamma_5 \psi_{d}(x),
\end{equation}
can be related to the measured life times of pions and kaons. The decay constants are finite in
the chiral limit, can be precisely measured in numerical simulations and are ideally suited
to set the scale in physical units. In order to achieve this with Wilson quarks one
needs to determine the correctly renormalized axial currents,
\begin{equation}
   \left(A_{\rm R}\right)_\mu^{f_1f_2}(x) = Z_{\rm A} A_\mu^{f_1f_2},\qquad 
\end{equation}
(with flavour indices $f_{1,2} = u,d,s$), which are to be inserted into the matrix elements.
Of course it is desirable that the error of the matrix elements is not dominated by
the uncertainty of the current normalization constant.

Over the last 30 years many efforts have been made to control the 
consequences of explicit chiral symmetry breaking with Wilson quarks. 
The main strategy consists in imposing continuum chiral symmetry relations as normalization conditions at 
finite lattice spacing~\cite{Bochicchio:1985xa,Luscher:1996jn}.  This is usually done using 
chiral Ward identities, which follow from an infinitesimal chiral change of variables in the QCD path integral. 
An example is the PCAC relation which determines the additive quark mass renormalization constant,
as the ``critical value'' of the bare mass parameter, where the axial current is conserved. 
The fact that chiral symmetry is fully recovered only in the continuum  limit implies that the 
choice of normalization condition matters at the cutoff level; at a fixed value of the lattice spacing 
the numerical results may occasionally differ substantially between any two such choices. 
Rather than interpreting this scatter as a systematic error, 
the modern approach consists in choosing a particular normalization condition and in fixing
all dimensionful parameters (such as momenta or distances or background fields) in terms of a physical scale. 
This defines a so-called ``line of constant physics'' (LCP), along which the continuum limit is taken.
As the lattice spacing $a$  (or, equivalently, the bare coupling, $g_0^2=6/\beta$), is varied, 
this defines a function $Z_{\rm A} = Z_{\rm A}(\beta)$.
Obviously, another choice for the LCP will result in a different function $Z'_{\rm A}(\beta)$.
However, their difference will be, within errors, a smooth function of $\beta$
which vanishes asymptotically $\propto a$ or $\propto a^2$ if O($a$) improvement is implemented. 
Hence, following a LCP ensures that cutoff effects are smooth functions of $\beta$ and
the choice of LCP becomes irrelevant in the continuum limit. Adopting this viewpoint, 
the relevant systematic error is therefore determined by the precision 
to which a chosen LCP can be followed.

In this paper we apply a recently developed method to lattice QCD with $\Nf=2$ and $\Nf=3$ flavours, matching
the lattice actions chosen by the CLS initiative~\cite{Fritzsch:2012wq,Bruno:2014jqa}.
Our method  is based on the chirally rotated Sch\"odinger functional ($\xSF$)~\cite{Sint:2005qz,Sint:2010eh}. 
The theoretical foundation of this framework has been explained in \cite{Sint:2010eh} and it has passed 
a number of perturbative and non-perturbative 
tests~\cite{Sint:2010xy,Lopez:2012as,Lopez:2012mc,Brida:2014zwa,Mainar:2016uwb,Brida:2016rmy}.
In contrast to the Ward identity method the axial current renormalization conditions follow
from a finite chiral rotation in the massless QCD path integral with Schr\"odinger functional (SF) 
boundary conditions. 
The renormalization constants are then obtained from ratios of simple 2-point functions. 
For the axial current, this represents a significant advantage over
the Ward identity method~\cite{Luscher:1996jn,DellaMorte:2005rd,Bulava:2016ktf} which involves 
3- and 4-point functions. Hence, we observe a dramatic improvement in the attainable statistical precision
for $Z_{\rm A}$ and some care is required to ensure that systematic errors are under control 
at a similar level of precision. We also discuss the normalization procedure for
the local vector current. While flavour symmetry remains unbroken on the lattice with (mass-degenerate) 
Wilson quarks, the corresponding Noether current lives on neighbouring lattice points connected 
by a gauge link, so that the use of the local vector current is often more practical. 

This paper is organized as follows: after a short reminder of the $\xSF$ correlation functions 
in the continuum and the normalization conditions derived from them in sect.~\ref{sec:xSF},
we define in sect.~\ref{sec:LCP} a couple of different LCPs which we have followed. 
We then present the $Z_{\rm A}$ and $Z_{\rm V}$ determinations for lattice QCD with $\Nf=2$ and 
$\Nf=3$ quark flavours in sects.~\ref{sec:Nf2} and \ref{sec:Nf3}, respectively, together with 
various tests we have carried out. Sect.~\ref{sec:Conclusions} contains a summary of the main 
results of this work and some concluding remarks. Finally, the paper ends with three technical 
appendices: appendix \ref{app:Simulations} collects the parameters and results of the simulations, 
appendix \ref{app:Syst} provides a detailed discussion on the systematic error estimates for
our determinations, and appendix \ref{app:Fits} gathers our set of chosen fit functions which
smoothly interpolate our $Z_{\rm A,V}$ results in $\beta$.

The main results for $\Nf=2$ are collected in table~\ref{tab:ZNf2}, while those for $\Nf=3$ are given
in tables \ref{tab:ZNf3L1} and \ref{tab:ZNf3L2}. These results can be directly applied to data obtained
from the CLS 2- and 3-flavour configurations, respectively~\cite{Fritzsch:2012wq,Bruno:2014jqa}. 
The $\Nf=3$ results have, in fact, already been used, and enabled the precision CLS scale setting 
in ref.~\cite{Bruno:2016plf} and the accurate quark-mass renormalization of ref.~\cite{Campos:2018ahf}.

\section{Renormalization conditions from universality relations}
\label{sec:xSF}

\subsection{The Schr\"odinger functional and chiral field rotations}
\label{subsec:xSF}

We start by considering massless two-flavour continuum QCD. 
The Euclidean space-time is taken to be a hyper-cylinder of volume $L^4$ with Schr\"odinger 
functional boundary conditions~\cite{Luscher:1992an,Sint:1993un}.
In particular, in the Euclidean time direction, the quark and anti-quark
fields satisfy,
\begin{equation}
  \label{eq:SFbc}
  P_+\psi(x)|_{x_0=0}= 0 =  \psibar(x)P_-|_{x_0=0},
\end{equation}
and similarly at time $x_0=L$ with the change $P_\pm \rightarrow P_\mp$.
The SU(2)$\times$SU(2) chiral and flavour symmetry leads to conserved
isovector Noether currents, given by
\begin{equation}
  A_\mu^a(x) = \psibar(x)\gamma_\mu\gamma_5\dfrac{\tau^a}{2} \psi(x),\qquad
  V_\mu^a(x) = \psibar(x)\gamma_\mu\dfrac{\tau^a}{2} \psi(x),
\end{equation}
with Pauli matrices $\tau^a$ and isospin index $a=1,2,3$. 
SF correlation functions of these currents with isovector boundary sources
$\Obs_5^a$ and $\Obs^a_k$ have been defined in \cite{Luscher:1996sc,Sint:1997jx} and are
given by
\begin{equation}
 \langle A^a_0(x) \mathcal{O}^{b}_5\rangle = -\delta^{ab} f_{\rm A}(x_0),\qquad
 \sum_{k=1}^3\langle V^a_k(x) \mathcal{O}^{b}_k\rangle = -3\,\delta^{ab} k_{\rm V}(x_0).
\end{equation}
Passing from the isospin notation to fields with definite flavour assignments,
\begin{equation}
 A^{f_1f_2}_\mu(x)=\psibar{}_{f_1}(x)\gamma_\mu\gamma_5\psi_{f_2}(x),\qquad
 V^{f_1f_2}_\mu(x)=\psibar{}_{f_1}(x)\gamma_\mu\psi_{f_2}(x),
\end{equation}
and similarly for the boundary sources, the correlation functions for isospin 
indices $a=1,2$, can be written in terms of the flavour off-diagonal fields,
\begin{equation}
 \label{eq:fA&kV}
 f_{\rm A}(x_0) = -{1\over2} 
 \langle A^{ud}_0(x) \mathcal{O}^{du}_5\rangle,\qquad
 k_{\rm V}(x_0) = -{1\over6} 
 \sum_{k=1}^3 \langle V_k^{ud}(x)\mathcal{O}^{du}_k\rangle\,.
\end{equation}
For the flavour diagonal fields in the isospin $a=3$ components, e.g.
\begin{equation}
  A_\mu^{3} = \frac12\left(A_\mu^{uu} - A_\mu^{dd}\right),
\end{equation}
one may use flavour symmetry to write
\begin{equation}
  f_{\rm A}(x_0) = -\frac12 
  \langle A^{uu'}_0(x) \mathcal{O}^{u'u}_5\rangle,\quad
\end{equation}
and analogously for $k_{\rm V}$. Note that the additional 
up-type flavour $u'$ is merely a notational device to indicate 
the fermionic contractions taken into account when applying Wick's theorem. 
Indeed, the sum of  all the disconnected contributions for the flavour diagonal 
$a=3$ components of SF correlation functions
cancels exactly due to flavour symmetry.

We now apply a flavour diagonal chiral rotation to the fields,
\begin{equation}
\label{eq:ChiralRotation}
  \psi\to \exp\left(i\dfrac{\alpha}{2}\gamma_5\tau^3\right)\psi,\qquad 
  \psibar\to\psibar \exp\left(i\dfrac{\alpha}{2}\gamma_5\tau^3\right).
\end{equation}
Choosing the rotation angle $\alpha=\pi/2$ then leads to the 
chirally rotated SF boundary conditions,
\begin{equation}
  \label{eq:chiSFbcs}
  \tilde{Q}_+\psi(x)|_{x_0=0}= 0 =  \psibar(x)\tilde{Q}_+|_{x_0=0},
\end{equation}
with projectors $\tilde{Q}_\pm = \frac12(1 \pm i\gamma_0\gamma_5\tau^3)$.
Analogous boundary conditions with reverted projectors are obtained at $x_0=L$.
Applying the same chiral field rotation to the axial currents,
\begin{equation}
  A^{ud}_\mu(x) \to -iV^{ud}_\mu(x),\qquad
  A^{uu}_\mu(x) \to A^{uu}_\mu(x),
\end{equation}
one obtains either a vector current or remains with
an axial current, depending on the flavour assignments. If the chiral rotation
of the field variables is performed as a change of variables in the functional integral,
one arrives at the formal continuum identities
\begin{equation}
 \label{eq:UniversalityRelations}
 f_{\rm A} = g^{uu'}_{\rm A} = -ig^{ud}_{\rm V},\qquad
 k_{\rm V} = l^{uu'}_{\rm V} = -il^{ud}_{\rm A},
\end{equation}
where the $g$- and $l$-functions are defined with $\chi$SF boundary
conditions, eqs.~(\ref{eq:chiSFbcs}), for instance
\begin{equation}
 g_{\rm A}^{f_1f_2}(x_0) = -\frac12
 \langle A^{f_1f_2}_0(x) \mathcal{Q}^{f_2f_1}_5\rangle_{({\tilde Q}_+)}\,.
\end{equation}
Here, the boundary operators $\mathcal{Q}^{f_1f_2}_5$ denote the 
chirally rotated versions of their SF counterparts, $\mathcal{O}^{f_1f_2}_5$.
For the complete expressions and further details we refer to ref.~\cite{Brida:2016rmy}.

Regarding the case of QCD with $\Nf=3$ quark flavours we note that the very
same steps can be taken provided the massless third quark does not take part
in the chiral rotation and thus remains with standard SF boundary conditions~\cite{Sint:2010eh}. 
Correlation functions are then considered for the doublet fields only, 
i.e.~the third quark never appears as a valence quark.

\subsection{Renormalization conditions} 
\label{subsec:RenCond}

In the lattice regularized theory with Wilson type quarks, 
relations such as (\ref{eq:UniversalityRelations}) can only be expected
to hold after renormalization and up to cutoff effects. One first has
to ensure that massless QCD with $\xSF$ boundary conditions has been correctly
regularized. This is achieved by tuning the bare mass parameter $m_0$ to its critical
value, $m_{\rm cr}$, where the axial current is conserved, and by tuning a boundary 
counterterm coefficient $z_f$ such that physical parity is restored (cf.~\cite{Brida:2016rmy} 
for more details). In terms of the bare $\chi$SF correlation functions 
one may choose the two conditions,
\begin{equation}
  \label{eq:mcrzf}
    m = {\tilde\partial_0 g_{\rm A}^{ud}(x_0)\over 2g_{\rm P}^{ud}(x_0)}\bigg|_{x_0=L/2} = 0, 
    \qquad
    g_{\rm A}^{ud}(L/2) = 0
\end{equation}
(with $\tilde{\partial}_0$ the symmetric lattice derivative). The division by
the pseudo-scalar correlation function $g_{\rm P}^{ud}$ is not really necessary, however
it is done for convenience, as
it gives rise to the definition of a (bare) PCAC quark mass $m$. Solutions
to these equations define $m_{\rm cr}$ and $z_f^*$ as functions of the bare coupling $g_0$, 
and the lattice size, $L/a$. 

Once the lattice regularization is correctly implemented, we expect e.g.
\begin{equation}
 Z_{\rm A} Z_{\zeta}^2 g^{uu'}_{\rm A}(x_0) =-i Z_{\rm V} Z_{\zeta}^2 g^{ud}_{\rm V}(x_0) +
 {\rm O}(a^2),
\end{equation}
where $Z_{\zeta}$ renormalizes a boundary quark or anti-quark
field~\cite{Sint:1995rb,Luscher:1996sc,Sint:2010eh} and $Z_{\rm A,V}$ are 
the current normalization constants of interest. Requiring such identities
to hold exactly at finite lattice spacing thus fixes the relative normalization
of axial and vector current. Replacing the latter by the exactly conserved lattice vector current
${\widetilde V}_\mu(x)$~(cf.~ref.~\cite{Brida:2016rmy}), for which  $Z_{\rm \widetilde V}=1$,
one may obtain $Z_{\rm A}$ from either one of the ratios
\begin{equation}
 \label{eq:RA}
 R^g_{\rm A} = {-ig_{\rm \widetilde V}^{ud}(x_0)\over
 \phantom{-i} g_{\rm A}^{uu'}(x_0)}\bigg|_{x_0=L/2}\qquad
 \text{or}\qquad
 R^l_{\rm A} = {il_{\rm \widetilde V}^{uu'}(x_0)\over
 \phantom{i} l_{\rm A}^{ud}(x_0)}\bigg|_{x_0=L/2}.
\end{equation}
Assuming that the parameters  $x_0$ (here set to $L/2$), the boundary angle $\theta$~\cite{Sint:1995ch}, 
the background gauge field~\cite{Luscher:1992an}, 
and the precise definition for the zero mass and $\alpha=\pi/2$ point~(\ref{eq:mcrzf})
are fixed, we define, on an $(L/a)^4$ lattice and for 
a given bare coupling $g_0^2=6/\beta$, 
\begin{equation}
  Z^{g,l}_{\rm A}(\beta,L/a) = R^{g,l}_{\rm A}.
\end{equation}
Finally the choice of a line of constant physics (cf.~section~3) 
defines a smooth function $(L/a)(\beta)$ such that the normalization constants 
become functions of $\beta$ alone, with the difference between any 
two definitions vanishing smoothly with a rate $\propto a^2$.

We also comment on the appearance of a second up-type flavour $u'$ in (\ref{eq:RA}). 
When applying the chiral rotation (\ref{eq:ChiralRotation}) to the diagonal components of 
$f_{\rm A}$, the disconnected diagrams are mapped to disconnected diagrams on the
$\chi$SF side which can be shown to add up to a pure cutoff effect. 
Their omission is thus perfectly legitimate, 
even if the formulation of the renormalization conditions then has an element of 
partial quenching to it. The situation is comparable with the Ward identity
method in two-flavour QCD~\cite{Luscher:1996jn,DellaMorte:2005rd}, where a fictitious 
$s$-quark can be introduced to eliminate the disconnected diagrams.

Even though there exists a conserved vector current, in practice
the local current is often used and then requires renormalization, too.
Its renormalization constant can be obtained from,
\begin{equation}
 \label{eq:RV}
 R^g_{\rm V} = {g_{\rm \widetilde V}^{ud}(x_0)\over 
 g_{\rm V}^{ud}(x_0)}\bigg|_{x_0=L/2}\qquad
 \text{or}\qquad
 R^l_{\rm V} = {l_{\rm \widetilde V}^{uu'}(x_0)\over
 l_{\rm V}^{uu'}(x_0)}\bigg|_{x_0=L/2}.
\end{equation}
The same remarks as for the axial current normalization apply here, and with definite choices
for all parameters we set,
\begin{equation}
  \label{eq:RenomalizationConditions}
  Z^{g,l}_{\rm V}(\beta,L/a) = R^{g,l}_{\rm V}.
\end{equation}
As in the case of the axial current normalization conditions, only 2-point functions are required, 
which connect the boundary quark bilinear sources with the currents in the bulk. This is
a major advantage over the Ward identity method~\cite{Luscher:1996jn,DellaMorte:2005rd}
where 3- and 4-point functions are required. Hence, one expects better statistical precision from
the simpler 2-point functions, and this will be confirmed below.
Furthermore, as discussed in \cite{Brida:2016rmy}, the cutoff effects in 
the ratios are O($a^2$), due to the mechanism of automatic O($a$) improvement~\cite{Frezzotti:2003ni},
even if the PCAC mass and the axial current are not O($a$) improved by the counterterm 
$\propto c_{\rm A}$~\cite{Luscher:1996sc}, or if the vector currents 
are not improved  by the corresponding counterterms 
$\propto c_{\rm V}, c_{\rm \widetilde{V}}$~\cite{Sint:1997jx,Frezzotti:2001ea}.

Finally, we emphasize that similar renormalization conditions can be devised for
other finite renormalization constants. An interesting example is 
the ratio $Z_{\rm P}/Z_{\rm S}$, where $Z_{\rm P}$ and $Z_{\rm S}$ are the pseudo-scalar
and scalar renormalization constant, respectively. We refer the reader to 
ref.~\cite{Brida:2016rmy} for more details.

\section{Lines of constant physics and choice of renormalization conditions}
\label{sec:LCP}

\subsection{General considerations}

A line of constant physics requires to specify a physical (length) scale $r$ which is
kept fixed as the continuum limit is taken. A typical choice would be the pion decay constant, $r=1/f_\pi$, either 
at the physical quark masses or in the chiral limit.  Once calculated  for a range of lattice spacings,
this scale defines a function $(r/a)(\beta)$ of the bare coupling $\beta=6/g_0^2$ which fixes the lattice spacing $a$
in units of the chosen physical scale. Choosing the spatial lattice extent $L/a$, at a given beta, such that
\begin{equation}
\dfrac{(L/a)(\beta)}{(r/a)(\beta)} = L/r = C_r
\label{eq:LCPgeneric}
\end{equation}
(with a numerical constant $C_r$) then fixes the spatial size of the finite volume system in units of $r$.
In practice we will choose $C_r$ such that the physical size of $L$ will be somewhat larger than half a femto metre.
Note that this equation can be read in two ways: first, if one fixes $C_r$ and then chooses a set of $\beta$-values for
which $r/a$ is known, one obtains a corresponding set of values $(L/a)(\beta)$, which 
will not necessarily be integers. To evaluate the normalization constants at these non-integer lattice sizes then 
requires some interpolation of results from neighbouring integer $L/a$-values at the same $\beta$. Alternatively, 
one could choose a set of integer $L/a$-values such that a choice for $C_r$
will imply a set of $\beta$-values. In general this means that the data for $r/a$ may have to be interpolated in $\beta$.
We will here choose the first option, with the set of $\beta$-values taken over from the large volume
simulations by the CLS project~\cite{Fritzsch:2012wq,Bruno:2014jqa}.

Having set the scale one needs to ensure the correlation functions are calculated
in the desired situation of massless QCD and for the chosen chirally rotated boundary conditions at $\alpha=\pi/2$.
This means one needs to tune the bare quark mass $am_0$ and $z_f$ as functions of $\beta$.
We will discuss this in more detail below. Finally, the correlation functions depend on kinematic parameters, 
such as $x_0$ or background field parameters such as $\theta$. 
We have already set $x_0=L/2$ in eqs.~(\ref{eq:RA},\ref{eq:RV}) and we choose
$\theta=0$ and work with vanishing SU(3) background field.

With these parameter choices we will have, for a given $r$ and $C_r$ in eq.~(\ref{eq:LCPgeneric}),
two definitions each for $Z_{\rm A}$ and $Z_{\rm V}$, namely 
\begin{equation}
   Z^{g,l}_{\rm A,V}(\beta) = R_{\rm A,V}^{g,l}(\beta,a/L)\big\vert_{L/r=C_r;\, m=0;\, \alpha=\pi/2}\,,
\end{equation}
either based on the $g$- or the $l$-ratios.
We then expect e.g.~that
\begin{equation}
   Z_{\rm A}^g(\beta) = Z_{\rm A}^l(\beta) +  \rmO(a^2),
   \label{eq:g-vs-l}
\end{equation}
where the $a^2$-effects are now expected to be smooth functions of the bare coupling.

\subsection{Perturbative subtraction of cutoff effects}
\label{sec:PTImprovement}

A possible refinement consists in using perturbation theory to reduce the cutoff effects perturbatively. 
This requires to compute the $R$-ratios (\ref{eq:RA},\ref{eq:RV}) perturbatively, with the exact 
same parameter choices as in the numerical simulations. We have performed this calculation to 1-loop order,
\begin{equation}
   R_{\rm A,V}^{g,l}(g_0^2,a/L)  = R_{\rm A,V}^{g,l(0)}(a/L) + g_0^2 R_{\rm A,V}^{g,l(1)}(a/L) + {\rm O}(g_0^4),
\end{equation}
and for the chosen parameters we always find $R_{\rm A,V}^{g,l(0)}(a/L)=1$, exactly. We may then define 
a 1-loop correction factor,
\begin{equation}
  r^{g,l}_{\rm A,V}(\beta,L/a) = \dfrac{1+g_0^2 R_{\rm A,V}^{g,l(1)}(0)}{1+g_0^2 R_{\rm A,V}^{g,l(1)}(a/L)}\,,
  \label{eq:rpert}
\end{equation}    
and results for the coefficients $R_{\rm A,V}^{g,l(1)}$ are collected 
in table~\ref{tab:RAV1}, for the relevant lattice resolutions $L/a$ and the two lattice gauge actions
used by CLS. Note that the 1-loop results are $\Nf$-independent and are thus obtained along the lines
of ref.~\cite{Brida:2016rmy}, the only difference being the form of the free gluon propagator in the 
case of the L\"uscher-Weisz gauge action~\cite{Aoki:1998qd}.
As an aside we note that our results converge to the known 1-loop results $Z_{\rm A,V}^{(1)}$ for an infinitely extended
lattice~\cite{Gabrielli:1990us,Gockeler:1996gu,Aoki:1998ar}, i.e.~for $a/L=0$. We also observe that the 1-loop 
cutoff effects for the $l$-definitions are generally much smaller than for the $g$-definitions.

\begin{table}[h]
\centering
\begin{tabular} {ccccc}
\toprule
\multicolumn{5}{c}{\bf Wilson gauge action}\\
\midrule
  $L/a$ 
  & $ R_{\rm A}^{g(1)}(a/L)$ 
  & $ R_{\rm A}^{l(1)}(a/L)$ 
  & $ R_{\rm V}^{g(1)}(a/L)$  
  & $ R_{\rm V}^{l(1)}(a/L)$  \\
\midrule
 6 & $  -0.104309 $ & $ -0.116808 $ & $ -0.118728 $ & $  -0.130549  $ \\ 
 8 & $  -0.109076 $ & $ -0.116640 $ & $ -0.122586 $ & $  -0.129838  $ \\ 
10 & $  -0.111857 $ & $ -0.116595 $ & $ -0.125088 $ & $  -0.129662  $ \\ 
12 & $  -0.113308 $ & $ -0.116564 $ & $ -0.126426 $ & $  -0.129588  $ \\ 
16 & $  -0.114714 $ & $ -0.116526 $ & $ -0.127747 $ & $  -0.129519  $ \\ 
\midrule
$\infty$ & \multicolumn{2}{c} {$Z_{\rm A}^{(1)}= -0.116458(2)$} & 
\multicolumn{2}{c}{$Z_{\rm V}^{(1)}= -0.129430(2)$}\\
\bottomrule
\toprule
\multicolumn{5}{c}{\bf L\"uscher-Weisz gauge action} \\
\midrule
  $L/a$ 
  & $ R_{\rm A}^{g(1)}(a/L)$ 
  & $ R_{\rm A}^{l(1)}(a/L)$ 
  & $ R_{\rm V}^{g(1)}(a/L)$  
  & $ R_{\rm V}^{l(1)}(a/L)$  \\
\midrule
 6 & $  -0.078368 $ & $ -0.091011 $ & $ -0.089760 $ & $  -0.101750  $ \\ 
 8 & $  -0.083286 $ & $ -0.090737 $ & $ -0.093819 $ & $  -0.101006  $ \\ 
10 & $  -0.085958 $ & $ -0.090650 $ & $ -0.096256 $ & $  -0.100811  $ \\ 
12 & $  -0.087374 $ & $ -0.090604 $ & $ -0.097578 $ & $  -0.100730  $ \\ 
16 & $  -0.088756 $ & $ -0.090557 $ & $ -0.098889 $ & $  -0.100657  $ \\ 
\midrule
$\infty$ & \multicolumn{2}{c} {$Z_{\rm A}^{(1)}= -0.090488(5)$} & \multicolumn{2}{c}{$Z_{\rm V}^{(1)}= -0.100567(2)$}\\
\bottomrule
\end{tabular}
\caption{Finite $L/a$ estimators for the current normalization constants at 1-loop order,
and our estimates for their asymptotic values; the latter agree with previous results in the 
literature~\cite{Gabrielli:1990us,Gockeler:1996gu,Aoki:1998ar}. All results are given for $\SUthree$.}
\label{tab:RAV1}
\end{table}

For given $L/a$ and $\beta$, the perturbatively improved current normalization constants are now defined by
\begin{equation}
   \label{eq:PThIZ}
   Z_{\rm A,V,\,sub}^{g,l}(\beta,L/a) = r_{\rm A,V}^{g,l}(\beta,L/a)\times  Z_{{\rm A,V}}^{g,l}(\beta,L/a),
\end{equation}
and, by construction, the O($a^2$) cutoff effects are subtracted to O($g_0^2$), reducing them
to O($a^2g_0^4$). The subtracted data for the $Z$-factors are then treated as before: 
a choice of a line of constant physics implies a set of $\beta$- and corresponding $L/a$-values 
to which the data must be interpolated.  We will see evidence for the effectiveness 
of this perturbative subtraction of cutoff effects in Sect.~\ref{sec:Nf2} and \ref{sec:Nf3}.

\subsection{Choices of LCP for $\Nf=2$ and $\Nf=3$}

In order to fix the physical scale $r$, we choose either the kaon decay 
constant $r=1/f_K$ ($\Nf=2$), or the gradient flow scale $r=\sqrt{8t_0}$ 
($\Nf=3$)~\cite{Luscher:2010iy}.%
\footnote{The choice of the scale from $f_K$ seems somewhat circular, as its measurement requires 
	the correctly normalized axial current. We use the results from ref.~\cite{Fritzsch:2012wq} which
	were obtained  using $Z_{\rm A}$ from a standard SF Ward identity determination.}
In order to fix the respective constants $C_r$ we proceed as follows. Given
the set of values $\beta_i$ for $i=1,2,\ldots$ (taken from CLS),
we choose as a reference value $\beta_{\rm ref}$ either the largest or
the smallest of the set. Choosing an integer lattice size $L/a$ at the
reference point $\beta_{\rm ref}$ now fixes $C_r$ through
\begin{equation}
   C_r = \dfrac{(L/a)(\beta_{\rm ref})}{(r/a)(\beta_{\rm ref})}\,.
\end{equation}
Having set the scale in this way, the $L/a$-values at the remaining
$\beta_i$ follow from eq.~(\ref{eq:LCPgeneric}).
For all our choices the physical size of our space-time extent will be 
$L \approx 0.6-0.7\, {\rm fm}$. As mentioned before, except at the chosen 
reference value for $\beta$ this requires interpolations of simulation results at 
integer $L/a$ and our current simulation code, which is based on the openQCD 
package~\cite{Luscher:2012av,LuscherWeb:2016}, requires that $L/a$ is also even. 

\subsection{Topology freezing}
\label{sec:Topology}

Numerical simulations of the SF by means of standard Monte Carlo algorithms are 
known to suffer from the topology freezing problem (see e.g. ref.~\cite{Luscher:2014kea} 
for a discussion). A possible solution is to follow the proposal of ref.~\cite{Luscher:2014kea}
and simulate the theory with open-SF boundary conditions. However, if for the given choice of 
parameters the problem is ``mild'', one can circumvent the issue in a straightforward manner
by simply imposing the renormalization conditions (\ref{eq:RenomalizationConditions}) \emph{and}
(\ref{eq:mcrzf}) within the trivial topological  sector~\cite{Fritzsch:2013yxa,DallaBrida:2016kgh}.
In a continuum notation, the correlation functions entering these definitions are modified
as follows,
\begin{equation}
g_{\rm A}^{ud}(x_0)\,
\to
\,
g_{\rm A,Q}^{ud}(x_0) =  
{-{1\over2}\langle A^{ud}_0(x) \mathcal{Q}^{du}_5  \delta_{Q,0}\rangle_{({\tilde Q}_+)}
	\over 
	\langle\delta_{Q,0}\rangle_{({\tilde Q}_+)}}, 
\end{equation}
and analogously in all other cases.%
\footnote{For ease of notation, in the following the subscript $Q$ is implicitly 
	understood, and we assume that all relevant correlation	functions are restricted
	to the $Q=0$ sector.}
Here, the Kronecker $\delta$ in the functional integral selects the gauge field
configurations with topological charge $Q=0$. Since relations based on chiral 
flavour symmetries should hold separately in each topological charge sector, 
this restriction to the trivial sector is a legitimate modification of the current
renormalization conditions. It provides a viable solution to the algorithmic problem of 
topology freezing in cases where this problem becomes marginally relevant; this means
when the fraction of topologically non-trivial gauge field configurations in the 
relevant ensembles is not too large. For our choices of parameters, the percentage
of gauge field configurations with $Q\neq0$ is generally below 10\%, and reaches 
approximately 30\% only in a couple of cases (cf. tables \ref{tab:Nf2Parms} and \ref{tab:Nf3Parms}).

On the lattice the topological charge is not unambiguously defined. We follow 
refs.~\cite{Fritzsch:2013yxa,DallaBrida:2016kgh} and define the trivial
topological sector as the set of gauge field configurations for which $|Q|<0.5$, 
where $Q$ is discretized in terms of the Wilson flow and the clover definition of 
the field strength tensor~\cite{Luscher:2010iy}. The flow time $t$ is then kept 
fixed in physical units by requiring $\sqrt{8t}=0.6\times L$.

\subsection{On the tuning of $am_0$ and $z_f$}
\label{sec:TuningExample}

The current normalization conditions require the $\chi$SF correlation functions
at zero quark mass and with a chiral twist angle of $\pi/2$. In practice this is achieved
by the simultaneous tuning of $m_0$ and $z_f$ such that eqs.~(\ref{eq:mcrzf})
are satisfied. In general a 2-parameter tuning can be quite involved.
However, here the non-perturbative O($a$) improvement of the action implies
that the O($a$) uncertainty of the zero mass point is very much reduced.
Since a change in $z_f$ merely re-defines the matrix element used to
define the PCAC mass, a variation of $z_f$ is expected to induce a small
variation of $m$ within this O($a$) uncertainty. The latter could in principle be
reduced to O($a^2$) by including the $c_{\rm A}$-counterterm to the axial current,
but this will not be pursued here. Another important observation is that, once
$m_0$ and $z_f$ are within O($a$) of their target values,  the sensitivity of the PCAC mass 
to a variation of $z_f$ is reduced to order $a^2$ (cf.~appendix~\ref{app:Syst}, discussion 
after eq.~(\ref{eq:dmLdzf})). One is therefore led to conclude that the PCAC mass
$m$ is to a good approximation independent of $z_f$, 
and the tuning of $m_0$ and $z_f$ thus becomes straightforward; 
given a reasonable guess for $z_f$, one  can first tune $m_0$,
and then turn to $z_f$.

\begin{figure}[hpbt]
 \centering
 \includegraphics[scale=0.95]{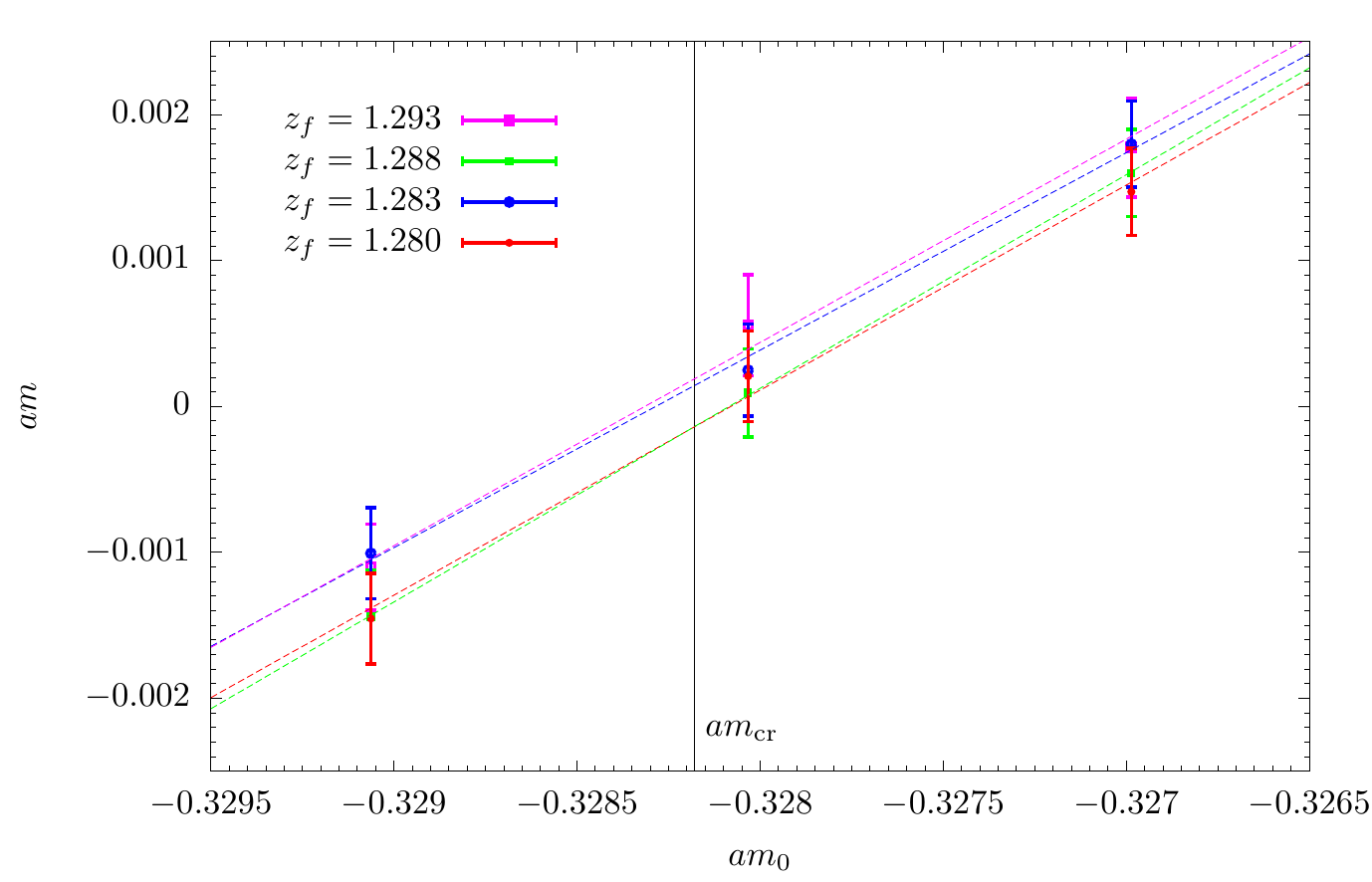}
 \caption{Results for the PCAC mass as a function of the bare quark mass,
  for different values of $z_f$. The dashed lines are linear fits to the data,
  while the solid vertical line indicates the location of our final estimate for
  $am_{\rm cr}(g_0,a/L)$ (s.~main text). The results are for $L/a=8$ and $\beta=5.3$.}
 \label{fig:mPCAC}
\end{figure}

As an illustration of this situation we discuss the $\Nf=2$ case for $L/a=8$, 
$\beta=5.3$. For the tuning we considered 3 values of $\kappa =1/(2am_0+8)$ and 
4 values of $z_f$. We then generated around 2000 gauge field configurations 
separated by 10 MDUs for each of the 12 ensembles, and measured the relevant 
correlation functions. Figure \ref{fig:mPCAC} collects the results for the PCAC mass
as a function of the bare quark mass, for the 4 different values of $z_f$. Within 
statistical errors, the PCAC mass depends linearly on $m_0$ and is essentially independent
of $z_f$. A linear fit of $m$ vs.~$m_0$ yields an estimate of $m_0=m_{\rm cr}(g_0,L/a)$
for which $m$ vanishes: these are collected in table \ref{tab:m0c}. The results are 
perfectly compatible with each other, and we take as our estimate for $m_{\rm cr}$
the result of a weighted average of these four.

\begin{table}[hpbt]
\centering
\begin{tabular}{lll}
\toprule
  $z_f$ & $am_{\rm cr}$ & $\kappa_{\rm cr}$ \\
\midrule
$1.280$ & $-0.32808(13)$ & $0.1361685(47)$ \\
$1.283$ & $-0.32828(14)$ & $0.1361761(51)$ \\                                             
$1.288$ & $-0.32808(12)$ & $0.1361687(45)$ \\
$1.293$ & $-0.32831(14)$ & $0.1361772(51)$  \\
\midrule
average & $-0.328179(65)$ & $0.1361722(24)$ \\
\bottomrule
\end{tabular}
\caption{Results for $am_{\rm cr}(g_0,L/a)$ for four different values of $z_f$,
	 for $L/a=8$ and $\beta=5.3$. The weighted average of the results is also
	 given in the last row of the table.}
\label{tab:m0c}
\end{table}

\begin{figure}[hpbt]
 \centering
 \includegraphics[scale=0.95]{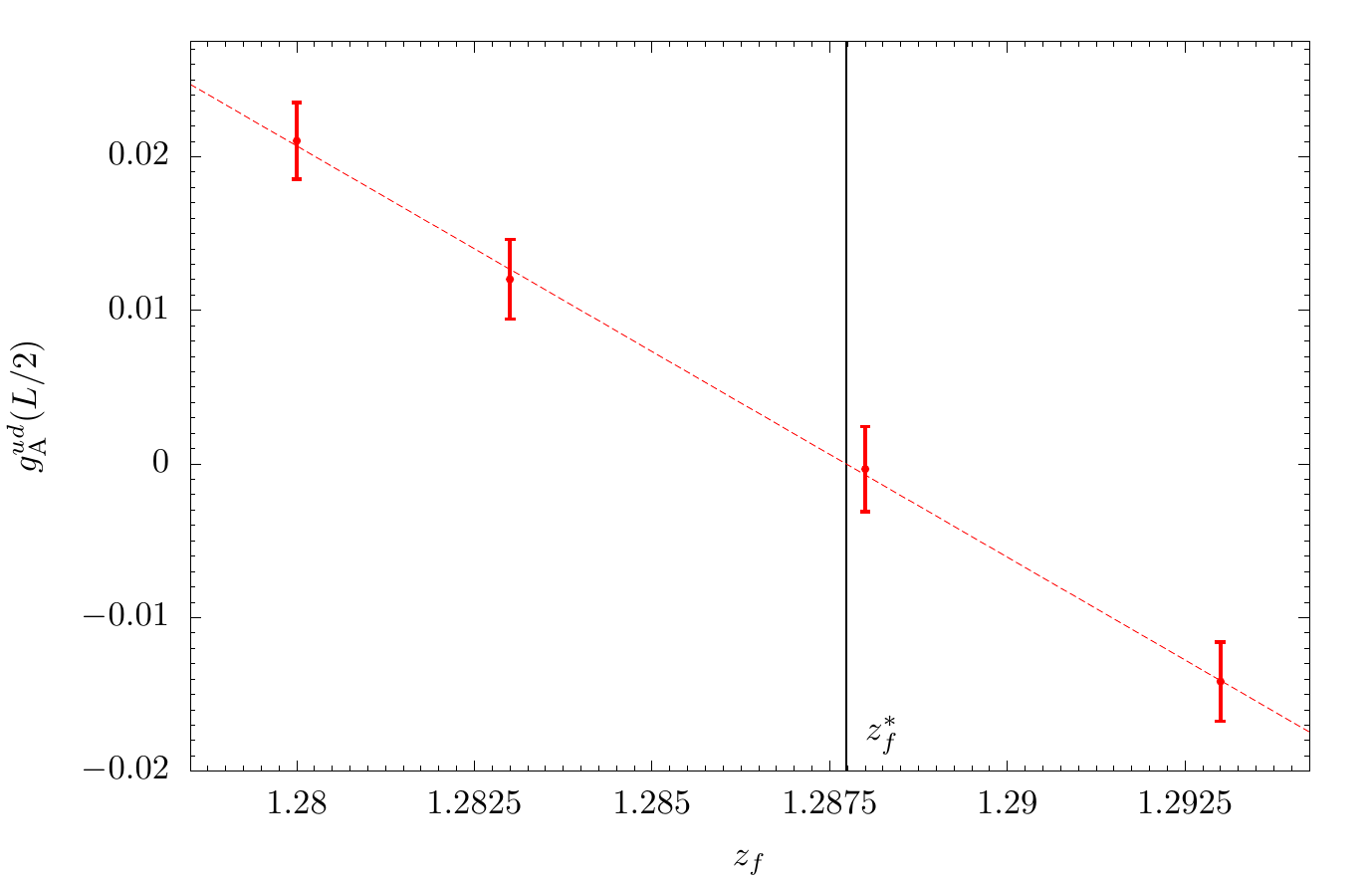}
 \caption{Results for $g_{\rm A}^{ud}(L/2)$ as a function of $z_f$.
 	The dashed line is a linear fit to the data, while the solid vertical
 	line indicates the location of our final estimate for $z_f^*$ (s.~main text). 
 	The values of $g_{\rm A}^{ud}(L/2)$ come from an interpolation 
 	to $\kappa=0.1361722$, and are for $L/a=8$ and $\beta=5.3$.}
 \label{fig:gAud}
\end{figure}

Once the critical bare mass is fixed, a smooth interpolation of $g_{\rm A}^{ud}(L/2)$ in $m_0$ 
gives the results shown in figure \ref{fig:gAud}. Over the chosen range, 
$g_{\rm A}^{ud}(L/2)$ so interpolated is perfectly linear in $z_f$, and it is thus
straightforward to determine the point $z_f^*$ where $g_{\rm A}^{ud}(L/2)$ vanishes i.e. 
$z^*_f= 1.2877(5)$ in this example.

The estimated values of $am_{\rm cr}$ and $z_f^*$ determined in this way turn out to be quite 
accurate in practice, cf.~table \ref{tab:Nf2Parms}.%
\footnote{Note that the $L/a=8$, $\beta=5.3$, simulations listed in 
 table \ref{tab:Nf2Parms}, use slightly different values for $am_{\rm cr}$ and 
 $z_f^*$ from a previous, less precise determination.}
We remark that results for $m_{\rm cr}$ could also be taken from a different source, for instance
from standard SF simulations. In this case  only $z_f$ needs to be tuned. The differences to the
above procedure would be O($a$) both in $m_{\rm cr}$  and in $z_f^*$ which, by the mechanism of 
automatic O($a$) improvement, induce O($a^2$) differences in observables such as the current 
normalization constants~\cite{Sint:2010eh,Brida:2016rmy}. One also expects that a precise tuning 
of $m_0$ is less crucial in the $\xSF$ than in the SF; the quark mass dependence 
of physical observables around the chiral limit is quadratic rather than linear~\cite{Sint:2007ug}.

\subsection{Sources of uncertainties}
\label{sec:Uncertainties}

Besides statistical errors directly affecting the estimators for the current normalization 
constants, the other source of uncertainty originates from the precision 
to which a line of constant physics can be followed. 
In principle also this latter effect is of a statistical nature, however, some
elements of modelling or estimates may be involved when propagating these errors to 
the normalization constants, so that it is partly justified labelling these
effects as systematic.

Our procedure consists of the following steps:

\begin{enumerate}
\item The LCP together with the set of values $\beta_i$ translates to 
target values $(L/a)(\beta_i)$. At each $\beta_i$ we 
choose lattices with even $L/a$  straddling the target values. 
We here anticipate that with our choices of LCPs 
the required lattice sizes are in the range $L/a=8$ to $L/a=16$. 
Note that all target values $(L/a)(\beta_i)$ come with statistical errors except for
$\beta=\beta_{\rm ref}$, where, by definition,  $L/a$ is given as an (even) integer.

\item For given $\beta$ and $L/a$ we determine the solutions $am_{0} = am_{\rm cr}$ and $z_f=z_f^*$ 
of eqs.~(\ref{eq:mcrzf}). In order to find their statistical errors which follow from 
the statistical uncertainties on $m$ and $g_{\rm A}^{ud}(L/2)$, we use estimates for the relevant
derivatives,
\begin{equation}
 \label{eq:m0zf2mgA}
 {\partial mL\over\partial m_0L},\quad
 {\partial mL\over\partial z_f},\quad
 {\partial g_{\rm A}^{ud}\over\partial m_0L},\quad
 {\partial g_{\rm A}^{ud}\over\partial z_f}.
\end{equation}

\item We then determine the induced error on the $Z$-factors by estimating their 
derivatives with respect to the bare parameters,
\begin{equation}
  \label{eq:dZdX}
  {\partial Z_{\rm A,V}\over\partial z_f},\quad
  {\partial Z_{\rm A,V}\over\partial m_0L}.
\end{equation}
It turns out that the derivatives (\ref{eq:m0zf2mgA}) and (\ref{eq:dZdX})
scale quite well with lattice size and lattice spacing, so that it is unnecessary
to evaluate them for all parameter choices. Some cross checks are sufficient. 
The errors coming from the uncertainties in $m_0$ and $z_f$ are then combined
in quadrature and added, again in quadrature, to the statistical error.

\item Where necessary, the results for $Z_{\rm A,V}$ at the different 
$L/a$-values and fixed $\beta_i$ are interpolated to the target $(L/a)(\beta_i)$; 
and the statistical error on $(L/a)(\beta_i)$ is propagated at this point. 
In the case where only one value of $L/a$ has been simulated, an estimate for the derivative
\begin{equation}
  \label{eq:dZdL}
  {\partial Z_{\rm A,V}\over\partial (L/a)}
\end{equation}
is used to assign a systematic error due to the difference 
$\Delta (L/a)\equiv L/a-(L/a)(\beta_i)$, also taking into account
the statistical uncertainty on $(L/a)(\beta_i)$. The resulting systematic
error is again added in quadrature.
\end{enumerate}
We emphasize that all systematic effects become essentially 
statistical errors provided enough data is produced to estimate 
the derivatives required to propagate the errors to the normalization constants.
In the following two sections we will present the lattice set-up and 
results for $\Nf=2$ and $\Nf=3$ lattice QCD. We will also come back to 
some of the above points.

\section{Numerical results for $\Nf=2$ flavours}
\label{sec:Nf2}

\subsection{Lattice set-up and parameter choices}
\label{sec:SetupNf2}

The CLS large volume simulations of 2-flavour QCD~\cite{Fritzsch:2012wq}  were performed using 
non-pertur\-batively O($a$) improved Wilson quarks and the Wilson gauge action. 
The matching to CLS data via the bare coupling requires that we use the same 
action in the $\chi$SF. As for the details of the action near the time boundaries 
we refer to ref.~\cite{Brida:2016rmy}. In particular the counterterm coefficients 
$c_{\rm t}(g_0)$ and $d_s(g_0)$ were set to their perturbative one-loop values using the results of that reference. 
In general, the incomplete cancellation of boundary O($a$) artefacts implies 
some remnant O($a$) effects in observables. However, for the estimators of the current normalization
constants, eqs.~(\ref{eq:RA},{\ref{eq:RV}), it can be shown
that such O($a$) effects only cause O($a^2$) differences~\cite{Brida:2016rmy}.

The CLS simulations were carried out for 3 values 
of the lattice spacing~\cite{Fritzsch:2012wq}, corresponding to the $\beta$-values $5.2$, $5.3$ and $5.5$.
For future applications we have added a finer lattice spacing corresponding to $\beta=5.7$.
We choose the smallest CLS-value $\beta=5.2$  as reference value and set 
\begin{equation}
   L/a=8
   \quad 
   {\rm at}
   \quad
   \beta=5.2,
\end{equation}
to define the starting point for the line of constant physics.
We then fix the space-time volume of the $\xSF$ simulations in terms
of the kaon decay constant, $f_K$, evaluated at physical quark masses. 
Taking $af_K$ from table~\ref{tab:LCPNf2} at $\beta=5.2$ yields
\begin{equation}
  \label{eq:fKL}
  f_KL = 0.4744(74),
\end{equation}
and corresponds to $L\approx0.6\,\fm$. Imposing this condition at the other $\beta$-values 
then leads to the (non-integer) $L/a$-values given in~table~\ref{tab:LCPNf2}. 
The quoted errors are a combination of statistical uncertainties, propagated
from eq.~(\ref{eq:fKL}) and the error on $af_K$ at the given $\beta$'s.

\begin{table}[hpbt]
  \centering
  \begin{threeparttable}
  \begin{tabular}{llll}
  \toprule
  $\beta$ & $af_K$ & $(L/a)(\beta)$  & $L/a$      \\
  \midrule                                                                                   
  5.2  & 0.0593(7)(6)  & 8         & 8 	 \\
  5.3  & 0.0517(6)(6)  & 9.18(21)  & 8, 10, 12 \\
  5.5  & 0.0382(4)(3)  & 12.42(25) & 12 	 \\
  5.7  & 0.0290(11)${}^\dag$ & 16.35(67) & 16 	 \\
  \bottomrule
  \end{tabular}
  \begin{tablenotes}
  \footnotesize 
  \item${}^\dag$This value is estimated using
  the perturbative running of the lattice 
  spacing (s. main text).
  \end{tablenotes}
  \end{threeparttable}
  \caption{Values of $af_K$ used to determine $(L/a)(\beta)$
  such as to satisfy the condition (\ref{eq:fKL}) for the given $\beta$. 
  The $\xSF$ simulations were performed at the neighbouring even integer $L/a$-values  
  given in the last column.}
  \label{tab:LCPNf2}
\end{table}

While the first 3 results for $af_K$ in table \ref{tab:LCPNf2} have been
directly measured~\cite{Fritzsch:2012wq} we have estimated $af_K$ at
the fourth value, $\beta=5.7$, as follows: with $af_K$ at $\beta=5.5$ taken 
as starting point  we used the three-loop $\beta$-function for the bare 
coupling~\cite{Bode:2001uz}, in order to determine the ratio of lattice spacings.
The error is obtained by summing (in quadrature) the statistical error propagated
from the result at $\beta=5.5$, and a systematic error due to the use of perturbation
theory. The latter is estimated as the difference between the non-perturbative 
result for $af_K$ at $\beta=5.5$, and the same perturbative procedure, applied
between $\beta=5.3$ and $\beta=5.5$. This systematic error is about 2.7
times larger than the statistical one, and thus dominates the error on $L/a$ at $\beta=5.7$.

Except for $\beta = 5.3$, the target values $(L/a)(\beta_i)$ resulting from condition~(\ref{eq:fKL}), 
are very close to even integer values of $L/a$, so that interpolations between simulations
at different $L/a$ can be avoided. At $\beta=5.3$ we simulated at the three $L/a$-values
given in the last column of table \ref{tab:LCPNf2} and interpolated to the target value
(see appendix \ref{app:InterpolationsNf2} for more details).
For each choice of $\beta$ and $L/a$, along the lines of the discussion in Sect.~\ref{sec:TuningExample}, 
we have carried out various tuning runs covering a range of $am_0$ and $z_f$, so as to determine 
the parameters satisfying the conditions (\ref{eq:mcrzf}). The values of the tuned parameters 
and the results for $m$ and $g_{\rm A}^{ud}(L/2)$ at these parameters are given in table \ref{tab:Nf2Parms}.

\subsection{Results and error budget}

In table \ref{tab:ZNf2} we collect the results for $Z_{\rm A,V}$, both $g$ and
$l$ definitions, at the four values of the lattice spacing. The statistics
range from $1,800$ to $12,000$ measurements depending on the 
ensemble, cf. table~\ref{tab:Nf2Parms}. The quoted uncertainties combine the statistical and
systematic errors. The statistical errors are at the level of $0.1-0.4\permil$, depending on
the $Z$-factor and ensemble considered. Hence a significant contribution to the error comes
from systematic uncertainties.

\begin{table}[hpbt]
\centering
\begin{tabular}{lllll}
\toprule
 $\beta$ & $Z_{\rm A}^g$ & $Z_{\rm A}^l$ & $Z_{\rm V}^g$ & $Z_{\rm V}^l$ \\
\midrule
$5.2$ & $ 0.78022(55)$ & $ 0.76944(94)$ & $ 0.74673(47)$ & $ 0.73849(97)$ \\
$5.3$ & $ 0.78411(61)$ & $ 0.77576(66)$ & $ 0.75220(70)$ & $ 0.74607(69)$ \\
$5.5$ & $ 0.7945(13)$ & $ 0.7895(13)$ & $ 0.7663(14)$ & $ 0.7625(14)$ \\
$5.7$ & $ 0.80526(97)$ & $ 0.80277(93)$ & $ 0.7800(11)$ & $ 0.77801(98)$ \\
\midrule
$\beta$ & $Z_{\rm A,\,sub}^g$ & $Z_{\rm A,\,sub}^l$ & $Z_{\rm V,\,sub}^g$ & $Z_{\rm V,\,sub}^l$ \\
\midrule
$5.2$ & $ 0.77262(54)$ & $ 0.76963(94)$ & $ 0.73986(47)$ & $ 0.73890(97)$ \\
$5.3$ & $ 0.77847(42)$ & $ 0.77591(66)$ & $ 0.74706(49)$ & $ 0.74636(70)$ \\
$5.5$ & $ 0.79138(89)$ & $ 0.7897(13)$ & $ 0.7634(11)$ & $ 0.7627(14)$ \\
$5.7$ & $ 0.80358(63)$ & $ 0.80283(93)$ & $ 0.77836(79)$ & $ 0.7781(10)$ \\
\bottomrule
\end{tabular}
\caption{Results for $Z_{\rm A,V}$, both $g$ and $l$ definitions, for $\Nf=2$
	 non-perturbatively O($a$) improved Wilson fermions and Wilson gauge
	 action. The lower part of the table contains the same results after subtraction of
	 the one-loop cutoff effects, cf.~eq.~(\ref{eq:PThIZ}).}
\label{tab:ZNf2}
\end{table}
As discussed in section \ref{sec:Uncertainties}, systematic errors  
result from uncertainties or deviations in following a chosen LCP, 
which correspond with statistical errors and deviations
from zero in $m$ and $g_{\rm A}^{ud}(L/a)$, as well as 
uncertainties in the target lattice extent $L/a$ and systematic errors
arising from inter- or extrapolations from the simulated lattices sizes,
if applicable. Tables \ref{tab:LCPNf2} and \ref{tab:Nf2Parms}  
contain the relevant information for the case $\Nf=2$. 
The propagation of these uncertainties to the $Z$-factors is then performed following
the steps outlined in Sect.~\ref{sec:Uncertainties}. We have carried out some additional 
simulations to estimate the derivatives in eqs.~(\ref{eq:m0zf2mgA},\ref{eq:dZdX}), 
and some perturbative calculation to check the expected scaling of the derivatives 
with the lattice size. We delegate a detailed discussion to appendix~\ref{app:Syst}.
Here we just note that with our statistics and our rather conservative approach, 
the propagated uncertainties are typically larger 
than the statistical errors for the $R$-estimators eqs.~(\ref{eq:RA},\ref{eq:RV})
(cf.~tables \ref{tab:ZAVRawNf2} and \ref{tab:ZAVRawNf2PT}).

\subsubsection{Effect of perturbative one-loop improvement}

As discussed in section~\ref{sec:PTImprovement}, we have also computed the relevant 
$\chi$SF correlation functions in perturbation theory to order $g_0^2=6/\beta$. 
Besides consistency checks and qualitative insight the main application
consists in the perturbative subtraction of cutoff effects from the data. Note that
this requires to emulate the non-perturbative procedure in all details, in particular
the determination of $am_{\rm cr}$ and $z_f^*$ according to eqs.~(\ref{eq:mcrzf}). 
The lower part of table \ref{tab:ZNf2} contains the results for $Z_{\rm A,V}$ after 
perturbative improvement. Comparing with the unimproved results in the upper part of
table \ref{tab:ZNf2}, one can see that the $g$-definitions are more affected, and are
brought closer to the corresponding $l$-definitions by the perturbative improvement
(cf.~also figure \ref{fig:ZrNf2}). In any case, the perturbative corrections are
at the level of 1 per cent at most.
\begin{figure}[hpbt]
 \includegraphics[width=0.5\textwidth]{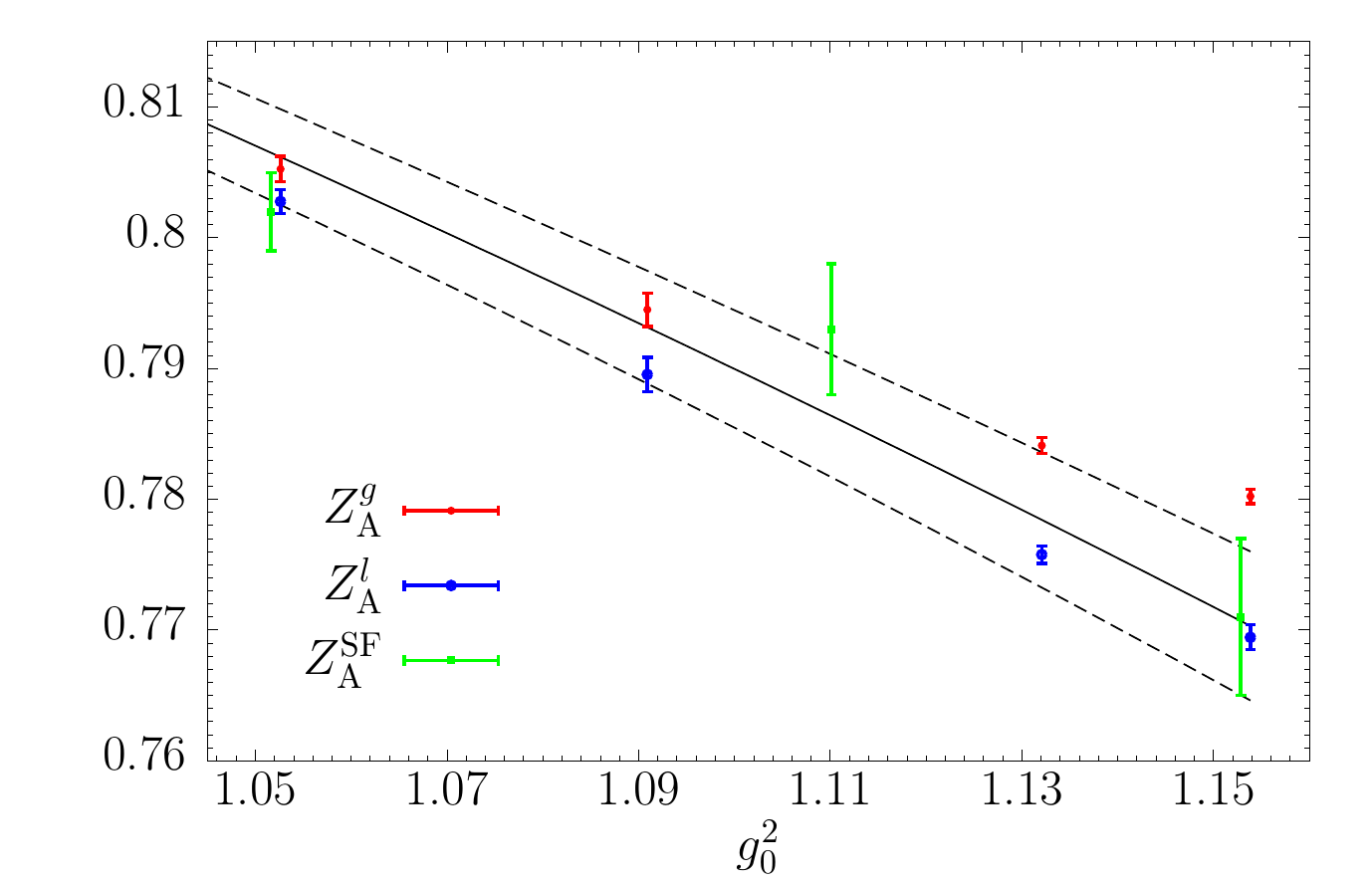}
 \includegraphics[width=0.5\textwidth]{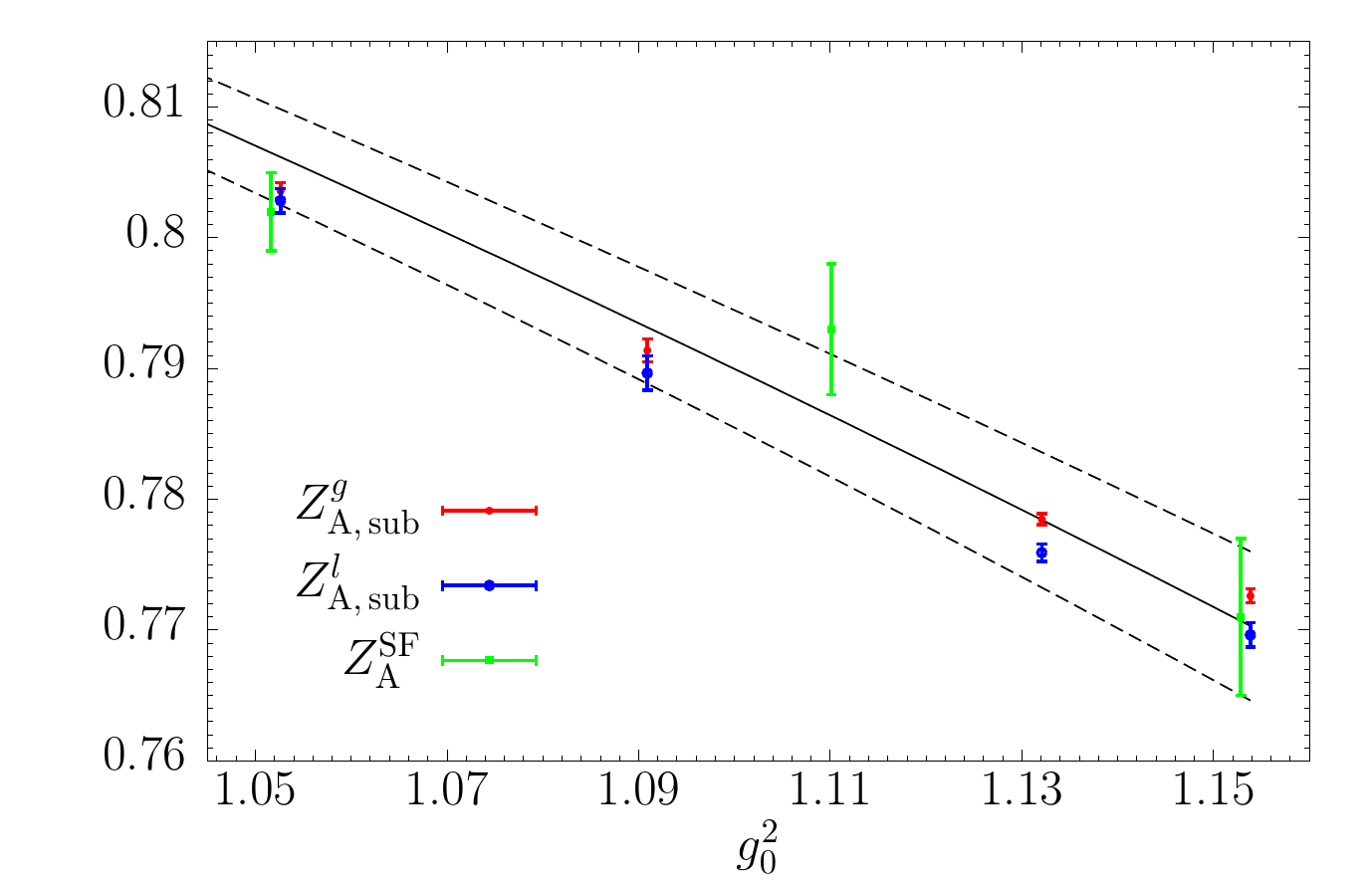}
 \caption{Comparison of different $Z_{\rm A}$ determinations for $\Nf=2$,  
 obtained from WIs in the standard SF and from universality relations in the $\xSF$. 
 The effect of the perturbative one-loop improvement of the $\xSF$ results is 
 also shown (right panel). The $\xSF$ results are those of table~\ref{tab:ZNf2}.
 The individual SF points are taken from refs.~\cite{DellaMorte:2008xb,Fritzsch:2012wq},
 and are slightly displaced on the $x$-axis for better clarity. The solid black line 
 corresponds to the SF results from the fit formula of ref.~\cite{Fritzsch:2012wq}, 
 and the dashed lines delimit the $1\sigma$ region of the fit. Note that the SF fit formula
 is obtained by considering additional points with $g_0^2<1$, here not shown, and by 
 enforcing the perturbative 1-loop behaviour for $g_0^2\to0$ (see ref.~\cite{Fritzsch:2012wq}
 for the details).}
 \label{fig:ZaNf2}
\end{figure}
In conclusion, our final results for $Z_{\rm A,V}$, either with or without
perturbative improvement, turn out to be very precise and improve significantly
on the standard SF determination based on chiral Ward 
identities (WIs)~\cite{DellaMorte:2005rd,DellaMorte:2008xb,Fritzsch:2012wq}. 
This is particularly true for the case of $Z_{\rm A}$, which can be appreciated 
in figure \ref{fig:ZaNf2} where the determinations of table \ref{tab:ZNf2} are 
compared with those of refs.~\cite{DellaMorte:2008xb,Fritzsch:2012wq}.  
In figure \ref{fig:ZvNf2} we show instead a comparison for the case of $Z_{\rm V}$, 
as obtained from the $\xSF$, cf.~table \ref{tab:ZNf2}, and from the standard SF 
(cf.~ref.~\cite{DellaMorte:2005rd}). We note that a relevant contribution to the
error of our results comes from propagating the uncertainties associated with 
maintaining the condition (\ref{eq:fKL}) i.e. keeping $L$ constant (cf.~table 
\ref{tab:ZAVRawNf2} and \ref{tab:ZAVRawNf2PT}). We anticipate that due to the 
much more accurate knowledge of the LCP in terms of $t_0$ (cf.~table \ref{tab:LCPNf3}),
and by using interpolations in $L/a$ at all relevant $\beta$ values, this source
of error will be essentially eliminated in the case of $\Nf=3$ (cf.~section~\ref{sec:Nf3}).

\begin{figure}[hpbt]
	\includegraphics[width=0.5\textwidth]{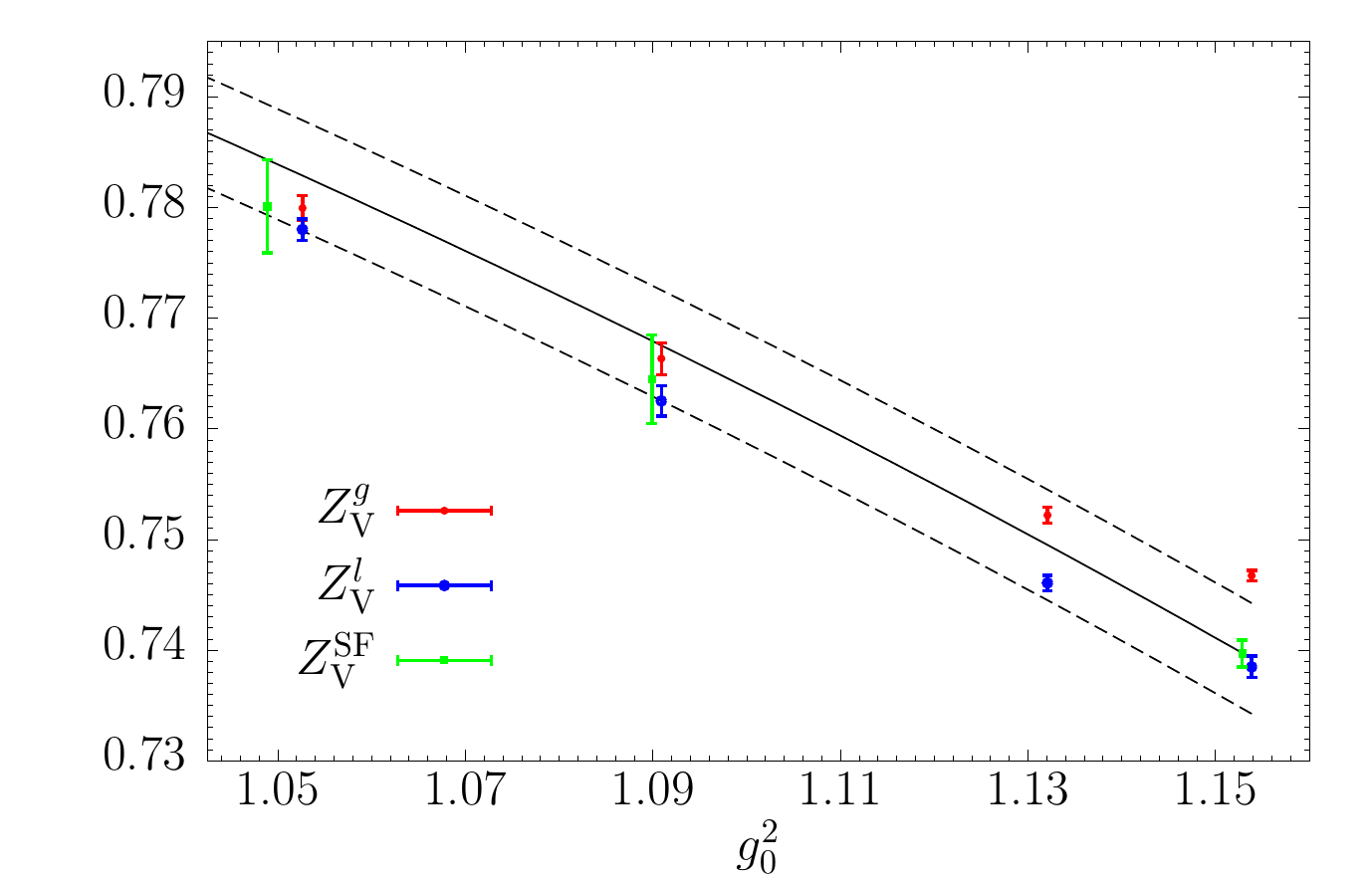}
	\includegraphics[width=0.5\textwidth]{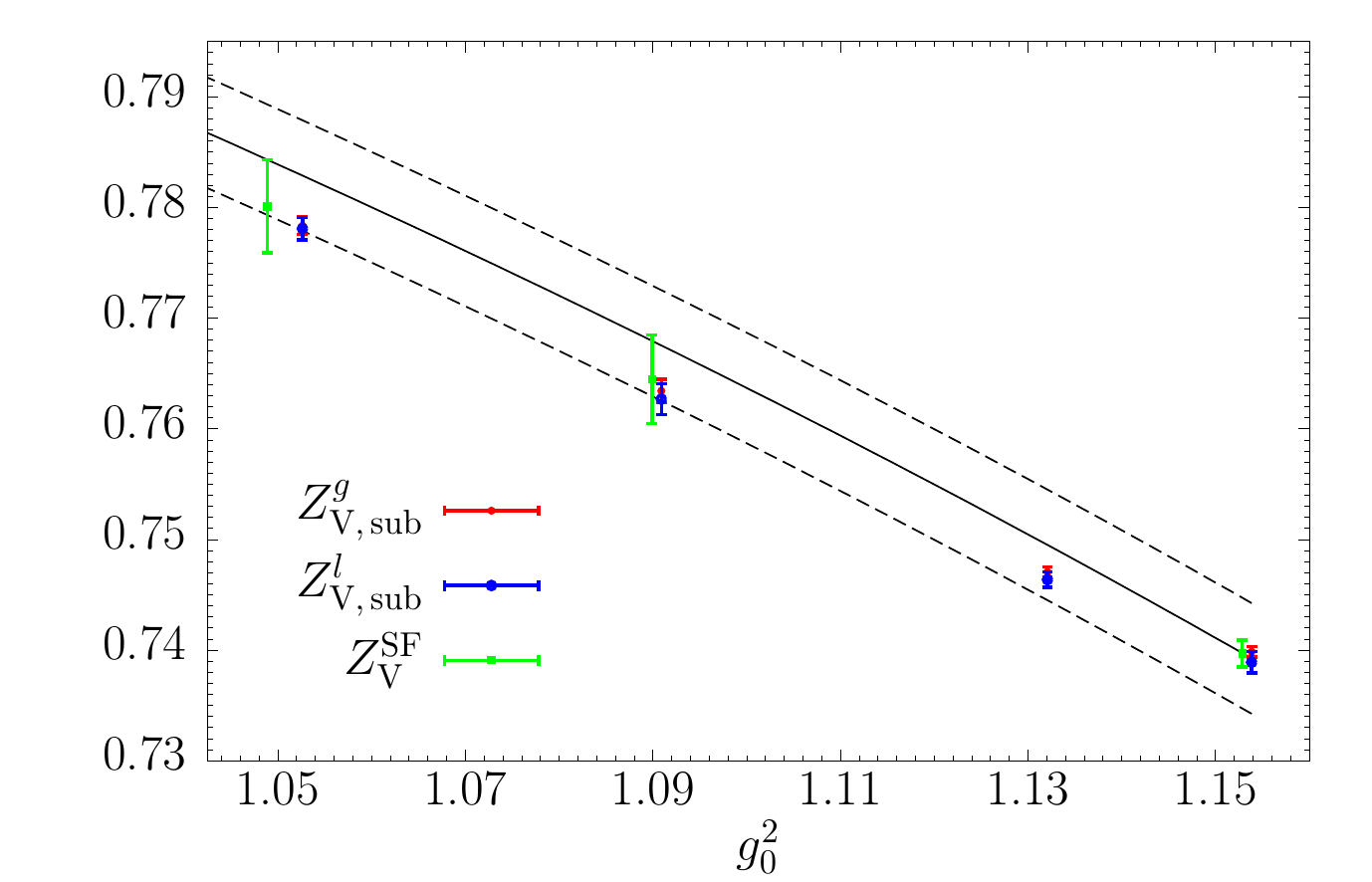}
	\caption{Comparison of different $Z_{\rm V}$ determinations for $\Nf=2$,  
		obtained from WIs in the standard SF and from universality relations in the $\xSF$. 
		The effect of the perturbative one-loop improvement of the $\xSF$ results is 
		also shown (right panel). The $\xSF$ results are those of table~\ref{tab:ZNf2}.
		The individual SF points are taken from refs.~\cite{DellaMorte:2005rd},
		and are slightly displaced on the $x$-axis for better clarity. The solid black line 
		corresponds to the SF results from the fit formula of ref.~\cite{DellaMorte:2005rd}, 
		and the dashed lines delimit the $1\sigma$ region of the fit. Note that the SF fit formula
		is obtained by considering additional points with $g_0^2<1$, here not shown, and by 
		enforcing the perturbative 1-loop behaviour for $g_0^2\to0$ (see ref.~\cite{DellaMorte:2005rd}
		for the details).}
	\label{fig:ZvNf2}
\end{figure}

\subsection{Universality and automatic O($a$) improvement}

\begin{figure}[hpbt]
 \includegraphics[width=0.5\textwidth]{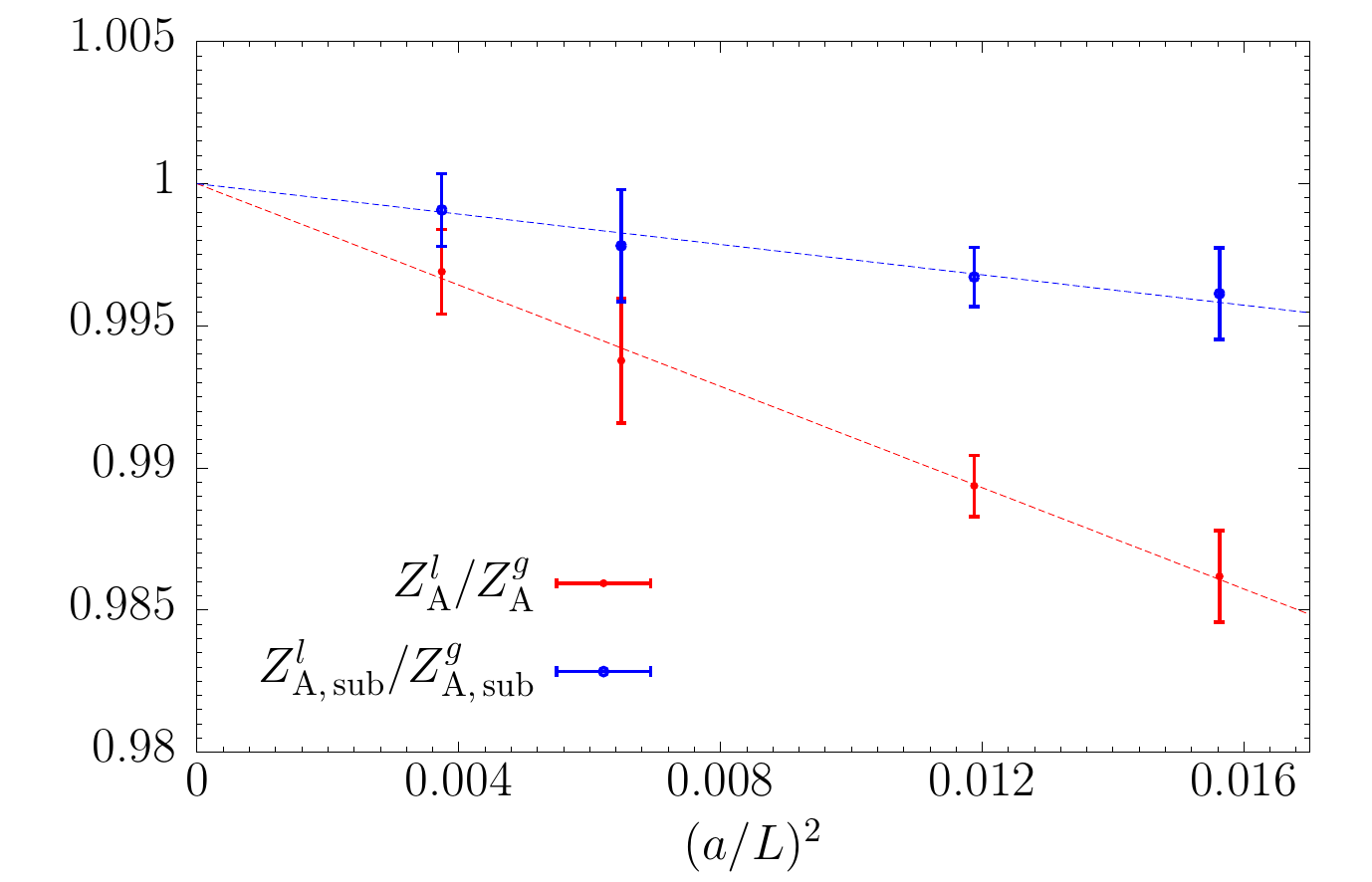}
 \includegraphics[width=0.5\textwidth]{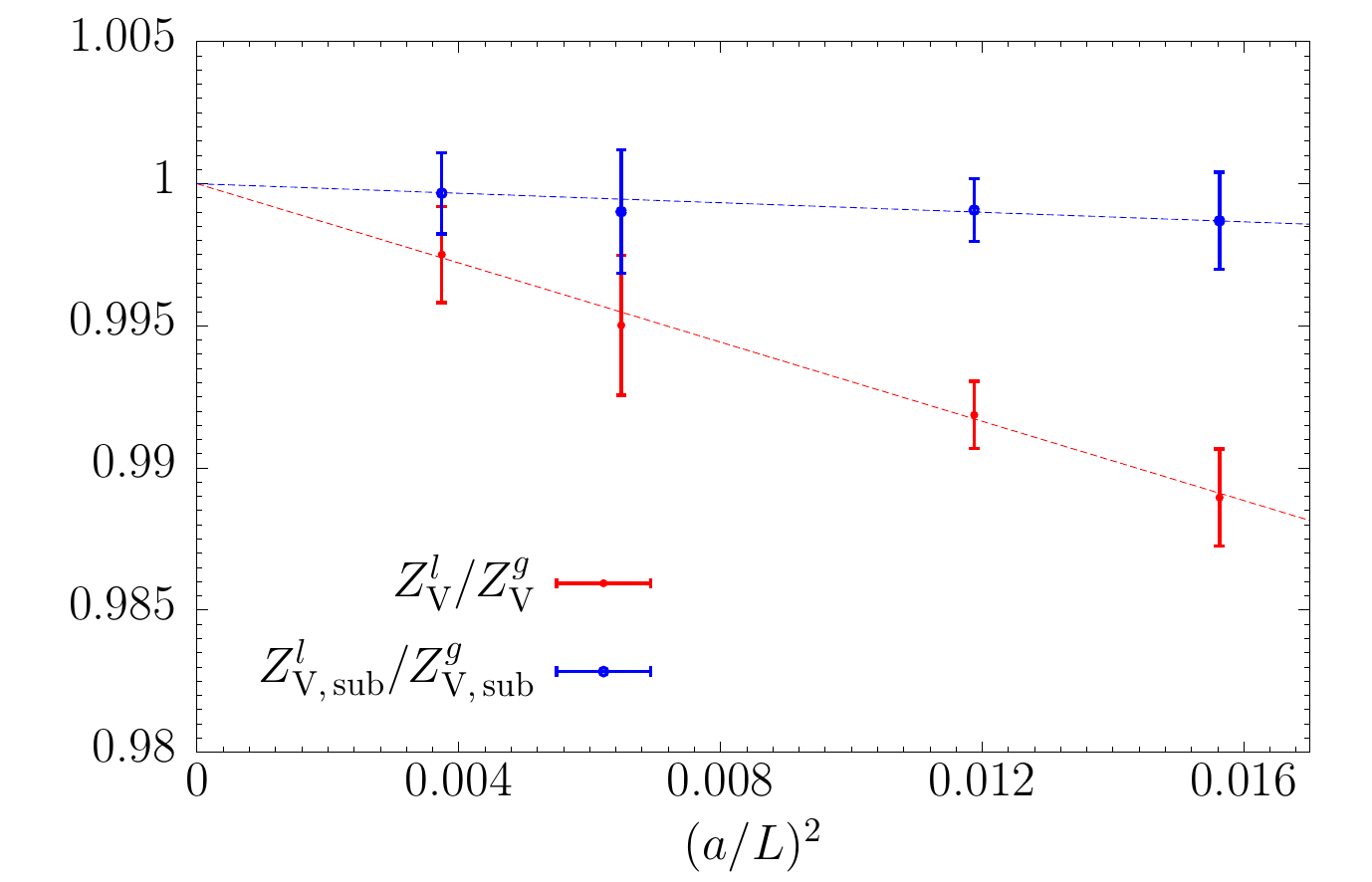}
 \caption{Continuum limit of the ratios between the $l$ and $g$ 
 	definitions of $Z_{\rm A}$ (left panel) and $Z_{\rm V}$ (right panel)
 	for the case of $\Nf=2$ quark-flavours; the effect of subtracting the
 	lattice artefacts from the $Z$-factors to O($g_0^2$) is also shown.
 	The dashed lines correspond to linear fits to the data, constrained to 
 	extrapolate to 1 for $a/L=0$.}
 \label{fig:ZrNf2}
\end{figure}

The $\xSF$ determinations (\ref{eq:RA}) and (\ref{eq:RV}) are expected to be
automatically O($a$) improved once the bare parameters $m_0$ and $z_f$
are properly tuned (cf.~section \ref{subsec:RenCond}). This means that neither
bulk nor boundary O($a$) counterterms are necessary to cancel O($a$) discretization
errors in these quantities. This was confirmed to one-loop order in 
perturbation theory~\cite{Brida:2016rmy} and should hold generally. To this end we now look 
at the ratios between $Z$-factors coming from the $g$- and $l$-definitions. The expectation
that these ratios converge to 1 with O($a^2$) corrections is indeed very well
borne out by the data, cf.~figure~\ref{fig:ZrNf2}, where we also include fits
to this expected behaviour. We emphasize that this is a non-trivial result: 
even though the bulk action is improved to match the CLS set-up, 
we did not O($a$) improve the currents entering the definitions 
(\ref{eq:RA},\ref{eq:RV}) \emph{and} (\ref{eq:mcrzf}). 
This result thus confirms automatic O($a$) improvement at the non-perturbative level, 
and, indirectly, the universality relations between the $\xSF$ and SF formulations. 
A direct way to test universality between the $\xSF$ and SF formulations would be
simply to study the continuum scaling of ratios of $Z$-factors as obtained from one and
the other formulation. Provided the SF determinations are properly improved, these
should also approach 1 in the continuum limit with O($a^2$) corrections. The large errors
on the SF determinations do not allow us for a precise test of this expectation. However, 
the results in figure \ref{fig:ZaNf2} and \ref{fig:ZvNf2} clearly show that our 
determinations are in fact compatible with the SF ones within errors.

\section{Numerical results for $\Nf=3$ flavours}
\label{sec:Nf3}

\subsection{Lattice set-up and parameter choices}
\label{sec:SetupNf3}

The CLS simulations with $\Nf=2+1$ flavours of non-perturbatively O($a$) improved Wilson
fermions~\cite{Bulava:2013cta} and L\"uscher-Weisz (LW)  gauge action, have been carried
out for 5 values of the lattice spacing, with $\beta$-values between $3.4$ and 
$3.85$~\cite{Bruno:2014jqa,Bruno:2016plf,Bruno:2017gxd}. For completeness we note that 
CLS has also tried to simulate at a coarser lattice spacing corresponding to $\beta=3.3$.
However, these ensembles have been discarded for the scale determination in~\cite{Bruno:2016plf}
due to very large cutoff effects observed e.g.~in $t_0$~\cite{Bruno:2014jqa}. For this reason
we will not consider this $\beta$-value in our study, however, we mention that it was adopted
as starting point for the Ward identity determination of $Z_{\rm A}$ in ref.~\cite{Bulava:2016ktf}.
Given the relatively large set of lattice spacings we here consider two different LCPs, with 
slightly different physical extent, $L_1$ and $L_2$, which we define through the gradient flow
time $t_0$~\cite{Luscher:2010iy}. The associated length scale $r=\sqrt{8t_0}$ can be interpreted
as a smoothing radius, and has been very precisely determined for the CLS $\beta$-values $\geq 3.4$
in \cite{Bruno:2016plf,Bruno:2017gxd}. Using this scale we impose the conditions
\begin{equation}
  \label{eq:Lovert0}
  L_1/\sqrt{8t_0}= 1.6719(16)  
  \quad
  {\rm and}
  \quad
  L_2/\sqrt{8t_0}= 1.5099(30),
\end{equation}
where the right hand sides were chosen in order to have exactly,
\begin{equation}
  L_1/a=8
  \quad
  {\rm at}
  \quad
  \beta=3.4
  \quad
  {\rm and}  
  \quad  
  L_2/a=16
  \quad
  {\rm at}
  \quad
  \beta=3.85,  
\end{equation}
respectively. Using the result for $t_0$ in physical units~\cite{Bruno:2016plf},
eqs.~(\ref{eq:Lovert0}) translate to $L_1\approx 0.7\,\fm$ and $L_2\approx 0.6\,\fm$.

\begin{table}[hpbt]
  \centering
  \begin{tabular}{lllll}
  \toprule
  $\beta$ & $t_0/a^2$ & $(L_1/a)(\beta)$ & $(L_2/a)(\beta)$ & $L/a$ \\
  \midrule
  $3.40$ & $ 2.8619(55)$ & $ 8$          & $ 7.225(16)$  & $6$, $8$,  $10$, $12$ \\
  $3.46$ & $ 3.662(13)$  & $ 9.049(18)$  & $ 8.172(22)$  & $6$, $8$,  $10$, $12$ \\
  $3.55$ & $ 5.166(17)$  & $ 10.748(21)$ & $ 9.706(25)$  & $8$, $10$, $12$, $16$ \\
  $3.70$ & $ 8.596(31)$  & $ 13.864(29)$ & $ 12.521(34)$ & $8$, $10$, $12$, $16$ \\
  $3.85$ & $ 14.036(57)$ & $ -$          & $ 16$         & $16$                  \\
  \bottomrule
  \end{tabular}  
  \caption{CLS $\beta$-values and corresponding results for $t_0/a^2$ in the 
  SU(3) flavour symmetric limit~\cite{Bruno:2016plf,Bruno:2017gxd}. The latter are used
  to determine the lattice sizes $(L_{1,2}/a)(\beta_i)$ which satisfy the conditions 
  (\ref{eq:Lovert0}). The $\xSF$ simulations are performed at the neighbouring $L/a$'s 
  given in the last column of the table.}
  \label{tab:LCPNf3}
\end{table}

In table~\ref{tab:LCPNf3} we collect the relevant $\beta$ values of the CLS
simulations and the corresponding results for $t_0/a^2$~\cite{Bruno:2017gxd}. 
The latter are evaluated for equal up-, down-, and strange-quark masses, which are 
close to the physical average quark mass (see refs.~\cite{Bruno:2016plf,Bruno:2017gxd}).
Table~\ref{tab:LCPNf3} also gives the lattice sizes $(L_{1,2}/a)(\beta)$
which satisfy the conditions (\ref{eq:Lovert0}). Compared to the $\Nf=2$ case 
(cf.~table~\ref{tab:LCPNf2}), it is obvious that these $\Nf=3$ LCPs are much more 
accurately determined. In order to exploit this higher precision, we performed 
simulations for several $L/a$-values at each $\beta$ (cf.~table \ref{tab:LCPNf3}). 
This allowed us to accurately interpolate the $Z$-factors to the target values 
(see appendix \ref{app:InterpolationsNf3} for more details).  Table~\ref{tab:Nf3Parms}
contains a summary of all simulations performed with the corresponding parameters.
Due to both technical and historical reasons, we do not use the finest lattice spacing for 
the LCP defined in terms of $L_1$. Following this LCP up to $\beta=3.85$ would have required
simulating lattices with $L/a=18,20$, which are particularly inconvenient to parellelize 
with our current simulation program. Note also that CLS simulations at $\beta=3.85$ 
are ongoing and currently limited to a single ensemble, so that the LCP with $L_1$ may
remain useful for a while. More importantly, however,  the comparison between both LCPs
allows us to perform additional tests on our results (cf.~section \ref{sec:UniversalityNf3}).

The lattice action we employ for the finite volume simulations matches the CLS action
in the bulk, i.e.~the L\"uscher-Weisz tree-level improved gauge action and 3 flavours
of non-perturbatively improved Wilson quarks~\cite{Bulava:2013cta}. Close to the time
boundaries of the lattice there is some freedom regarding the implementation of Schr\"odinger
functional boundary conditions. For the gauge fields we choose option~B of ref.~\cite{Aoki:1998qd};  
we refer the reader to this reference for the details. Regarding the fermions, two quark
flavours satisfy $\xSF$ boundary conditions~(option $\tau=1$ of \cite{Sint:2010eh}), 
while the third one obeys the standard SF boundary conditions~\cite{Sint:1993un}. 
In general, such a mixed set-up increases the number of O($a$) improvement 
coefficients which need to be tuned in order to eliminate O($a$) discretization errors from
the time boundaries. As in the $\Nf=2$ case, however, one can show that the corresponding counterterms
affect the renormalization constants $Z_{\rm A,V}$ only at O($a^2$). For definiteness we  
have used the one-loop estimate $c_{\rm t} = 1 + g_0^2 c_{\rm t}^{(1)}$, where the 
one-loop coefficient decomposes as follows,
\begin{equation}
 c_{\rm t}^{(1)} =  c_{\rm t}^{(1,0)} + 2\times c_{\rm t}^{(1,1)}(\text{$\chi$SF}) + 1\times c_{\rm t}^{(1,1)}(\text{SF}).
\end{equation}
The pure gauge contribution is taken from ref.~\cite{Takeda:2003he},
the fermionic $\xSF$ contribution from ref.~\cite{Brida:2016rmy} and the SF contribution from ref.~\cite{Sint:1995ch}.%
\footnote{Even though the fermionic contributions were calculated with the Wilson gauge action, 
to this order the calculation only depends on the gauge background field, which is not modified
when using the LW action with option~B of~\cite{Aoki:1998qd}.}
Furthermore, we use the tree-level values $d_s=1/2$~\cite{Brida:2016rmy} and 
$\tilde{c}_{\rm t}=1$~\cite{Luscher:1996sc}.

\subsection{Results and error budget}
\label{sec:ResultsNf3}

In table \ref{tab:ZNf3L1} and \ref{tab:ZNf3L2} we collect the results for $Z_{\rm A,V}$,
corresponding to the $L_1$- and $L_2$-LCP, respectively. The statistics we accumulated 
for the different ensembles ranges between 3,200 and 31,000 measurements, with exact numbers
given in table~\ref{tab:Nf3Parms}. The corresponding statistical precision on the 
$Z$-factors is between $0.1-0.55\permil$, depending on the exact quantity and ensemble. 
The errors quoted in the tables then combine the statistical errors with the systematic
errors originating from the uncertainties on the LPCs.

\begin{table}[hpbt]
\centering
\begin{tabular}{llllllllll}
\toprule
 $\beta$ & $Z_{\rm A}^g$ & $Z_{\rm A}^l$ & $Z_{\rm V}^g$ & $Z_{\rm V}^l$ \\
\midrule
$3.40$ & $ 0.76847(35)$ & $ 0.75446(68)$ & $ 0.72923(27)$ & $ 0.71940(70)$ \\
$3.46$ & $ 0.77128(44)$ & $ 0.76018(80)$ & $ 0.73392(36)$ & $ 0.72637(77)$ \\
$3.55$ & $ 0.77703(30)$ & $ 0.76879(42)$ & $ 0.74261(23)$ & $ 0.73758(44)$ \\
$3.70$ & $ 0.78831(30)$ & $ 0.78327(43)$ & $ 0.75833(31)$ & $ 0.75521(44)$ \\
\midrule
$\beta$ & $Z_{\rm A,\,sub}^g$ & $Z_{\rm A,\,sub}^l$ & $Z_{\rm V,\,sub}^g$ & $Z_{\rm V,\,sub}^l$ \\
\midrule
$3.40$ & $ 0.75702(35)$ & $ 0.75485(68)$ & $ 0.71882(27)$ &  $ 0.72008(70)$ \\
$3.46$ & $ 0.76245(44)$ & $ 0.76048(80)$ & $ 0.72578(36)$ &  $ 0.72683(77)$ \\
$3.55$ & $ 0.77103(29)$ & $ 0.76900(42)$ & $ 0.73701(23)$ &  $ 0.73789(44)$ \\
$3.70$ & $ 0.78485(30)$ & $ 0.78340(43)$ & $ 0.75506(31)$ &  $ 0.75538(44)$ \\
\bottomrule
\end{tabular}
\caption{$\Nf=3$ results for $Z_{\rm A,V}$ using the $L_1$-LCP, both for $g$ and $l$ definitions.
The lower part of the table
	contains the results after subtraction of the one-loop cutoff effects, cf.~eq.~(\ref{eq:PThIZ}).}
\label{tab:ZNf3L1}
\end{table}

\begin{table}[hpbt]
\centering
\begin{tabular}{llllllllll}
\toprule
 $\beta$ & $Z_{\rm A}^g$ & $Z_{\rm A}^l$ & $Z_{\rm V}^g$ & $Z_{\rm V}^l$ \\
\midrule
$3.40$ & $ 0.77129(39)$ & $ 0.75592(72)$  & $ 0.73368(30)$  & $ 0.72164(74)$ \\
$3.46$ & $ 0.77371(51)$ & $ 0.76132(93)$  & $ 0.73721(42)$  & $ 0.72782(89)$ \\
$3.55$ & $ 0.77856(31)$ & $ 0.76953(43)$  & $ 0.74468(25)$  & $ 0.73846(46)$ \\
$3.70$ & $ 0.78925(31)$ & $ 0.78362(47)$  & $ 0.75936(33)$  & $ 0.75552(48)$ \\
$3.85$ & $ 0.79985(31)$ & $ 0.79657(47)$  & $ 0.77304(33)$  & $ 0.77061(49)$ \\
\midrule
$\beta$ & $Z_{\rm A,\,sub}^g$ & $Z_{\rm A,\,sub}^l$ & $Z_{\rm V,\,sub}^g$ & $Z_{\rm V,\,sub}^l$ \\
\midrule
$3.40$ & $ 0.75741(38)$ & $ 0.75642(72)$ & $ 0.72120(29)$ & $ 0.72259(74)$ \\
$3.46$ & $ 0.76288(51)$ & $ 0.76169(93)$ & $ 0.72732(41)$ & $ 0.72845(89)$ \\
$3.55$ & $ 0.77115(31)$ & $ 0.76979(43)$ & $ 0.73780(24)$ & $ 0.73886(46)$ \\
$3.70$ & $ 0.78499(31)$ & $ 0.78378(47)$ & $ 0.75534(33)$ & $ 0.75574(48)$ \\
$3.85$ & $ 0.79734(31)$ & $ 0.79667(47)$ & $ 0.77065(33)$ & $ 0.77074(49)$ \\
\bottomrule
\end{tabular}
\caption{Same as table~\ref{tab:ZNf3L1} but for the $L_2$-LCP.}
\label{tab:ZNf3L2}
\end{table}
        
Like in the $\Nf=2$ case, the high statistical precision 
requires a careful assessment of the systematic errors in order to arrive at reliable
error estimates. Tables \ref{tab:LCPNf3} and \ref{tab:Nf3Parms}  contain information on the
accuracy with which the chosen LCPs are realized for our simulation parameters. 
Our estimates for the systematic uncertainties due to deviations from the chosen LCP were then 
obtained analogously to the case of $\Nf=2$; we refer the reader to appendix~\ref{app:Syst}
for the details. Here it is worth noting that, similarly to this case, the propagated uncertainties
are typically larger than the statistical errors for the $R$-estimators, eqs.~(\ref{eq:RA},\ref{eq:RV}),
cf.~table~\ref{tab:ZAVRawNf3}.

\subsubsection{Effect of perturbative one-loop improvement}

In the lower halves of tables \ref{tab:ZNf3L1} and \ref{tab:ZNf3L2} we give the results 
for $Z_{\rm A,V}$ after perturbatively subtracting the lattice artefacts to one-loop order. 
The results have been obtained by first improving the $Z_{\rm A,V}$ determinations for
each $L/a$ and $g_0$ value, and then interpolating to the proper $(L_{1,2}/a)(\beta)$
(see appendix \ref{app:InterpolationsNf3}). 

Comparing the results for $Z_{\rm A,V}$ before and after perturbative improvement, one sees
that the $g$-definitions are the most affected, and are brought closer to the corresponding 
$l$-definitions. All in all, the effect of the perturbative improvement is at most at the level of
a couple of percent (cf. figure  \ref{fig:ZrNf3}). Hence, not too surprisingly perhaps, the 
situation is very much the same as for the $\Nf=2$ case.

\begin{figure}[hpbt]
	\includegraphics[width=0.5\textwidth]{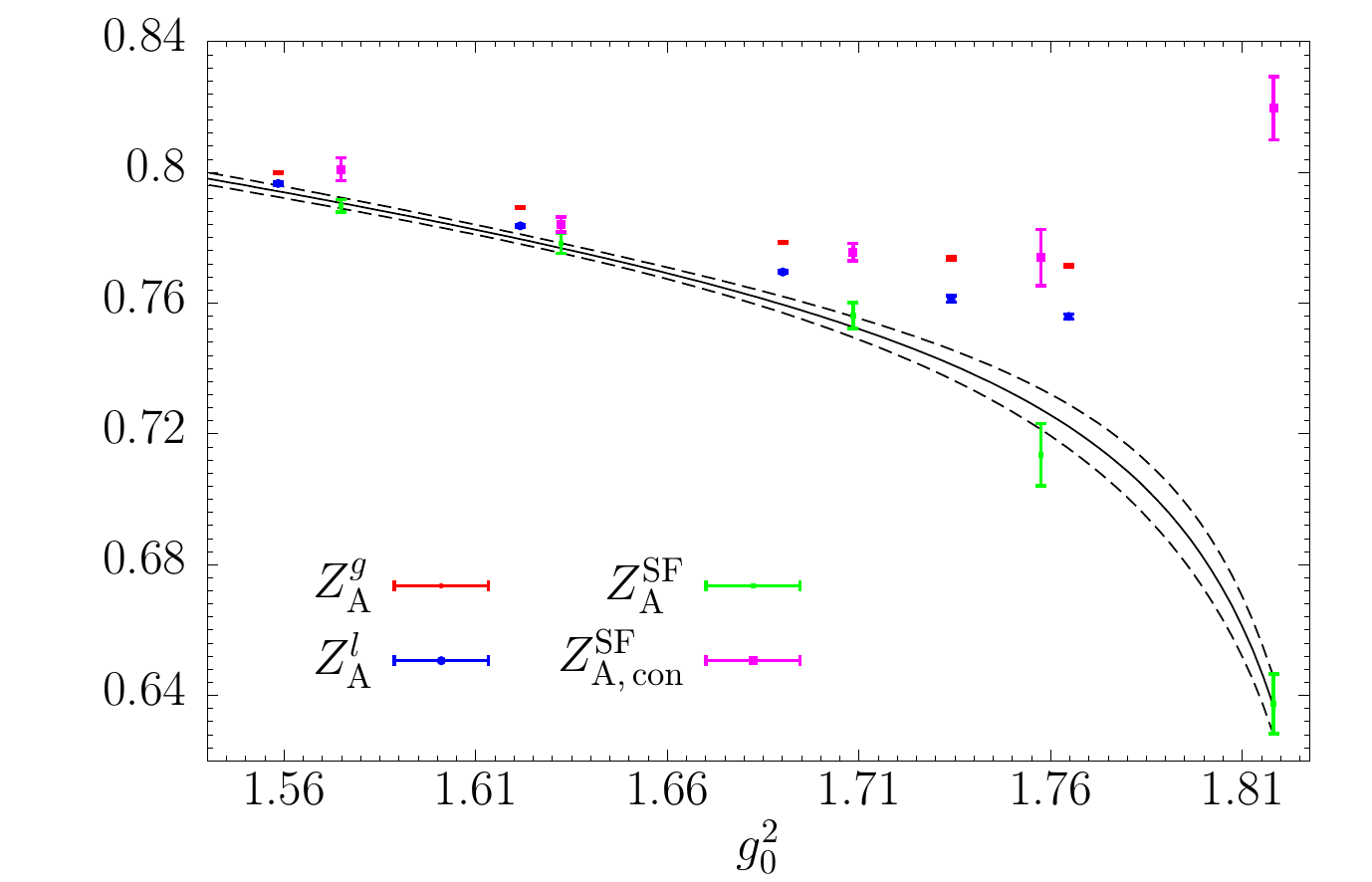}
	\includegraphics[width=0.5\textwidth]{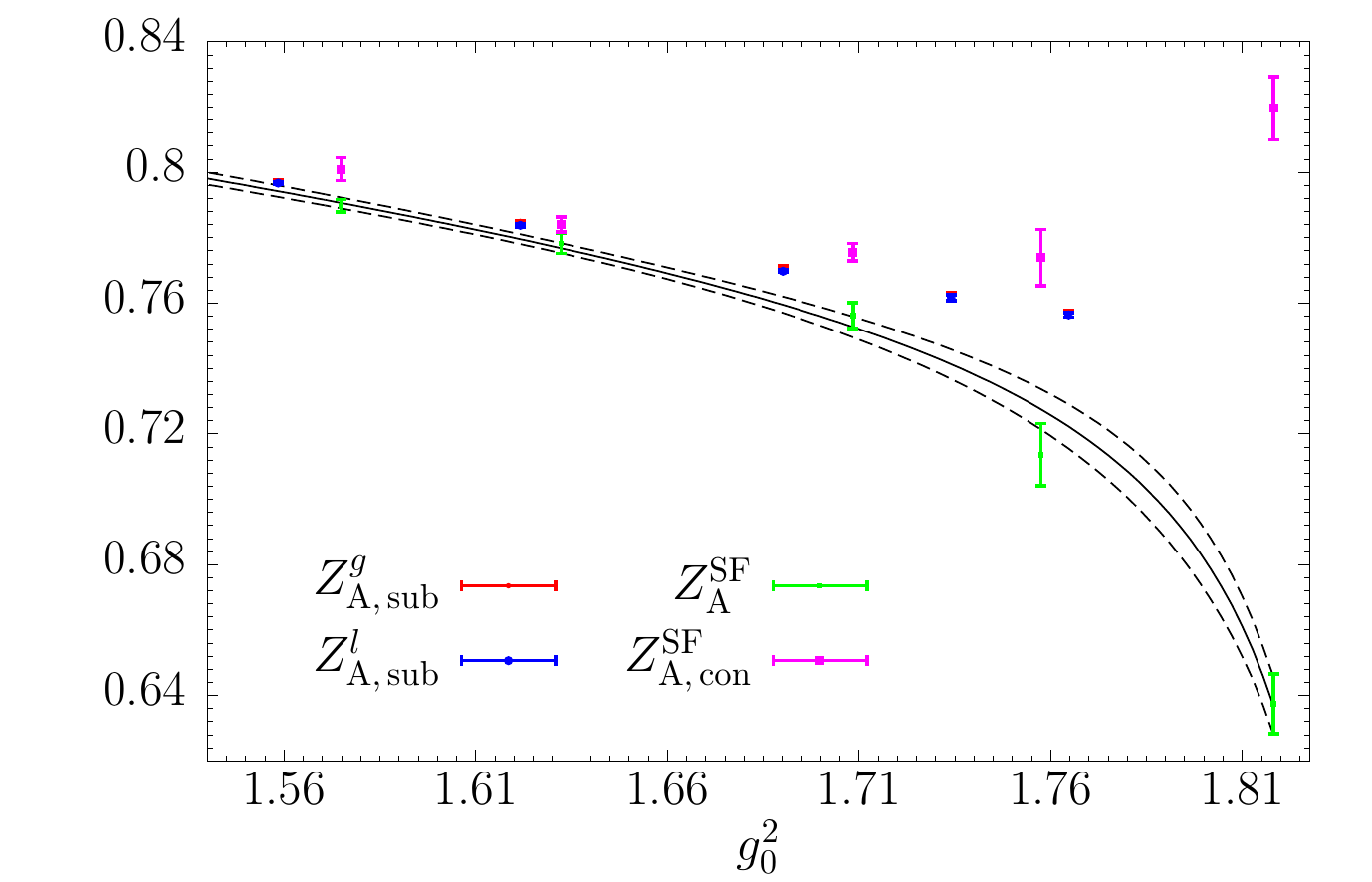}
	\caption{Comparison between different $Z_{\rm A}$ determinations for $\Nf=3$, 
		obtained either from WIs in the standard SF or from universality relations in the $\xSF$. 
                The $\xSF$ results are taken from table~\ref{tab:ZNf3L2} and 
		the effect of the perturbative one-loop improvement is 
		shown in the right panel. 
		The individual SF points labelled $Z_{\rm A}^{\rm SF}$  and  $Z_{\rm A, con}^{\rm SF}$ 
		are taken from ref.~\cite{Bulava:2016ktf} and correspond to the definitions $Z_{\rm A,0}$ and 
                $Z_{\rm A,0}^{\rm con}$, respectively, of that reference. 
		The solid black line is the fit formula to $Z_{\rm A}^{\rm SF}$ 
		also given in~\cite{Bulava:2016ktf} and the dashed lines delimit the $1\sigma$ 
		region of the fit. Note that this fit function enforces the perturbative 1-loop 
		behaviour for $g_0^2\to0$.}
	\label{fig:ZaNf3}
\end{figure}

In conclusion, our final results for $Z_{\rm A,V}$ are very precise for both LCPs. Similarly to
the $\Nf=2$ case, the results for $Z_{\rm A}$ are significantly more accurate than the standard
SF determination based on Ward identities~\cite{Bulava:2016ktf}. This can be appreciated in 
figure \ref{fig:ZaNf3}, where the results from table~\ref{tab:ZNf3L2} are displayed together with
the 2 alternative definitions $Z_{\rm A,0}$ and $Z_{\rm A,0}^{\rm con}$ of ref.~\cite{Bulava:2016ktf}.

\subsection{Universality and automatic O($a$) improvement}        
\label{sec:UniversalityNf3}

Given our estimates for $Z_{\rm A,V}$ we can study the approach to the continuum 
limit of the ratio between different definitions. We begin with figure \ref{fig:ZrNf3} where
the ratios between the $g$- and $l$-definitions are considered for the $L_1$- and
 $L_2$-LCPs; both the results before and after perturbative improvement are shown. 
The conclusions are very much the same as for the $\Nf=2$ case. Considering the
results before perturbative improvement, along both LCPs, the $g$ and $l$ definitions deviate 
by at most a couple of per-cent. These differences then perfectly scale with $a^2$ to zero as the 
continuum limit is approached. If perturbative improvement is 
implemented, these differences almost vanish even at the coarsest lattice spacings.
There is no significant deviation from $a^2$ scaling, however, some small admixture of
higher order effects cannot be excluded either.
\begin{figure}[hptb]
	
	\includegraphics[scale=0.5]{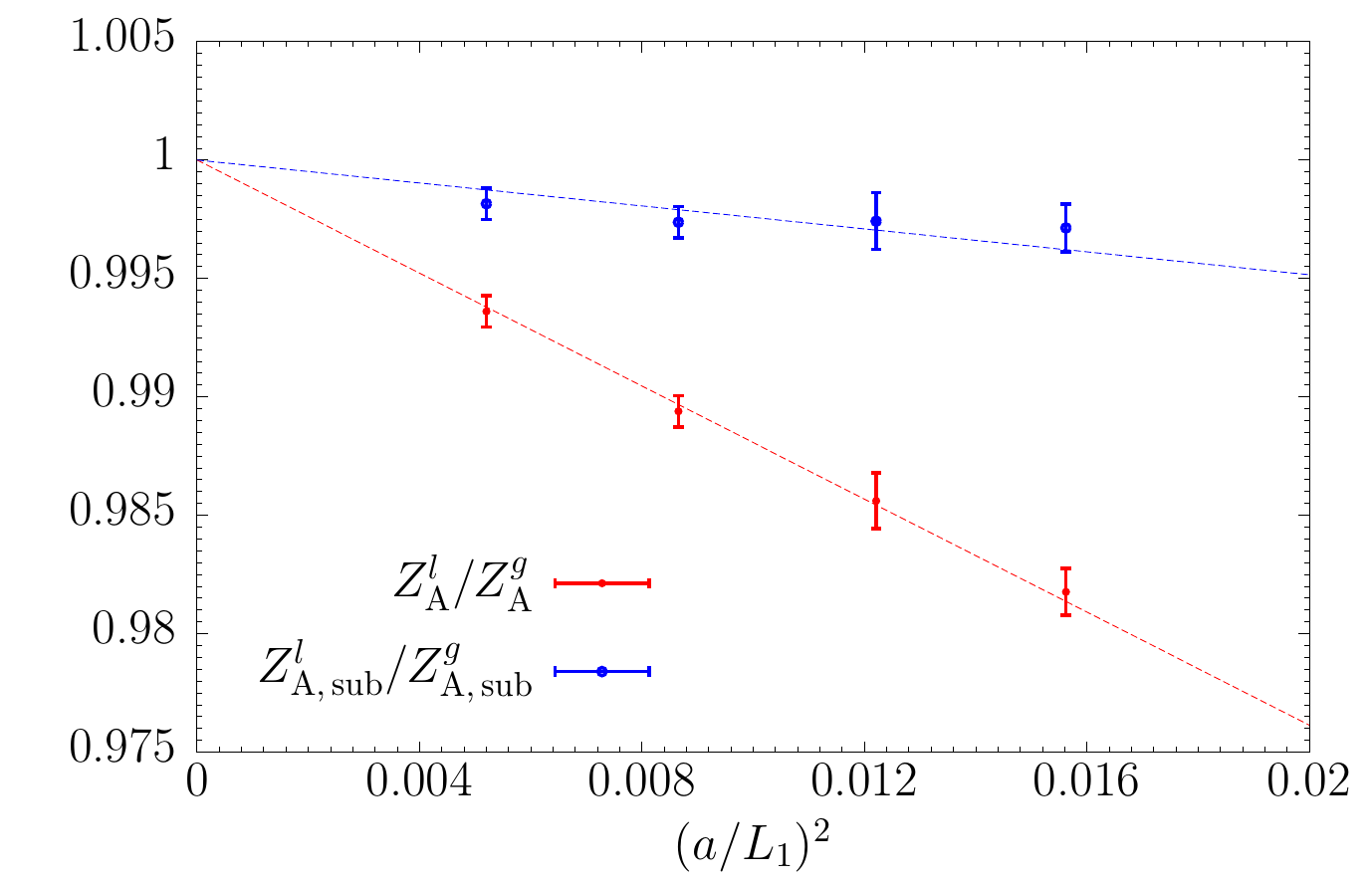}
	\includegraphics[scale=0.5]{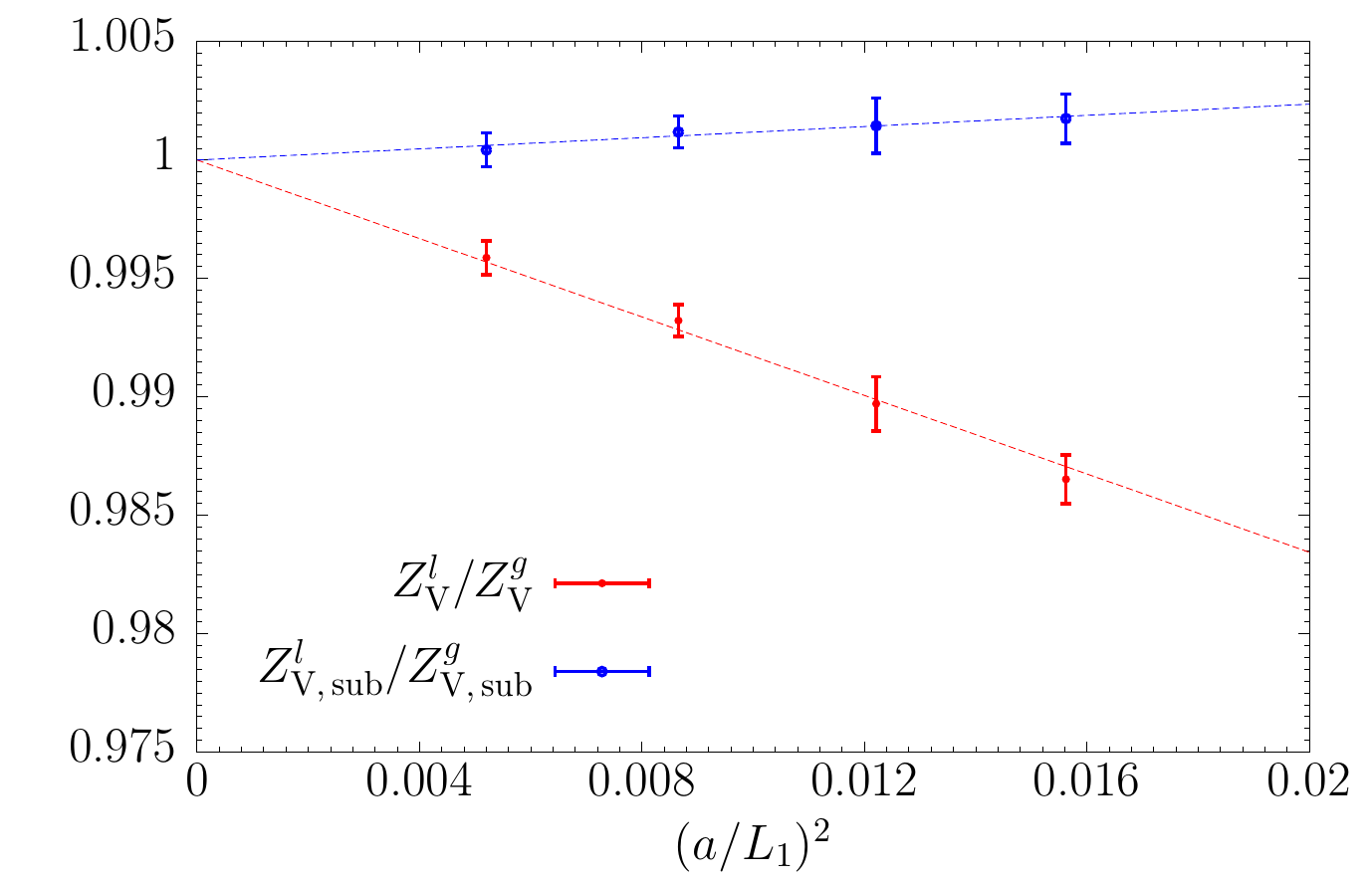}\\[1ex]
	\includegraphics[scale=0.5]{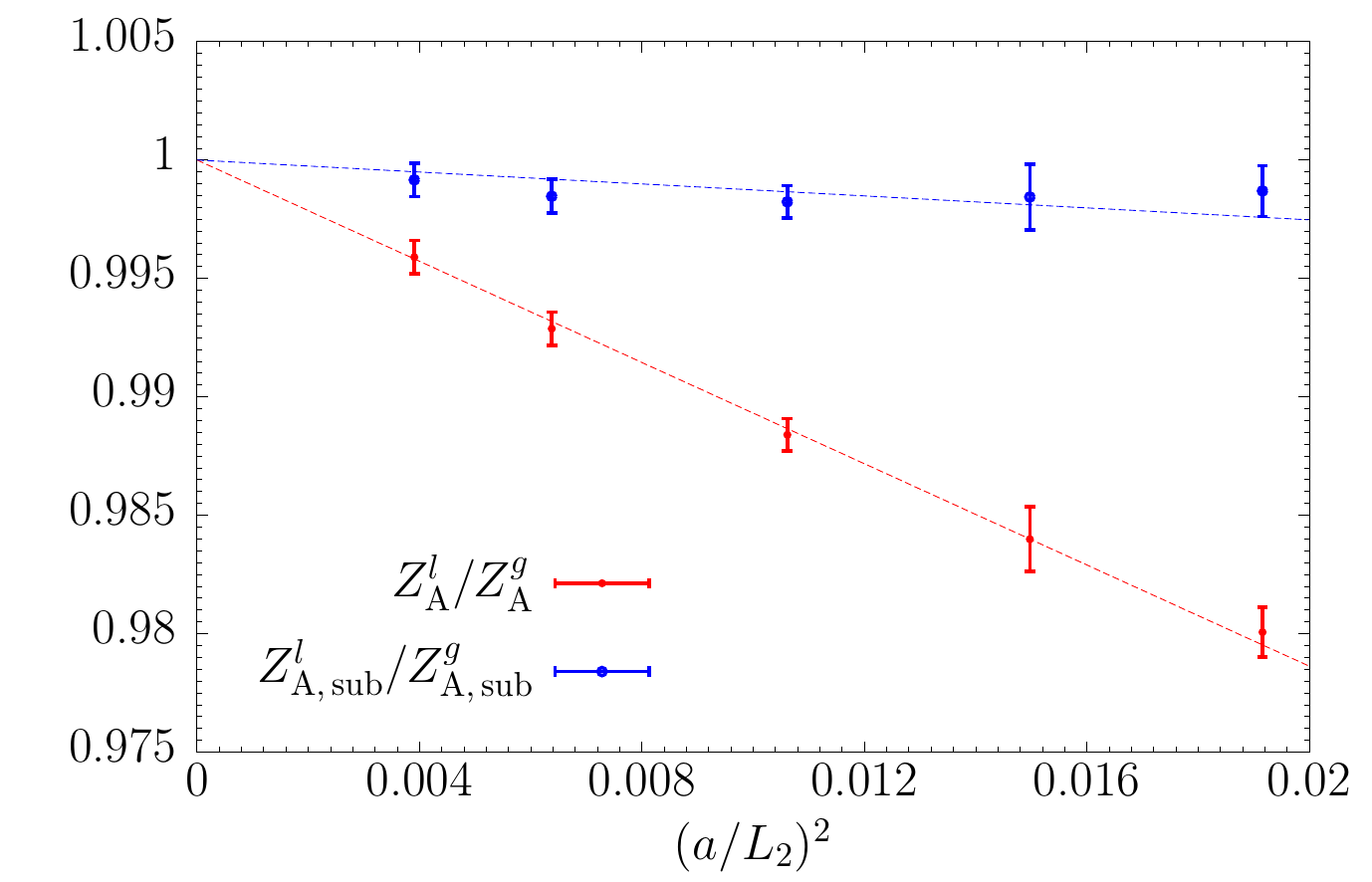}
	\includegraphics[scale=0.5]{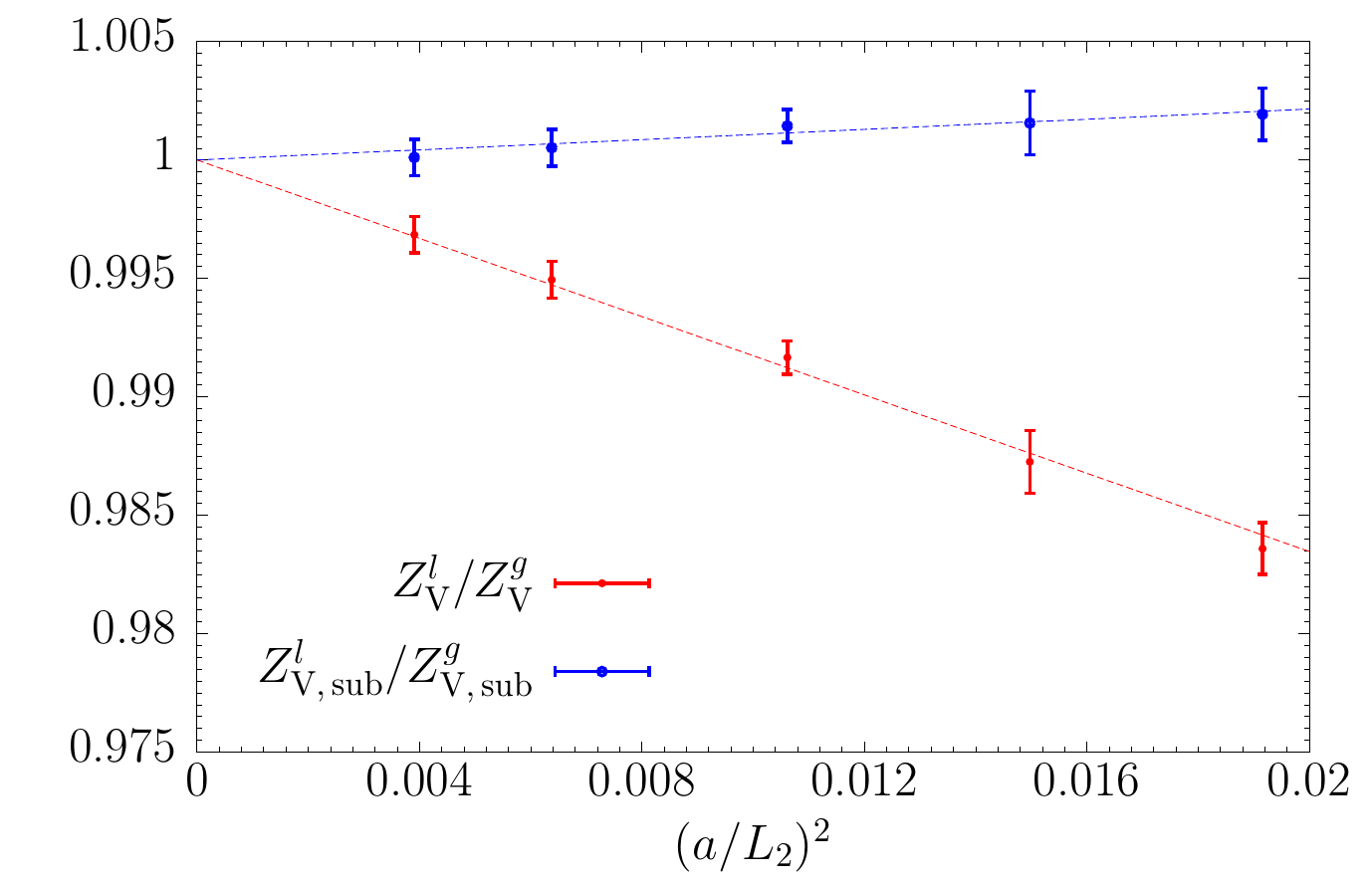}
	\caption{Continuum limit of the ratios between the $l$ and $g$ definitions 
		of $Z_{\rm A}$ (left panels) and $Z_{\rm V}$ (right panels) for the case of 
		$\Nf=3$ quark-flavours; the effect of subtracting the lattice artefacts from
		the $Z$-factors to O($g_0^2$) is also shown. The upper panels show the 
		$L_1$-LCP results while the lower ones show those of the $L_2$-LCP. In all 
		cases, the dashed lines correspond to linear fits to the data constrained to 
		extrapolate to 1 for $a/L_{1,2}=0$. Note that the (tiny) effect of the 
		statistical correlation between numerator and denominator has been neglected
		in these ratios.}
	\label{fig:ZrNf3}
\end{figure}         
It is also interesting to consider the continuum limit of the ratio between the definitions
belonging to different LCPs i.e. the $L_1$- and $L_2$-LCP. An example of such a ratio is 
shown in figure \ref{fig:ZrLCP}. Also in this case, the continuum scaling of this ratio is the one 
expected, and the initial difference is at the 2 per cent level. Apart from 
providing an important check of universality and automatic O($a$) improvement, these results show
that considering one definition or the other for the renormalization of matrix elements of the 
axial and vector currents, will only introduce small O($a^2$) differences over the whole range of
lattice spacings covered.
\begin{figure}[hpbt]
	\centering
	\includegraphics[scale=0.95]{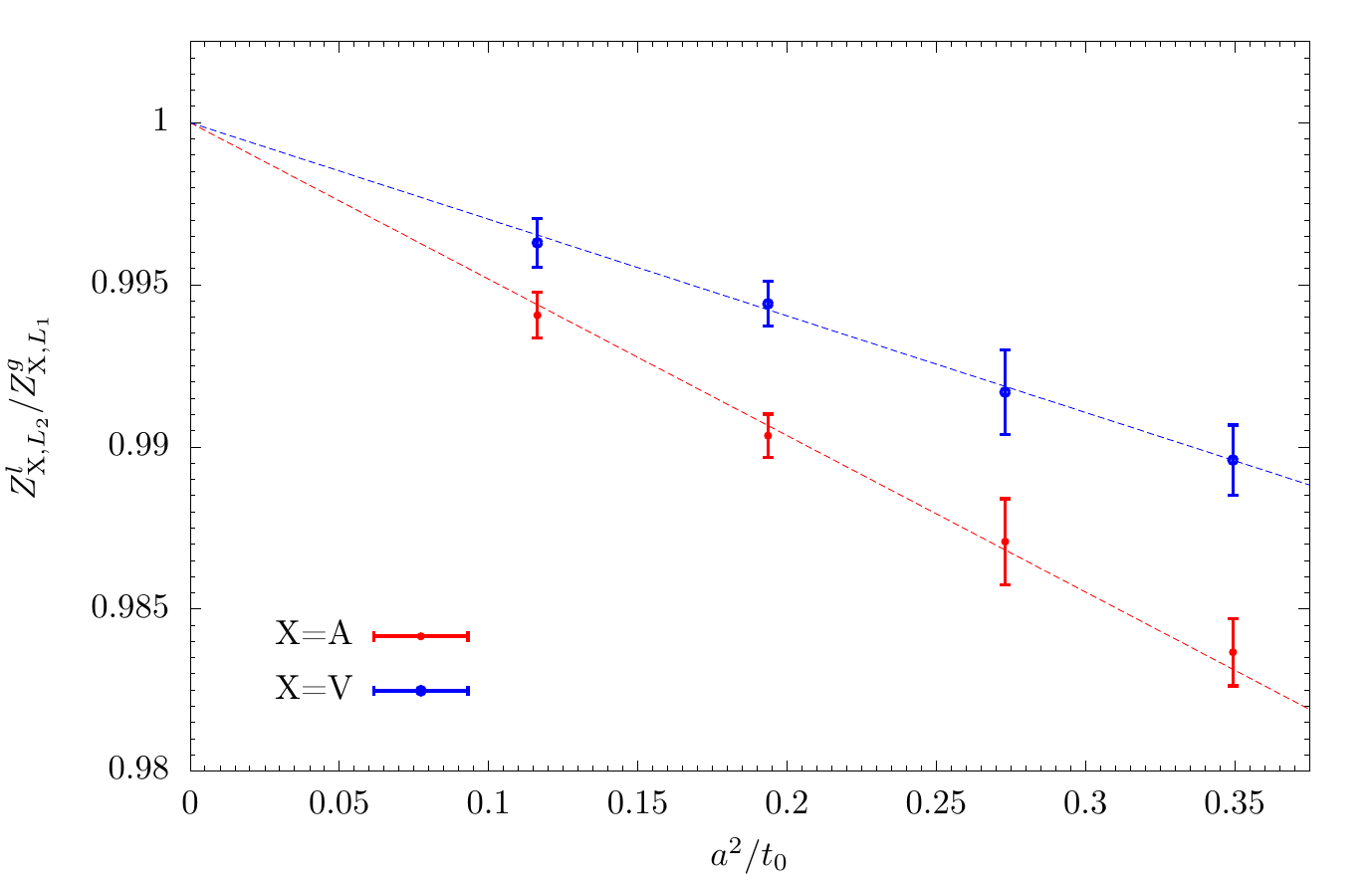}
	\caption{Continuum limit of the ratio between the $Z^l_{{\rm X},\,L_2}$ definitions, X=A,V, 
		corresponding to the $L_2$-LCP, and the $Z^g_{{\rm X},\,L_1}$ definitions corresponding
		to the $L_1$-LCP. The dashed lines correspond to linear fits to 
		the data constrained to extrapolate to 1 for $a^2/t_0=0$.}
	\label{fig:ZrLCP}
\end{figure}    

Finally, we look at ratios between $\xSF$ and standard SF determinations. 
Towards the continuum limit these should also scale like $1+{\rm O}(a^2)$, if the SF 
determinations are O($a$) improved. In figure~\ref{fig:ZASFrNf3} we show the continuum 
limit of the ratios between the standard SF determinations of ref.~\cite{Bulava:2016ktf} 
and the $\xSF$ results of table \ref{tab:ZNf3L2}. We here consider both 
definitions of this reference, and label them as $Z_{\rm A}^{\rm SF} = Z_{\rm A,0}$ and 
$Z_{\rm A,\,con}^{\rm SF}  = Z_{\rm A,0}^{\rm con}$, respectively  (cf.~\cite{Bulava:2016ktf}
for the exact definitions). 

As one can  see in figure~\ref{fig:ZASFrNf3}, for their preferred definition, $Z^{\rm SF}_{\rm A}$,
the expected scaling is only setting in around $a^2/t_0<0.2$, where the SF and  $\xSF$ determinations
differ by a couple of per cent. At the coarsest lattice spacing, corresponding
to $\beta=3.4$, the deviation from the O($a^2$) scaling is significant. The results for $Z_{\rm A}^g$
show the largest deviation from the SF determination, which is about 6\%. Considering the perturbatively
improved $\xSF$ results this difference is somewhat reduced to 4-5\%, but O($a^2$) scaling is
not observed either. If we consider instead the alternative definition, $Z^{\rm SF}_{\rm A,\,con}$,
the deviation is reduced to about 2 per cent at the coarsest lattice spacing for 
$Z_{\rm A}^l$, while, remarkably, the results for $Z_{\rm A}^g$ and $Z^{\rm con}_{\rm A,0}$
are compatible within errors. In particular, the difference between this SF and both our 
$\xSF$ determinations is perfectly compatible with an O($a^2$) effect over the whole range of lattice 
spacings considered. While discretization effects can only be defined with respect to some
reference definition, we conclude that the alternative SF definition $Z_{\rm A,\,con}^{\rm SF}$
is, within errors, perfectly scaling with $a^2$ for $\beta\ge 3.4$ relative to all $\xSF$ definitions,
whereas the preferred definition $Z_{\rm A}^{\rm SF}$ of ref.~\cite{Bulava:2016ktf}
requires much finer lattices before this expected asymptotic behaviour sets in.
With hindsight,  $Z_{\rm A,\,con}^{\rm SF}$ seems to be a better choice within the SF framework 
and also has been the preferred SF definition within the $\Nf=2$ 
setup of refs.~\cite{Fritzsch:2012wq,DellaMorte:2008xb}.

\begin{figure}[hptb]
	
	\includegraphics[scale=0.5]{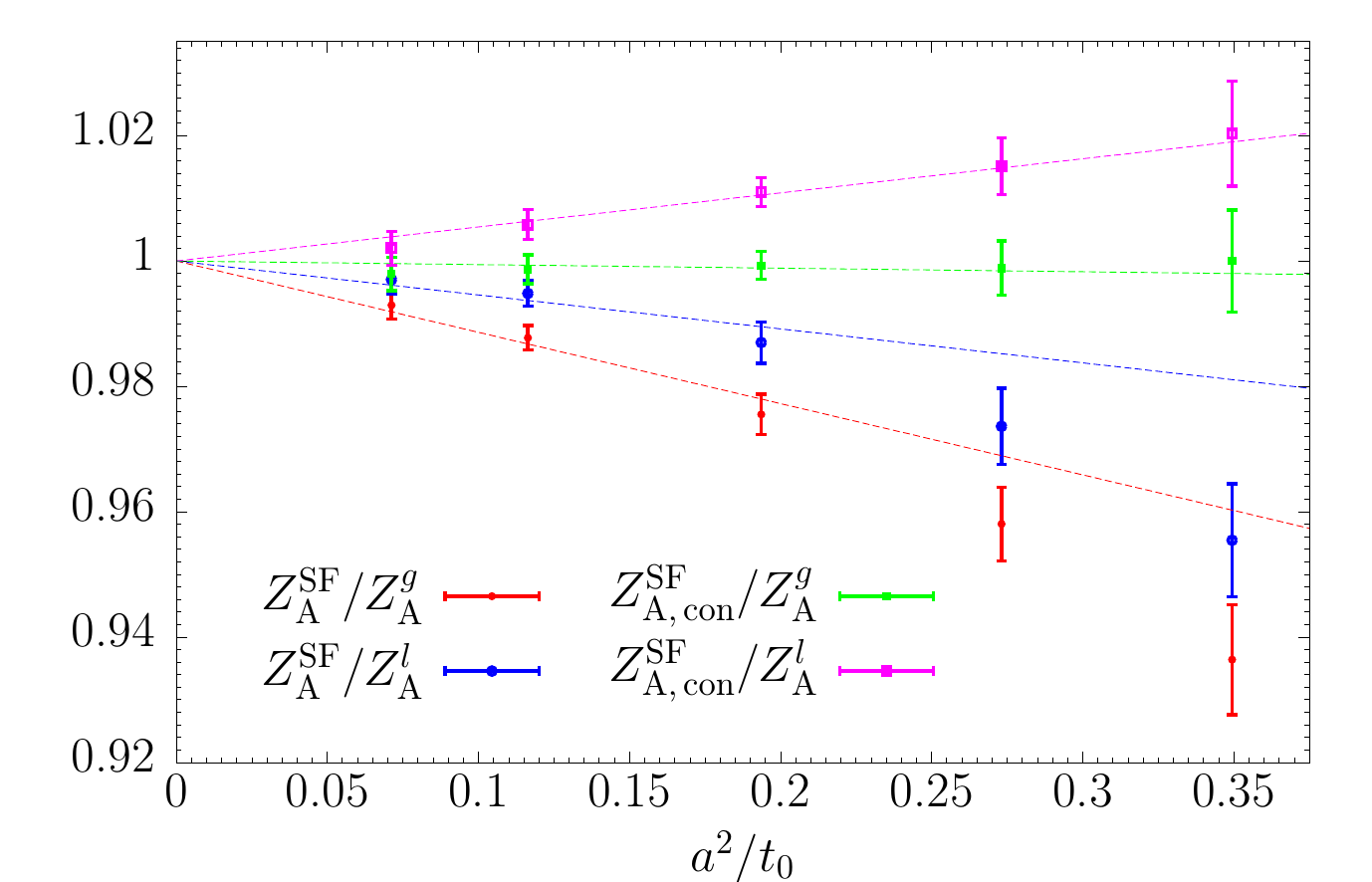}
	\includegraphics[scale=0.5]{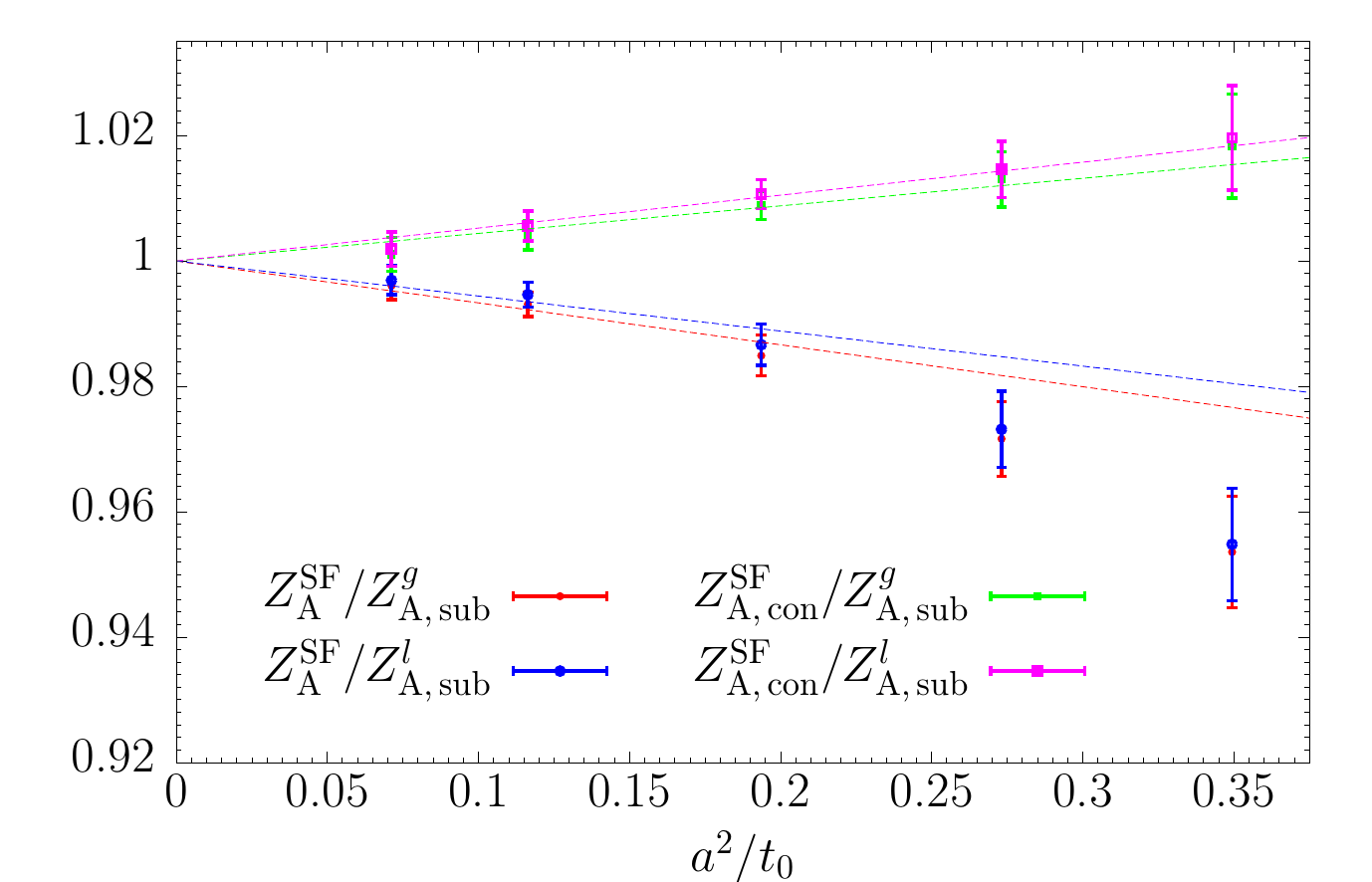}
	\caption{Continuum limit of the ratios between the $\Nf=3$ WI determinations of 
		$Z_{\rm A}$ of ref.~\cite{Bulava:2016ktf}, and the $\xSF$ determinations 
		$Z_{\rm A}^{g,l}$ (left panel) and $Z_{\rm A,\,sub}^{g,l}$ (right panel)
		of table~\ref{tab:ZNf3L2}. The $Z_{\rm A}^{\rm SF}$ results are from the fit 
		formula provided in ref.~\cite{Bulava:2016ktf}, and correspond to their 
		preferred, $Z_{\rm A,0}$, definition. The associated dashed lines (red and blue lines)
		are linear fits to the data with $a^2/t_0<0.2$, constrained to extrapolate to 1
		for $a^2/t_0=0$. The $Z_{\rm A,\,con}^{\rm SF}$ results come instead from a fit
		of the results for the alternative, $Z_{\rm A,0}^{\rm con}$, definition considered 
		in ref.~\cite{Bulava:2016ktf}. The latter fit was obtained using the same fit
		ansatz used in ref.~\cite{Bulava:2016ktf} for $Z_{\rm A,0}$. The associated dashed 
		lines (green and magenta lines) are linear fits to all data, constrained to
		extrapolate to 1 for $a^2/t_0=0$.}
	\label{fig:ZASFrNf3}
\end{figure}

\section{Summary and conclusions}
\label{sec:Conclusions}

We have used a new method~\cite{Brida:2016rmy} based on the chirally rotated
Schr\"odinger functional~\cite{Sint:2010eh} to obtain high precision results
for the normalization constants of the Noether currents corresponding to 
non-singlet chiral and flavour symmetries. The matrix elements of these axial 
and vector currents play a crucial r\^ole in various contexts of hadronic physics.
Our method differs from the traditional Ward identity method~\cite{Bochicchio:1985xa,Luscher:1996jn} 
in that it compares correlation functions which are related by finite chiral or flavour rotations, 
rather than infinitesimal ones. The major advantage compared to the Ward identity
method consists in the avoidance of 3- and 4-point functions in favour of simple 2-point functions. 
This very significantly improves on the precision achieved in previous 
determinations~\cite{DellaMorte:2005rd,DellaMorte:2008xb,Fritzsch:2012wq,Bulava:2016ktf,Heitger:2017njs}. 
In particular, for the case of $Z_{\rm A}$, we obtain a reduction of the error by up to an order
of magnitude (cf.~figure~\ref{fig:ZaNf2} and \ref{fig:ZaNf3})}. 
The relatively poor precision obtained for $Z_{\rm A}$ with the traditional Ward identity 
methods~\cite{DellaMorte:2005rd,DellaMorte:2008xb,Fritzsch:2012wq,Bulava:2016ktf} 
(around the percent level at the coarsest lattice spacings of interest), has now become
a limiting factor in several applications.  For this reason, our results are in high 
demand and have already been used in several works~\cite{Bruno:2016plf,Campos:2018ahf,Koponen:2018}. 
In particular, the precise $\Nf=2+1$ scale setting from a linear combination of $f_K$ and
$f_\pi$ in ref.~\cite{Bruno:2016plf} crucially relies on our values of $Z_{\rm A}^l$ in 
table~\ref{tab:ZNf3L1} and the associated uncertainty is negligible compared
to the statistical error of the bare hadronic matrix elements. 
In turn, the precise scale setting result of~\cite{Bruno:2016plf} 
is entering almost all studies done with CLS gauge configurations: 
in particular it has enabled the precise result for the 3-flavour QCD 
$\Lambda$-parameter and thus $\alpha_s(m_Z)$ by the ALPHA-collaboration~\cite{Brida:2016flw,
DallaBrida:2016kgh,Bruno:2017gxd,DallaBrida:2018rfy}. Further applications of 
our $Z_{\rm A}$-results include the  non-perturbative quark mass renormalization factor 
in~\cite{Campos:2018ahf} and the related determination of the light and strange 
quark masses~\cite{Koponen:2018}. 
Regarding the $\Nf=2$ case, the potential improvement of the scale 
setting in ref.~\cite{Fritzsch:2012wq} due to our $Z_{\rm A}$-results would be very significant, too. 
Tentative estimates anticipate a gain by a factor $3-6$ in precision, when going from the finest to 
the coarsest lattice spacing~\cite{Lottini:2014}.

In order to maximize the usefulness of our results we have chosen the same actions and the
same $\beta$-values for $\Nf=2$ and $\Nf=3$ lattice QCD as used by the CLS
initiative~\cite{Fritzsch:2012wq,Bruno:2014jqa}. Hence, anyone working with CLS gauge 
configurations will be able to directly use our results: for $\Nf=2$ we recommend to use
$Z_{\rm A,V,\,sub}^l$ from table~\ref{tab:ZNf2}, and for $\Nf=3$ we recommend
using $Z_{\rm A,V,\,sub}^l$ either of table \ref{tab:ZNf3L1} or \ref{tab:ZNf3L2}. 
Although the results for $Z_{\rm A,V,\,sub}^l$ are slightly less precise than those for 
$Z_{\rm A,V,\,sub}^g$, their $L/a$-interpolations turn out to be more robust. Furthermore,
the effect of the perturbative subtraction of cutoff effects is rather small and
only marginally significant with current errors.
While the precise choice of the $\xSF$ results for the $Z$-factors is not crucial, 
it is however very important to be consistent and to not switch definitions 
when changing $\beta$. Only then cutoff effects are guaranteed to vanish smoothly at a rate
$\propto a^2$.

Our determination of $Z_{\rm A,V}(\beta)$ was carried out for each $\beta$-value independently, 
in order to avoid adding statistical correlation between physics results at different lattice
spacings. However, it is straightforward to fit our $Z$-factors to a smooth function
of $\beta$ (or $g_0^2$), which interpolates to any intermediate $\beta$-value. We have included a 
few such fits in appendix \ref{app:Fits} to our preferred definitions  $Z_{\rm A,V,\,sub}^l$.
We also include fits which incorporate the expected perturbative
behaviour to 1-loop order. However, the high precision obtained in the $\beta$-range covered
by the data cannot be guaranteed outside this range. If a similar precision is required
at higher $\beta$, an extension of our non-perturbative determination will be required.
If $t_0/a^2$ was known for higher $\beta$-values one could extend our chosen line of constant physics 
covering another factor of 2 or so in the lattice spacing. The required simulations of the $\chi$SF for 
lattice sizes up to $L/a=32$ would be feasible with current resources. Going beyond this range it
may be advisable to choose a different line of constant physics from a finite volume observable,
or at least estimate the errors incurred by deviating from the original choice.

In applications to hadronic physics one would also like to control the O($a$) effects cancelled
by  the counterterms to the currents. Close to the chiral limit,
one essentially requires the counterterm coefficients $c_{\rm A,V}$~\cite{Luscher:1996sc,Sint:1997jx}.
We emphasize that our method of determining the $Z$-factors does not rely on any assumptions about
these counterterms and can therefore be combined with results for $c_{\rm A,V}$ 
from other studies, e.g.~\cite{Bulava:2015bxa,Heitger:2017njs}. The same remark applies to the 
$b$-coefficients multiplying O($am$) counterterms, which have recently been determined 
for the vector current in ref.~\cite{Fritzsch:2018zym}.

Looking beyond direct applications of our results in the CLS context, it is quite obvious that 
the precision gains of this method are generic and could be implemented with any other choice of
Wilson type fermions. One would need to implement the $\chi$SF boundary conditions following 
ref.~\cite{Sint:2010eh}, as well as the $\chi$SF correlation functions~\cite{Brida:2016rmy}. 
We also note that the computer resources required are rather modest: in fact our largest lattice
size was $16^4$; indeed, the main work for the present results went into painstakingly following
lines of constant physics and the determination of the corresponding uncertainties and their 
propagation to the $Z$-factors. We have reported many technical details in the hope
that any further applications of the method will be able to benefit from our experience. 
One possible improvement we did not explore was to measure the derivatives 
(\ref{eq:m0zf2mgA},\ref{eq:dZdX}) by computing the corresponding operator insertions into the 
correlation functions directly on the tuned ensembles; this was done e.g.~in refs.~\cite{Bruno:2016plf,
Bruno:2017gxd} for the PCAC mass, $t_0$, and other observables, and this would certainly allow one 
to further improve on the precision, as no assumptions on the derivatives need to be made. 

Possible future applications of the $\chi$SF include the determination of the ratio between pseudo-scalar
and scalar renormalization constants, $Z_{\rm P}/Z_{\rm S}$. Advantages of the 
$\chi$SF are also expected for scale-dependent problems, such as the renormalization of 
4-quark operators, where the contamination by O($a$) effects could be significantly reduced by the
mechanism of automatic O($a$) improvement~\cite{Mainar:2016uwb}. Finally the $\chi$SF offers new 
methods for the determination of O($a$) improvement coefficients, which we hope to explore in the 
future.

\section{Acknowledgments}

We would like to thank Rainer Sommer and Stefano Lottini for useful discussions, and 
Patrick Fritzsch, Tim Harris, and Alberto Ramos for helpful comments on the analysis
of the data. We are moreover thankful to the computer centers at ICHEC, LRZ (project id pr84mi), 
and DESY-Zeuthen for the allocated computer resources and support. The code we used for 
the simulations is based on the openQCD package developed at CERN~\cite{LuscherWeb:2016}.

\clearpage

\appendix

\section{Simulation parameters and results}
\label{app:Simulations}

\begin{table}[htp!]
	\centering
	\begin{tabular}{llllllll}
	\toprule
	 $L/a$ & $\beta$ & $\kappa$ & $z_f$ & $mL\times10^{3}$ & $g_{\rm A}^{ud}(L/2)\times10^{3}$ & $P_{\rm Q}$ & $N_{\rm ms}$\\
	\midrule
	$ 8$ & $5.2$ & $0.1356450$ & $1.28300$ & $-0.8(1.4)$ & $-2.9(2.2)$ & $99.8\%$ & $8002$ \\
	$ 8$ & $5.3$ & $0.1361712$ & $1.28680$ & $\phantom{-} 0.9(1.0)$ & $\phantom{-} 1.0(1.5)$ & $99.9\%$ & $12002$ \\
	$10$ & $5.3$ & $0.1362811$ & $1.30900$ & $-3.1(1.7)$ & $\phantom{-} 6.6(2.3)$ & $98.6\%$ & $4004$ \\
	$12$ & $5.3$ & $0.1363310$ & $1.32280$ & $\phantom{-} 0.1(2.3)$ & $\phantom{-} 7.5(2.8)$ & $93.8\%$ & $2403$ \\
	$12$ & $5.5$ & $0.1367093$ & $1.31120$ & $\phantom{-} 1.8(1.6)$ & $-2.0(1.9)$ & $99.4\%$ & $2403$ \\
	$16$ & $5.7$ & $0.1367058$ & $1.30600$ & $-0.2(1.4)$ & $\phantom{-} 1.4(1.6)$ & $100\%$  & $1803$ \\
	\bottomrule
	\end{tabular}
	\caption{Parameters of the $\Nf=2$ ensembles and corresponding results for $mL$ and $g_{\rm A}^{ud}(L/2)$. 
	 The total number of measurements we collected is given by $N_{\rm ms}$; 
	 these are spaced by 10 MDUs. In the table we also give the percentage $P_Q$ of gauge fields with $Q=0$.}
\label{tab:Nf2Parms}
\end{table}

\begin{table}[hbpt!]
	\centering
	\begin{tabular}{llllllll}
	\toprule
	$L/a$ & $\beta$ & $\kappa$ & $z_f$ & $mL\times10^{3}$ & $g_{\rm A}^{ud}(L/2)\times10^{3}$ & $P_Q$ & $N_{\rm ms}$ \\
	\midrule
	$ 6$ & $3.40$ & $0.1364794$ & $1.33331$ & $-0.19(99)$ & $\phantom{-} 3.1(1.6)$ & 99.9\% & 31208 \\
	$ 8$ & $3.40$ & $0.1366405$ & $1.37100$ & $-0.3(1.0)$ & $-1.4(1.6)$ & 98.4\% & 18620\\
	$10$ & $3.40$ & $0.1367529$ & $1.40741$ & $-2.0(1.8)$ & $-2.7(2.5)$ & 92.3\% & 14416\\
	$12$ & $3.40$ & $0.1368158$ & $1.43650$ & $\phantom{-} 0.5(2.0)$ & $-2.0(2.8)$ & 72.9\% & 21685\\
	$ 6$ & $3.46$ & $0.1367283$ & $1.33580$ & $\phantom{-} 0.06(84)$ & $-0.9(1.5)$ & 99.95\% & 31208 \\
	$ 8$ & $3.46$ & $0.1368457$ & $1.36250$ & $\phantom{-} 1.2(1.5)$ & $\phantom{-} 0.2(2.1)$ & 99.4\% & 8002\\
	$10$ & $3.46$ & $0.1369430$ & $1.39170$ & $-3.7(1.5)$ & $\phantom{-} 3.4(2.1)$ & 96.1\% & 12441\\
	$12$ & $3.46$ & $0.1369731$ & $1.40600$ & $\phantom{-} 2.9(2.2)$ & $-0.5(2.8)$ & 83.1\% & 5109 \\
	$ 8$ & $3.55$ & $0.1370247$ & $1.35500$ & $-0.3(1.3)$ & $-0.4(1.8)$ & 99.85\% & 8002\\
	$10$ & $3.55$ & $0.1370827$ & $1.37200$ & $-0.45(95)$ & $\phantom{-} 0.7(1.3)$ & 99.1\% & 7712 \\
	$12$ & $3.55$ & $0.1371100$ & $1.38320$ & $-0.6(1.6)$ & $\phantom{-} 2.0(2.1)$ & 96.4\% & 4004 \\
	$16$ & $3.55$ & $0.1371487$ & $1.40530$ & $-0.3(1.7)$ & $-3.3(1.9)$ & 71.2\% & 6404\\
	$ 8$ & $3.70$ & $0.1370673$ & $1.34250$ & $-0.55(99)$ & $\phantom{-} 0.8(1.5)$ & 100\% & 8002 \\
	$10$ & $3.70$ & $0.1370938$ & $1.35164$ & $\phantom{-} 0.9(1.6)$ & $-3.9(2.4)$ & 99.4\% & 3208 \\
	$12$ & $3.70$ & $0.1371160$ & $1.35860$ & $-2.2(1.3)$ & $\phantom{-} 0.0(2.0)$ & 99.9\% & 8000\\
	$16$ & $3.70$ & $0.1371370$ & $1.36790$ & $-1.0(1.2)$ & $\phantom{-} 0.7(1.3)$ & 91.9\% & 4004\\
	$16$ & $3.85$ & $0.1369595$ & $1.34540$ & $\phantom{-} 0.6(1.0)$ & $\phantom{-} 1.4(1.1)$ & 99.1\% & 4354\\
	\bottomrule
	\end{tabular}
	\caption{Parameters of the $\Nf=3$ ensembles and corresponding results for $mL$ and $g_{\rm A}^{ud}(L/2)$.
    The total number of measurements we collected is given by $N_{\rm ms}$; 
	these are spaced by 10 MDUs. In the table we also give the percentage $P_Q$ of gauge fields with $Q=0$.}
	\label{tab:Nf3Parms}
\end{table}

\begin{sidewaystable}[htpb!]
	\scalebox{0.9}{	
	\centering
	\begin{tabular}{llllllll}
	\toprule
	$L/a$ & $\beta$ & $\kappa$ & $z_f$ & $Z_{\rm A}^g$ & $Z_{\rm A}^l$ & $Z_{\rm V}^g$ & $Z_{\rm V}^l$ \\\midrule
	$8$ & $5.2$ & $0.1356450$ & $1.28300$ & $  0.78022 (24) (44) (22) $ & $  0.76944 (19) (43) (82) $ & $  0.746732 (128) (443) (94) $ & $  0.73849 (21) (46) (83) $ \\
	$8$ & $5.3$ & $0.1361712$ & $1.28680$ & $  0.78767 (16) (34) (15) $ & $  0.77754 (13) (33) (55) $ & $  0.756611 (93) (346) (62) $ & $  0.74806 (14) (36) (55) $ \\
	$10$ & $5.3$ & $0.1362811$ & $1.30900$ & $  0.78204 (22) (76) (36) $ & $  0.77505 (18) (74) (135) $ & $  0.75000 (12) (77) (15) $ & $  0.74539 (19) (80) (136) $ \\
	$12$ & $5.3$ & $0.1363310$ & $1.32280$ & $  0.77957 (30) (55) (33) $ & $  0.77331 (22) (54) (124) $ & $  0.74639 (15) (56) (14) $ & $  0.74327 (25) (58) (126) $ \\
	$12$ & $5.5$ & $0.1367093$ & $1.31120$ & $  0.79450 (18) (57) (24) (111) $ & $  0.78955 (13) (56) (88) (79) $ & $  0.76634 (12) (58) (10) (131) $ & $  0.76253 (15) (60) (89) (86) $ \\
	$16$ & $5.7$ & $0.1367058$ & $1.30600$ & $  0.80526 (12) (35) (16) (88) $ & $  0.802768 (86) (345) (587) (624) $ & $  0.779955 (88) (358) (67) (1039) $ & $  0.778014 (99) (371) (595) (676) $ \\
	\bottomrule
	\end{tabular}}
	\caption{Renormalization constants $Z_{\rm A,V}^{g,l}$ for the different $\Nf=2$ ensembles. The four errors refer to (cf.~appendix \ref{app:Syst}): statistical, systematic coming from the uncertainty on $am_{\rm cr}$ ($\Delta_{m_0} Z_{\rm A,V}$), systematic coming from the uncertainty on $z^*_f$ ($\Delta_{z_f} Z_{\rm A,V}$), and systematic coming from maintaining the condition (\ref{eq:fKL}) ($\Delta_{x} Z_{\rm A,V}$); the latter is only relevant for the last two ensembles.}
	\label{tab:ZAVRawNf2}
	
	\vspace*{25mm}
		
	\scalebox{0.9}{
	\centering
	\begin{tabular}{llllllll}
	\toprule
	$L/a$ & $\beta$ & $\kappa$ & $z_f$ & $Z_{\rm A,\,sub}^g$ & $Z_{\rm A,\,sub}^l$ & $Z_{\rm V,\,sub}^g$ & $Z_{\rm V,\,sub}^l$ \\\midrule
	$8$ & $5.2$ & $0.1356450$ & $1.28300$ & $  0.77262 (24) (44) (22) $ & $  0.76963 (19) (43) (82) $ & $  0.739864 (127) (443) (94) $ & $  0.73890 (21) (46) (83) $ \\
	$8$ & $5.3$ & $0.1361712$ & $1.28680$ & $  0.78016 (15) (34) (15) $ & $  0.77772 (13) (33) (55) $ & $  0.749804 (92) (346) (62) $ & $  0.74846 (14) (36) (55) $ \\
	$10$ & $5.3$ & $0.1362811$ & $1.30900$ & $  0.77738 (22) (76) (36) $ & $  0.77519 (18) (74) (135) $ & $  0.74570 (12) (77) (15) $ & $  0.74562 (19) (80) (136) $ \\
	$12$ & $5.3$ & $0.1363310$ & $1.32280$ & $  0.77638 (30) (55) (33) $ & $  0.77342 (22) (54) (124) $ & $  0.74343 (15) (56) (14) $ & $  0.74342 (25) (58) (126) $ \\
	$12$ & $5.5$ & $0.1367093$ & $1.31120$ & $  0.79138 (18) (57) (24) (62) $ & $  0.78965 (13) (56) (88) (80) $ & $  0.76343 (12) (58) (10) (88) $ & $  0.76268 (15) (60) (89) (89) $ \\
	$16$ & $5.7$ & $0.1367058$ & $1.30600$ & $  0.80358 (12) (35) (16) (49) $ & $  0.802833 (86) (345) (587) (631) $ & $  0.778359 (87) (358) (67) (693) $ & $  0.778097 (99) (371) (595) (699) $ \\
	\bottomrule
	\end{tabular}}
	\caption{Renormalization constants $Z_{\rm A,V,\,sub}^{g,l}$ for the different $\Nf=2$ ensembles. Lattice artefacts have been subtracted at O($g_0^2$) in perturbation theory. The four errors refer to (cf.~appendix \ref{app:Syst}): statistical, systematic coming from the uncertainty on $am_{\rm cr}$ $(\Delta_{m_0} Z_{\rm A,V})$, systematic coming from the uncertainty on $z^*_f$ $(\Delta_{z_f} Z_{\rm A,V})$, and systematic coming from maintaining the condition (\ref{eq:fKL}) ($\Delta_{x} Z_{\rm A,V}$); the latter is only relevant for the last two ensembles.
	Comparing with the results of table \ref{tab:ZAVRawNf2}, only the mean values, statistical errors, and errors associated with maintaining the condition (\ref{eq:fKL}), are different. The mean values and corresponding statistical errors can be obtained from those of table \ref{tab:ZAVRawNf2} through eq.~(\ref{eq:PThIZ}). For determining the systematic errors associated with the uncertainty on $am_{\rm cr}$ and $z^*_f$ we use the same estimates for the relevant derivatives as for the results in table \ref{tab:ZAVRawNf2}: we thus obtain the same values.
	The systematic errors associated with maintaining the condition (\ref{eq:fKL}), instead, involve different derivatives for the data in table \ref{tab:ZAVRawNf2} and	the perturbatively improved ones (cf.~appendix \ref{app:Syst}).}
	\label{tab:ZAVRawNf2PT}	
\end{sidewaystable}

\clearpage

\begin{sidewaystable}[htpb!]
	\centering
	\begin{tabular}{llllllll}
	\toprule
	$L/a$ & $\beta$ & $\kappa$ & $z_f$ & $Z_{\rm A}^g$ & $Z_{\rm A}^l$ & $Z_{\rm V}^g$ & $Z_{\rm V}^l$ \\\midrule
	$6$ & $3.40$ & $0.1364794$ & $1.33331$ & $  0.77905 (18) (38) (15) $ & $  0.75899 (15) (29) (69) $ & $  0.74294 (11) (32) (11) $ & $  0.72534 (18) (33) (71) $ \\
	$8$ & $3.40$ & $0.1366405$ & $1.37100$ & $  0.768471 (184) (286) (98) $ & $  0.75446 (14) (32) (58) $ & $  0.729230 (84) (253) (35) $ & $  0.71940 (15) (37) (58) $ \\
	$10$ & $3.40$ & $0.1367529$ & $1.40741$ & $  0.76315 (28) (68) (20) $ & $  0.75190 (20) (77) (120) $ & $  0.721224 (93) (603) (72) $ & $  0.71552 (27) (88) (119) $ \\
	$12$ & $3.40$ & $0.1368158$ & $1.43650$ & $  0.76227 (41) (54) (18) $ & $  0.75037 (22) (61) (105) $ & $  0.716639 (76) (478) (63) $ & $  0.71220 (33) (70) (104) $ \\
	$6$ & $3.46$ & $0.1367283$ & $1.33580$ & $  0.784318 (159) (294) (57) $ & $  0.76501 (13) (22) (44) $ & $  0.750160 (100) (263) (46) $ & $  0.73259 (15) (26) (44) $ \\
	$8$ & $3.46$ & $0.1368457$ & $1.36250$ & $  0.77445 (25) (46) (17) $ & $  0.76154 (21) (60) (76) $ & $  0.738058 (142) (416) (69) $ & $  0.72806 (22) (60) (70) $ \\
	$10$ & $3.46$ & $0.1369430$ & $1.39170$ & $  0.76852 (21) (75) (27) $ & $  0.75933 (18) (98) (126) $ & $  0.730708 (84) (677) (114) $ & $  0.72552 (21) (97) (116) $ \\
	$12$ & $3.46$ & $0.1369731$ & $1.40600$ & $  0.76711 (41) (81) (27) $ & $  0.75760 (29) (105) (123) $ & $  0.72743 (14) (72) (11) $ & $  0.72293 (27) (104) (113) $ \\
	$8$ & $3.55$ & $0.1370247$ & $1.35500$ & $  0.78311 (20) (35) (12) $ & $  0.77108 (16) (31) (46) $ & $  0.749621 (124) (301) (68) $ & $  0.73993 (17) (35) (46) $ \\
	$10$ & $3.55$ & $0.1370827$ & $1.37200$ & $  0.77803 (13) (29) (10) $ & $  0.76936 (10) (26) (38) $ & $  0.743999 (68) (249) (56) $ & $  0.73813 (11) (29) (38) $ \\
	$12$ & $3.55$ & $0.1371100$ & $1.38320$ & $  0.77595 (25) (45) (17) $ & $  0.76786 (21) (40) (64) $ & $  0.740868 (125) (393) (94) $ & $  0.73692 (17) (46) (65) $ \\
	$16$ & $3.55$ & $0.1371487$ & $1.40530$ & $  0.77378 (42) (47) (19) $ & $  0.76676 (28) (41) (69) $ & $  0.73723 (10) (40) (10) $ & $  0.73456 (20) (47) (70) $ \\
	$8$ & $3.70$ & $0.1370673$ & $1.34250$ & $  0.796626 (144) (279) (61) $ & $  0.78571 (12) (25) (38) $ & $  0.766997 (103) (304) (62) $ & $  0.75716 (12) (27) (38) $ \\
	$10$ & $3.70$ & $0.1370938$ & $1.35164$ & $  0.79223 (18) (44) (11) $ & $  0.78454 (16) (39) (71) $ & $  0.76247 (12) (48) (12) $ & $  0.75626 (18) (43) (71) $ \\
	$12$ & $3.70$ & $0.1371160$ & $1.35860$ & $  0.789547 (139) (539) (96) $ & $  0.78386 (11) (47) (60) $ & $  0.759810 (87) (588) (98) $ & $  0.75576 (13) (53) (61) $ \\
	$16$ & $3.70$ & $0.1371370$ & $1.36790$ & $  0.787298 (140) (372) (70) $ & $  0.78283 (15) (33) (44) $ & $  0.757196 (76) (405) (71) $ & $  0.754818 (99) (363) (440) $ \\
	$16$ & $3.85$ & $0.1369595$ & $1.34540$ & $  0.799848 (82) (294) (62) $ & $  0.796571 (88) (259) (386) $ & $  0.773042 (58) (321) (63) $ & $  0.770607 (70) (287) (387) $ \\
	\bottomrule
	\end{tabular}
	\caption{Renormalization constants $Z_{\rm A,V}^{g,l}$ for the different $\Nf=3$ ensembles. The three errors refer to (cf.~appendix \ref{app:Syst}): statistical, systematic coming from the uncertainty on $am_{\rm cr}$ $(\Delta_{m_0} Z_{\rm A,V})$, and systematic coming from the uncertainty on $z^*_f$ $(\Delta_{z_f} Z_{\rm A,V})$. The corresponding results with the O($g_0^2$) lattice artefacts subtracted are obtained from those above by applying eq.~(\ref{eq:PThIZ}) to the mean values and corresponding statistical errors. The systematic uncertainties are instead left unchanged (cf.~table \ref{tab:ZAVRawNf2PT}).}
	\label{tab:ZAVRawNf3}
\end{sidewaystable}

\clearpage

\section{Error propagation, systematic error estimates and comparison with 
perturbation theory}
\label{app:Syst}

In this appendix we describe in some detail the elements required to carry
out the steps 2,3 and 4 sketched in subsection~\ref{sec:Uncertainties}, for
the propagation of uncertainties. We first report on the numerical estimates
of the various derivatives (steps 2,3) for both $\Nf=2$ and $\Nf=3$. These
estimates are obtained on smaller lattices $L/a=8$ and $L/a=6,8$ respectively,
and then used for all lattice sizes. 
We therefore also summarize the expected scaling with $L/a$ of these 
derivatives and confirm this to first non-trivial order in perturbation theory. 
Finally, the interpolation of the $Z$-factors in $L/a$ (step 4, where necessary)
is discussed, first for $\Nf=2$, where this step is almost avoidable, and then 
for $\Nf=3$, where interpolations are necessary at most $\beta$-values, thus requiring
a more thorough analysis.

\subsection{Estimating the derivatives: $\Nf=2$}

As described in section \ref{sec:Uncertainties}, to estimate the uncertainty in
$Z_{\rm A,V}$ originating from those of $am_{\rm cr}$ and $z_f^*$,
we require the derivatives (\ref{eq:m0zf2mgA}) and (\ref{eq:dZdX}). 
We estimated these through dedicated simulations at $L/a=8$ and $\beta=5.2$, 
measuring the relevant quantities for several different values of $\kappa$ ($z_f$) at fixed $z_f$ 
($\kappa$), straddling the tuned values given in table \ref{tab:Nf2Parms}. The results we obtained
for the derivatives (\ref{eq:m0zf2mgA}) are:
\begin{align}
 \nonumber
 {\partial mL\over\partial m_0L}&=1.251(68),
 &
 {\partial mL\over\partial z_f}&=0.048(73),\\
 \label{eq:m0zf2mgANf2}
 {\partial g^{ud}_{\rm A} \over\partial m_0L}&=-2.719(96),
 &
 {\partial g^{ud}_{\rm A}\over\partial z_f}&=-2.39(10).
\end{align}
We observe that the $z_f$-derivative of $mL$ vanishes within an uncertainty much smaller than 
the values of the other derivatives, so that we may safely set it to zero. 
These results were then used for all other ensembles and lattice sizes listed in table \ref{tab:Nf2Parms}.
The uncertainty in the tuning of $mL$ and $g_{\rm A}^{ud}(L/2)$ is thus converted
to one in $m_{\rm cr}L$ and $z_f^*$. Specifically, one has that,
\begin{equation}
 \Bigg(
 \begin{array}{c}
  \Delta (m_{\rm cr}L) \\ 
  \Delta z_f^*
 \end{array}\Bigg)
 = A^{-1}
 \Bigg(
 \begin{array}{c}
 \Delta (mL) \\
 \Delta g^{ud}_{\rm A}
 \end{array}\Bigg)
 \quad
 \text{where}
 \quad
  A=
  \Bigg(
 \begin{array}{cc}
  {\partial mL\over\partial m_0L} &
  {\partial mL\over\partial z_f} \\
  {\partial g^{ud}_{\rm A}\over\partial m_0L} &
  {\partial g^{ud}_{\rm A}\over\partial z_f}
 \end{array}\Bigg),
\end{equation}
and $\Delta(mL)$, $\Delta g_{\rm A}^{ud}$, $\Delta (m_{\rm cr}L)$ and $\Delta z_f^*$,
are the uncertainties in $mL$, $g_{\rm A}^{ud}(L/2)$, $m_{\rm cr}L$ and $z_f^*$,
respectively. For the uncertainties $\Delta(mL)$ and $\Delta g_{\rm A}^{ud}$
we took the (absolute) values of $mL$ and $g_{\rm A}^{ud}(L/2)$ measured on 
the given ensemble (cf.~table \ref{tab:Nf2Parms}), plus 2 times the corresponding 
statistical errors. The corresponding systematic errors on the $Z$-factors can 
then be estimated as,
\begin{equation}
 \label{eq:SystNf2}
 (\Delta_{m_0} Z_{{\rm X}})^2 + 
 (\Delta_{z_f} Z_{{\rm X}})^2 = 
 \bigg({\partial Z_{\rm X}\over\partial m_0L}\bigg)^2 (\Delta m_{\rm cr}L)^2 +
 \bigg({\partial Z_{\rm X}\over\partial z_f}\bigg)^2 (\Delta z_f^*)^2,
 \quad
 {\rm X=A,V}.
\end{equation}
Through the very same simulations used to determine the derivatives 
(\ref{eq:m0zf2mgANf2}) we obtained,
\begin{align}
 \nonumber
 {\partial Z_{\rm A}^g \over\partial m_0L} &=  0.126(11),
 &
 {\partial Z_{\rm A}^l \over\partial m_0L} &= -0.1266(88),\\
 {\partial Z_{\rm V}^g \over\partial m_0L} &=  0.1376(60),
 &
 {\partial Z_{\rm V}^l \over\partial m_0L} &= -0.1356(98),
 \label{eq:dZdm0LNf2}
\end{align}
and
\begin{align}
 \nonumber
 {\partial Z_{\rm A}^g \over\partial z_f}&=-0.011(12),
 &
 {\partial Z_{\rm A}^l \over\partial z_f}&=-0.1092(93),\\
 {\partial Z_{\rm V}^g \over\partial z_f}&=-0.0017(65),
 &
 {\partial Z_{\rm V}^l \over\partial z_f}&=-0.108(11).
 \label{eq:dZdzfNf2}
\end{align}
We then used these values at all other $L/a$- and $\beta$-values of table \ref{tab:Nf2Parms}. 
More precisely, we propagated the errors according to eq.~(\ref{eq:SystNf2})
using the measured absolute  mean values to which we added twice the statistical errors.
Taking these results at the coarsest available lattice spacing 
is a conservative choice which should be safe, also given that 
some favourable expected scaling of the derivative towards larger $L/a$ (s.~below) 
is not taken advantage of. Complementary information obtained during
the tuning runs for finding $m_{\rm cr}$ and $z_f^*$, further corroborates
this assumption. Looking at eqs.~(\ref{eq:dZdm0LNf2},\ref{eq:dZdzfNf2})
it is clear that the $l$-definitions have in general larger systematic
uncertainties due to their larger sensitivity to $z_f$. This is then reflected in
the $Z$-factors in tables \ref{tab:ZAVRawNf2} and \ref{tab:ZAVRawNf2PT} where the 
uncertainties are listed separately.

\subsection{Estimating the derivatives: $\Nf=3$}

For $\Nf=3$ we proceeded in very much the same way, except that we carried out a more
complete study of $L/a=8$ lattices at all $\beta$-values, except $\beta=3.85$, and also
included additional $L/a=6$ lattices at $\beta=3.4$ and $3.46$. As before for the derivatives
(\ref{eq:m0zf2mgA}) we used their mean values and set $\rmd mL/\rmd z_f=0$, while for the 
derivatives (\ref{eq:dZdX}) we considered their (absolute) mean values plus twice their 
statistical errors. However, here we did this separately for all $\beta$-values, with $\beta=3.7$
results also applied at $\beta=3.85$. Table \ref{tab:ZAVRawNf3} contains the results for 
$Z_{\rm A,V}^{g,l}$ for all ensembles of table \ref{tab:Nf3Parms}, including the different 
estimated uncertainties. As with $\Nf=2$, the soundness of our assumption is supported by 
experience gained during the tuning runs to find $m_{\rm cr}$ and $z_f^*$.

\subsection{Expected scaling with $L/a$ and perturbative calculations}

\subsubsection{Expected $L/a$-scaling}

Using general arguments based on the Symanzik expansion and $P_5$ parity~\cite{Brida:2016rmy}
one may obtain the expected scaling with the lattice spacing $a$ 
of the derivatives of the  PCAC mass and $g_{\rm A}^{ud}$ with respect to $m_0$ and $z_f$, 
\begin{equation}
  \dfrac{\partial g_{\rm A}^{ud}}{\partial z_f} = \rmO(1),\qquad
  \dfrac{\partial g_{\rm A}^{ud}}{\partial m_0L}=\rmO(1),
\end{equation}
and
\begin{equation}
   \dfrac{\partial mL}{\partial m_0L} = \rmO(1),\qquad 
   \dfrac{\partial mL}{\partial z_f} = \rmO(a^2).
\label{eq:dmbydzf}
\end{equation}
To explain how one arrives at these scaling properties we go through the example
of the PCAC mass:
\begin{equation}
   \label{eq:dmLdzf}
   \dfrac{\partial mL}{\partial z_f} = L \dfrac{\partial}{\partial z_f}  
         \left(\dfrac{\tilde{\partial}_0 g_{\rm A}^{ud}} {2 g_{\rm P}^{ud}}\right) 
   = L\left(m'-m\right) \times \dfrac{g_{{\rm P};z_f}^{ud}}{g_{\rm P}^{ud}},
\end{equation}
where we have introduced the modified PCAC mass $m'$ through the equation
\begin{equation}
 \tilde{\partial}_0 g_{{\rm A};z_f}^{ud} = 2m' g_{{\rm P};z_f}^{ud},
\end{equation}
and the notation $;z_f$ indicates differentiation with respect to $z_f$~\cite{Brida:2016rmy}.
The crucial point to note is that this differentiation merely modifies
the fields at the time boundaries and thus produces a different matrix element for the PCAC
relation and hence the modified PCAC mass $m'$.
Since the difference $m'-m$ between two PCAC masses is of O($a$) in general, 
and the $z_f$-derivative of the $P_5$-even correlation function $g_{\rm P}^{ud}$ is $P_5$-odd and
thus of O($a$), we arrive at O($a^2$) for the complete expression. It is re-assuring to see 
that this derivative is indeed found to be small in the simulations.

Similar arguments lead to
\begin{equation}
  \dfrac{\partial Z_{\rm A,V}^{g,l}}{\partial m_0L} = \rmO(a),  \qquad  
  \dfrac{\partial Z_{\rm A,V}^{g,l}}{\partial z_f} = \rmO(a),
  \label{eq:Z-scaling}
\end{equation}
where the derivatives are taken at fixed $\beta,z_f$ and $\beta, am_0$, respectively.

\subsubsection{Comparison with perturbation theory}

Perturbation theory confirms all of these expected scaling properties.%
\footnote{Note that perturbative results without explicit group factors 
assume gauge group SU(3) and fermions in the fundamental representation.} 
The derivatives (\ref{eq:m0zf2mgA}) have only been considered to tree-level,
which gives,
\begin{align}
 \nonumber
 {\partial mL\over\partial m_0L}&=1 + \rmO(g_0^2),
 &
 {\partial mL\over\partial z_f}&= \rmO(g_0^2),\\
 {\partial g^{ud}_{\rm A} \over\partial m_0L}&=-3 + \rmO(g_0^2),
 &
 {\partial g^{ud}_{\rm A}\over\partial z_f}&= -6 + \rmO(g_0^2).
\end{align}
For the mass sensitivity of the axial current we have, to one-loop order,
\begin{eqnarray}
    \dfrac{\partial Z_{\rm A}^{g}}{\partial m_0L} &\approx& \dfrac{a}{L}\times
    \begin{cases} 1 - 0.47 \times g_0^2 +\ldots &  \text{(Wilson action)}, \cr
                  1 - 0.43 \times g_0^2 +\ldots &  \text{(LW action)},      \end{cases}\\
    \dfrac{\partial Z_{\rm A}^{l}}{\partial m_0L} &\approx& \dfrac{a}{L}\times
                 \left(- 0.16 \times g_0^2+\ldots \right) \quad\text{(LW \& Wilson action)}, 
\end{eqnarray}
and, for the vector current,
\begin{eqnarray}
    \dfrac{\partial Z_{\rm V}^{g}}{\partial m_0L} &\approx& \dfrac{a}{L}\times
    \begin{cases} 1 - 0.48 \times g_0^2 +\ldots &  \text{(Wilson action)}, \cr
                  1 - 0.44 \times g_0^2 +\ldots &  \text{(LW action)},      \end{cases}\\
   \dfrac{\partial Z_{\rm V}^{l}}{\partial m_0L} &\approx& \dfrac{a}{L}\times
    \begin{cases} - 0.20 \times g_0^2 +\ldots &  \text{(Wilson action)}, \cr
                  - 0.19 \times g_0^2 +\ldots &  \text{(LW action)}.      \end{cases}
\end{eqnarray}
Here, the one-loop coefficients are the values at $L/a=12$ and are stable within 3-10 per 
cent for the range $L/a$ from 8 to 16. Incidentally, this result resolves qualitatively a
puzzle posed by the non-perturbative results, eq.~(\ref{eq:dZdm0LNf2}), where the
derivatives of the $g$- and $l$-definitions are almost the same in magnitude and opposite
in sign, whereas the tree level results are 1 and 0, respectively, as first noticed
in~\cite{Brida:2016rmy}. 

The $z_f$-sensitivity of the $Z$-factors is easily described: all $z_f$-derivatives vanish
at tree level and the one-loop coefficients for the $g$-definitions are very small 
and vanish with a rate roughly proportional to $a^3$ for $Z^g_{\rm A,V}$ and both gauge actions.
On the other hand, the $l$-definitions behave as expected (s.~above): very similar
numbers are obtained which are, for both gauge actions, within a few percent given by
\begin{equation}
   \dfrac{\partial Z_{\rm A,V}^{l}}{\partial z_f} = -0.32 \times g_0^2 \dfrac{a}{L} +\rmO(g_0^4).
\end{equation}
For completeness we report the perturbative results for $am_{\rm cr}$ and $z_f^*$ we
obtain from the same computation. For the critical mass to order $g_0^2$, the known 
values~\cite{Wohlert:1987rf,Luscher:1996vw,Aoki:1998ar,Aoki:1998qd} 
(with  $C_F=4/3$ for gauge group SU(3)),
\begin{equation}
  am_{\rm cr} = g_0^2 C_{\rm F} \times \begin{cases}-0.2025565(3),  & \text{(Wilson action),} \cr
                                                    -0.1509201(1),  & \text{(LW action)},
                                        \end{cases}
\end{equation}
are already reproduced to 4-5 digits on lattices with $L/a$ in the range from 8 to 16. 
As for $z_f^*$, we have, to order $g_0^2$ and for $a/L\rightarrow 0$,
\begin{equation}
  z_{f}^{*} = 1+ g_0^2C_{\rm F} \times \begin{cases} 0.16759(1),  & \text{(Wilson action)}, \cr
                                                     0.12923(5),  & \text{(LW action)},
                                        \end{cases}
\end{equation}
where the Wilson action result is from ref.~\cite{Brida:2016rmy}, whereas the LW action value is
the one of $L/a=16$ with a generous guess for the error. Also in this case the values at
finite $L/a$ from 8 and 16 coincide with these numbers to 4-5 digits precision.

We observe that the quantitative comparison of non-perturbative data with bare perturbation
theory to order $g_0^2$ works quite well in certain cases. For instance, for $\Nf=2$ at 
$\beta=5.2$, we compare the non-perturbative value $am_{\rm cr} = -0.3282$ to $am_{\rm cr} = -0.3116
+ \rmO(g_0^4)$, and similarly for $z_f^* =1.288$ we need to compare to $z_f^*=1.258 + \rmO(g_0^4)$. 
Also the non-perturbative current normalization constants themselves are reproduced by one-loop 
perturbation theory at the 5-10 percent level (compare table~\ref{tab:RAV1} with tables~\ref{tab:ZNf2}
and \ref{tab:ZNf3L1}, \ref{tab:ZNf3L2}). On the other hand, the majority of the derivatives differ very 
significantly, for instance,
\begin{equation}
  \left.\dfrac{\partial Z_{\rm A}^{g}}{\partial m_0L}\right\vert_{\beta=5.2, L/a=8} =  0.126(11) 
  \quad
  \text{vs.}
  \quad
   0.057 + \rmO(g_0^4),
\end{equation}
is off by a factor 2, and for the $l$-definition the comparison is between $-0.127(9)$ and $-0.023$,
which differs by a factor 5. For the $z_f$-derivatives we note that perturbation theory correctly
predicts the smallness of the sensitivity in the $g$-definition. Quantitatively, the 
perturbative $z_f$-derivatives of $Z_{\rm A,V}^l$ at $\beta=5.2$ and $L/a=8$ 
are $-0.046 +\rmO(g_0^4)$, to be compared with eq.~(\ref{eq:dZdzfNf2}), so again we observe
a difference by a factor 2. 

Finally, for the interpolations in $L/a$ of the data (s.~below) we are also interested in the derivatives
of the $Z$-factors with respect to $x=(a/L)^2$ at fixed bare coupling. In perturbation theory we can
obtain approximate results from table~\ref{tab:RAV1}.
Expanding
\begin{equation}
    R_{\rm A,V}^{g,l}  =   1 + g_0^2 \left( Z_{\rm A,V}^{(1)}  + k_{\rm A,V}^{g,l}\times\dfrac{a^2}{L^2} + \rmO(a^3)\right) + \rmO(g_0^4), 
\end{equation}
the $x$-derivatives to order $g_0^2$ are approximately given by 
\begin{equation}
	\label{eq:PT-L/a-dependence}
   \dfrac{\partial Z_{\rm A,V}^{g,l}}{\partial x} \approx g_0^2 \times  k_{\rm A,V}^{g,l},
\end{equation}
provided higher order cutoff effects are small. We find that this is quite well satisfied, 
with very similar coefficients for both Wilson and LW actions, given approximately by
\begin{equation}
   k_{\rm A}^{g}\approx 0.45,\qquad k_{\rm V}^{g} \approx 0.43, 
\end{equation}
whereas $k_{\rm A,V}^l$ are ca.~20 to 30 times smaller in magnitude, for axial and vector cases, respectively,
and come with the opposite sign. Comparing this with the $\beta=5.3$ non-perturbative data in eqs.~(\ref{eq:dZdLNf2})
we see again that for the $g$-definitions these derivatives are reproduced by perturbation theory
up to a factor 2, while for the $l$-definitions, perturbation theory to O($g_0^2$) is clearly
missing the bulk of the effect. While, as expected, the non-perturbative derivatives
are smaller than for the $g$-definitions, it seems that the smallness  of the O($g_0^2$) term is
an accident and higher orders are dominating at these values of $\beta$.

To conclude this comparison, perturbation theory often gives valuable qualitative information
and may provide reasonable starting values for the tuning of $am_0$ and $z_f$.
However, quantitatively, the agreement with non-perturbative data at lattice
spacings of interest for hadronic physics hugely varies for different observables.
Hence, the main practical use of perturbation theory consists in the perturbative 
subtraction of cutoff effects. Here, even a qualitative agreement, which may be quantitatively 
off by a factor 2, still means a welcome reduction of cutoff effects by 50 percent,
and our data analysis does indeed point to such benefits.

Given this situation, we have refrained from using perturbative data in our estimates of the derivatives,
and we have decided to ignore the favourable O($a$) scaling, eq.~(\ref{eq:Z-scaling}), 
when applying the results obtained at $L/a=8$ at all other $L/a$-values, too. 
In this respect, the tuning runs for $am_0$ and $z_f$,
both for $\Nf=2$ and $\Nf=3$, provided some consistency checks which make us confident that the 
chosen procedure is indeed sound and rather conservative.

\subsection{Interpolation in $L/a$ for $\Nf=2$}
\label{app:InterpolationsNf2}

Once the uncertainties associated with the conditions (\ref{eq:mcrzf}) have been propagated
to $Z_{\rm A,V}$ at given $\beta$ and $L/a$, one needs to keep the LCP condition (\ref{eq:fKL})
and also take into account the corresponding uncertainties. The choice (\ref{eq:fKL}) is
made such that the $L/a=8$ results at $\beta=5.2$ satisfy this condition by definition, 
while at $\beta=5.3$ we needed to interpolate the results for $L/a=8,10,12$ to the target value 
$(L/a)(5.3)=9.18(21)$  (cf.~table \ref{tab:LCPNf2}). We have performed a simple linear interpolation
in $(a/L)^2$ which describes the data very well, similarly to the case $\Nf=3$ which will be 
discussed in more detail below. For $\beta=5.5$ and $5.7$, the simulated $L/a$ values are, within
errors, compatible with the target values in table~\ref{tab:LCPNf2}. Systematic errors related to
the condition (\ref{eq:fKL}) were then estimated by using the slope of the interpolation in 
$x=(a/L)^2$ at $\beta=5.3$ also for the higher $\beta$-values. Note that we interpolate results 
with bare parameters $am_{\rm cr}$ and $z_f^*$ tuned at the given $\beta$- and $L/a$-values. 
Hence the derivative defines the sensitivity to a change of the  physical size of the system. This 
is a pure cutoff effect of O($a^2$) on the $Z$-factors. Since, by the choice of the variable $x$, 
a factor $a^2$ is also divided out, the $x$-derivative is expected to be of O(1) and we expect a 
smooth dependence of this derivative on $\beta$;%
\footnote{We recall that at leading order in PT the $x$-derivatives of the $Z$-factors are of 
		  O($g^2_0$) (cf.~eq.~(\ref{eq:PT-L/a-dependence})). They are thus expected to diminish, 
		  and eventually vanish, as $g_0\to0$.}%
this expectation is in fact confirmed by the results for $\Nf=3$ where the slope shows a very mild 
$\beta$-dependence over the whole range (cf.~table \ref{tab:dZdxNf3}). As we are looking 
at an O($a^2$) effect in disguise, it is no surprise that the results depend on whether or not the
cutoff effects have been subtracted perturbatively.

Without perturbative subtraction we obtained the results,
\begin{align}
 \nonumber
 {\partial Z_{\rm A}^g \over\partial x} &=  0.946(89),
 &
 {\partial Z_{\rm A}^l \over\partial x} &=  0.48(16),\\
 {\partial Z_{\rm V}^g \over\partial x} &=  1.177(77),
 &
 {\partial Z_{\rm V}^l \over\partial x} &=  0.54(17),
 \label{eq:dZdLNf2}
\end{align}
while for the perturbatively improved ones we obtain,
\begin{align}
 \nonumber
 {\partial Z_{\rm A}^g \over\partial x} &=  0.447(89),
 &
 {\partial Z_{\rm A}^l \over\partial x} &=  0.49(16),\\
 {\partial Z_{\rm V}^g \over\partial x} &=  0.734(77),
 &
 {\partial Z_{\rm V}^l \over\partial x} &=  0.56(17).
 \label{eq:dZdLNf2PT}
\end{align}
The corresponding systematic error is then simply taken to be,
\begin{equation}
 (\Delta_x Z_{{\rm X}})^2 = \bigg({\partial Z_{\rm X}\over\partial x}\bigg)^2 (\Delta x)^2,
 \qquad
 {\rm X=A,V},
\end{equation}
which is summed in quadrature to (\ref{eq:SystNf2}) and the statistical 
error from the Monte Carlo simulations. The uncertainly $\Delta x$ was estimated as:
\begin{equation}
 \Delta x=|x - x(\beta)| + 2\,\sigma(x(\beta)),
\end{equation}
where $x(\beta)=((a/L)(\beta))^2$ and $\sigma(x(\beta))$ is the associated 
error. As a further safeguard we 
took for the derivatives (\ref{eq:dZdLNf2}) and (\ref{eq:dZdLNf2PT})
the (absolute) mean value plus twice their statistical error. 
We observe that the $l$-definitions have a milder $L/a$-dependence 
than the $g$-based ones, unless perturbative improvement is implemented.

\subsection{Interpolation in $L/a$ for $\Nf=3$}
\label{app:InterpolationsNf3}

Once the systematic errors deriving from the tuning of $am_0$ and $z_f$ have been taken into
account, the results for $Z_{\rm A,V}$ at different $L/a$ and fixed $\beta$ must be interpolated
to either $(L_{1}/a)(\beta)$ or $(L_{2}/a)(\beta)$, depending on the LCP; for completeness the 
values of $Z_{\rm A,V}$ prior to interpolation are given in table \ref{tab:ZAVRawNf3}. We have 
considered three types of interpolation in $x=(a/L)^2$, these are: linear using all 4 available
values of $L/a$  (cf.~table~\ref{tab:LCPNf3}), linear using only the 3 closest $L/a$-values to 
the target $(L_{1,2}/a)(\beta)$, and quadratic using all 4 $L/a$-values. Given this choice, the 
interpolations needed for the $L_1$- and $L_2$-LCPs only differ for $\beta=3.4$, where, by definition, 
$L_1/a=8$ is exact, and in the case of linear interpolations with 3 points at $\beta=3.55$. 
Recall that for  $\beta=3.85$ no interpolation is required, as $L_2/a=16$ is exact and this 
$\beta$-value has been excluded for the $L_1$-LCP.

\begin{figure}[hpbt]
	\centering
	\includegraphics[scale=0.95]{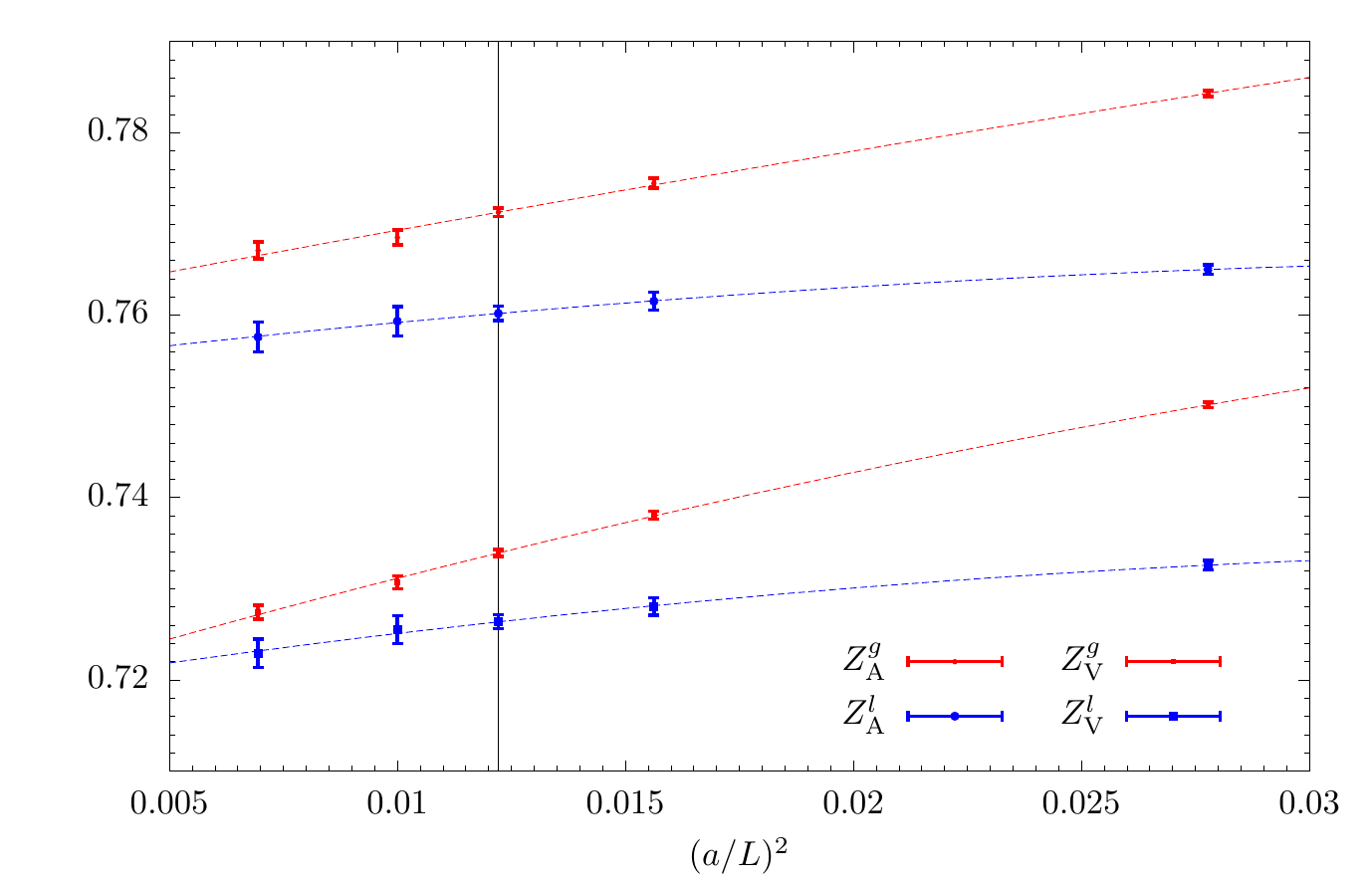}
	\caption{$L/a$-interpolations for $\beta=3.46$ and 
		the $L_1$-LCP. The upper two sets of points correspond 
		to the $Z^{g,l}_{\rm A}$ results while the lower two
		sets are the $Z^{g,l}_{\rm V}$ results. The 
		dashed lines are our preferred, quadratic, fits to the data,
		and the interpolation points are marked by a black vertical line.}
	\label{fig:ZXint-LCP1}
\end{figure}  

\begin{figure}[hpbt]
	\centering
	\includegraphics[scale=0.95]{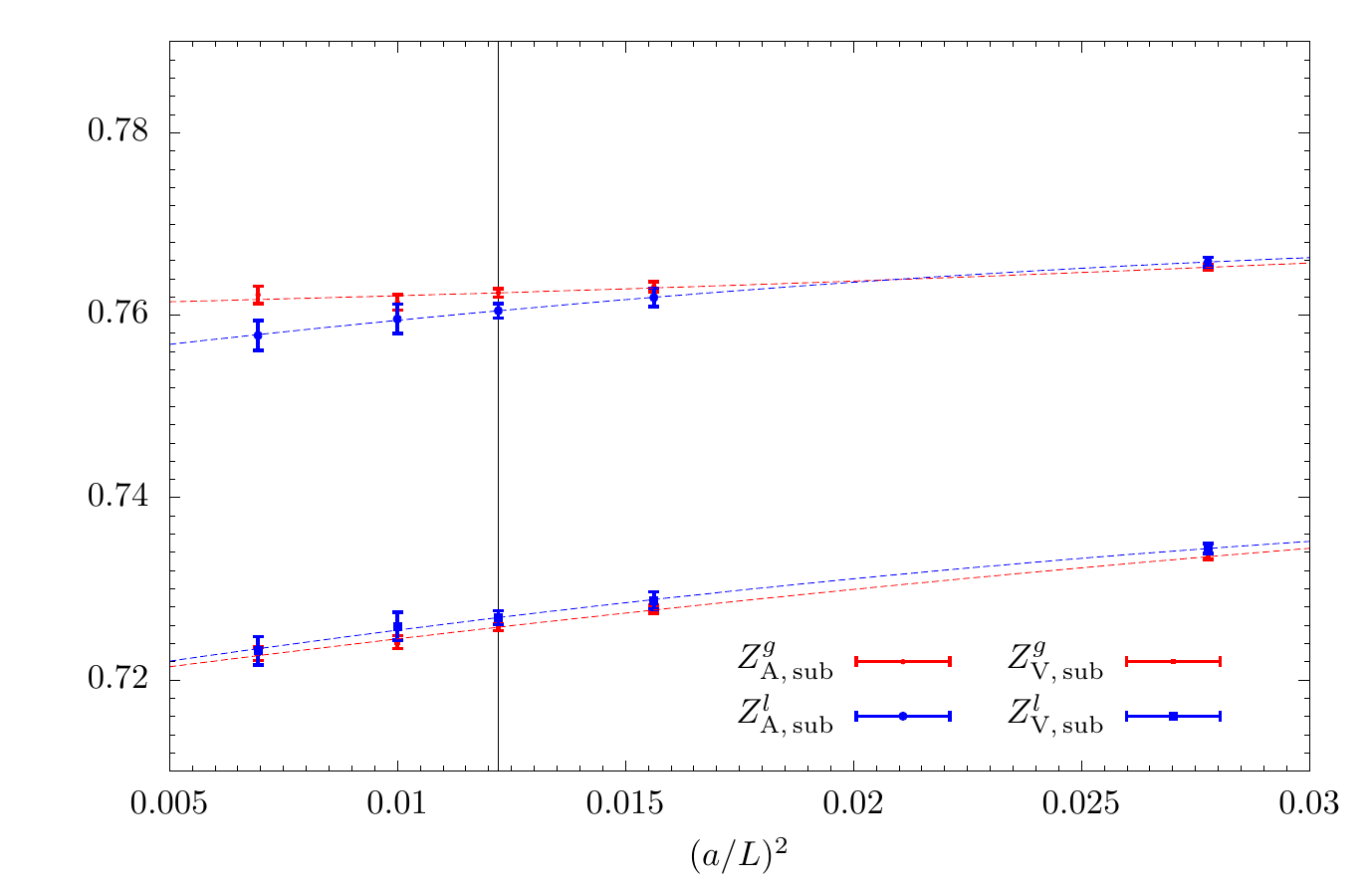}
	\caption{Same as figure~\ref{fig:ZXint-LCP1}, for the $Z$-factors with perturbative
	subtraction of the cutoff effects.}
	\label{fig:ZXint-LCP1-PT}
\end{figure}  

Starting with the $L_1$-LCP, the different interpolations describe the data quite well
in general, particularly so for the results at the two smallest lattice spacings and for
definitions based on the $l$-correlators. The most relevant exception is given indeed 
by the linear interpolation of $Z_{\rm V}^g$ at $\beta=3.46$ using all 4 values of $L/a$,
for which we find a $\chi^2{\rm/d.o.f}\approx2.2$. It should be noted, however, that the
$\chi^2$-criterion does not come with the usual probability interpretation due to the 
errors being dominated by systematics. In any case, the interpolated values are generally
compatible at the 1$\sigma$ level.

Considering the perturbatively improved data, the quality of the interpolations is 
generally improved, and all fits have excellent $\chi^2$. The beneficial effect of the 
perturbative improvement can be appreciated by comparing figure \ref{fig:ZXint-LCP1} and 
\ref{fig:ZXint-LCP1-PT}, where the $Z_{\rm A,V}$ interpolations at $\beta=3.46$ are shown
for the cases before and after perturbative improvement, respectively. This example also 
illustrates the general feature that, before perturbative improvement, the $l$-definitions
have a significantly milder $L/a$- and hence $x$-dependence. In addition, it is interesting 
to note that the $x$-dependence of the $Z$-factors does not change significantly over the 
range of $\beta$ considered, but seems in general to diminish,  as expected, as $\beta\to\infty$ 
(cf.~table \ref{tab:dZdxNf3}). Based on these observations, we take as our final estimates
for the $Z$-factors the results of the quadratic fits, which have the largest errors.

\begin{table}[!bhtp]
\centering
\begin{tabular}{lrrrr}
\toprule
$\beta$ & $\partial Z_{\rm A}^{g}/\partial x$ & $\partial Z_{\rm A}^{l}/\partial x$  & 
$\partial Z_{\rm V}^{g}/\partial x$ & $\partial Z_{\rm V}^{l}/\partial x$   \\
\midrule                                                                                   
$3.46$  & $0.90(11)$ & $0.44(21)$ & $1.25(09)$  & $0.56(20)$ \\
$3.55$  & $0.69(11)$ & $0.44(15)$ & $1.10(08)$  & $0.56(16)$ \\
$3.70$  & $0.81(10)$ & $0.29(16)$ & $0.87(11)$ & $0.24(17)$ \\
\midrule
$\beta$ & $\partial Z_{\rm A,\,sub}^{g}/\partial x$ & $\partial Z_{\rm A,\,sub}^{l}/\partial x$  & 
$\partial Z_{\rm V,\,sub}^{g}/\partial x$ & $\partial Z_{\rm V,\,sub}^{l}/\partial x$   \\
\midrule
$3.46$ & $0.16(11)$  & $0.46(21)$ & $0.58(09)$  & $0.61(20)$ \\
$3.55$ & $-0.01(11)$ & $0.46(15)$ & $0.45(08)$  & $0.60(16)$ \\
$3.70$ & $0.12(10)$  & $0.31(16)$ & $0.23(11)$ & $0.28(17)$ \\
\bottomrule
\end{tabular}
\caption{Results for $\partial Z_{\rm A,V}^{g,l}/\partial x$, where $x=(a/L)^2$, as a function
of $\beta$ for $\Nf=3$ quark-flavours. The derivatives are estimated along the $L_1$-LCP from 
the linear fits using the 3 closest $L/a$-values to the target $(L_1/a)(\beta)$.}
\label{tab:dZdxNf3}		
\end{table}

Regarding the $L_2$-LCP, the situation is more complicated due to the fact that we need to interpolate
the data at the coarsest lattice spacing, $\beta=3.4$. The quality of
the interpolations is still good in general, but there are a few significant exceptions. We note that all
these cases involve $g$-definitions:  indeed, we obtain a pretty large $\chi^2/$d.o.f.~for the linear 
fits of $Z_{\rm V}^g$ at $\beta=3.4$ and $3.46$,  around $7.8$ and $5$ respectively. 
Also the quadratic fit for $Z_{\rm A}^g$ at $\beta=3.4$ has a large 
$\chi^2{\rm/d.o.f}\approx1.8$. While in this case, however, the results of the interpolation are compatible
with those of the linear fits within less than one standard deviation, in the case of $Z_{\rm V}^g$
at $\beta=3.4$ the discrepancy between the linear and quadratic interpolations is close to 3 standard
deviations. The situation definitely improves when the perturbatively improved data are considered. In this
case, with the exception of the linear interpolations with 4 points of $Z^g_{\rm V,A,\,sub}$ at $\beta=3.4$,
all fits have very good $\chi^2$, and give compatible results within one standard deviation or so. 
As in the case of the $L_1$-LCP, we take as our final estimates for $Z_{\rm A,V}$ the 
results of the quadratic fits, which have the best $\chi^2$-values and the largest errors.
\clearpage

\section{Fit formulas for $Z_{\rm A,V}(g_0^2)$}
\label{app:Fits}

In this appendix we collect some useful fit formulas for 
the $Z_{\rm A,V}$ results, both for $\Nf=2$ and $\Nf =3$.
We will focus on the data for $Z_{\rm A,V,\,sub}^l$,
cf.~the discussion in sect.~\ref{sec:Conclusions}.

\subsection{$\Nf=2$}

For $\Nf=2$ the final $Z_{\rm A,V}$ results are given in
table~\ref{tab:ZNf2}. Over the whole range of $\beta\in[5.2,5.7]$,
the data for $Z_{\rm A,\,sub}^l$ is well described by a simple 
linear fit function,
\begin{gather}
	\nonumber
	Z^l_{\rm A,\,sub} = c_1 + c_2 g_0^2,\\[1.5ex]
	c_{1,2}=
	\begin{pmatrix}
	\phantom{+}1.15183  \\
	-0.33176
	\end{pmatrix}
	\qquad
	{\rm Cov}=
	10^{-3}
	\times
	\begin{pmatrix}
	\phantom{+} 0.17228874 & -0.15443145 \\
	-0.15443145 & \phantom{+} 0.13858044
	\end{pmatrix},
\end{gather}
which has a $\chi^2/{\rm d.o.f.}=0.759/2$.

Similarly, for the vector current renormalization,
$Z_{\rm V,\,sub}^l$, a good description of the 
data is given by, 
\begin{gather}
	\nonumber
	Z^l_{\rm V,\,sub} = c_1 + c_2 g_0^2,\\[1.5ex]
	c_{1,2}=
	\begin{pmatrix}
	\phantom{+}1.18984 \\
	 -0.39138
	\end{pmatrix}
	\qquad
	{\rm Cov}= 
	10^{-3}
	\times
	\begin{pmatrix}
	\phantom{+}   0.19505967  & -0.17469696  \\
	-0.17469696 & \phantom{+}  0.15663400
	\end{pmatrix},
\end{gather}
which has a $\chi^2/{\rm d.o.f.}= 0.866/2$.
 
\subsubsection{Matching with perturbation theory}

It is also interesting to consider fit functions with the  
correct perturbative 1-loop behaviour for $g_0^2\to0$ (cf.~sect.~\ref{sec:PTImprovement}). 
In the case of $Z_{\rm A,\,sub}^l$ this is possible using a 2-parameter polynomial fit,
\begin{gather}
\nonumber
Z^l_{\rm A,\,sub} = 1 -0.116458\,g_0^2 + c_1 g_0^4 + c_2 g_0^6,\\[1.5ex]
c_{1,2}=
\begin{pmatrix}
-0.015248  \\
-0.049793  
\end{pmatrix}
\qquad
{\rm Cov}=
10^{-3}
\times
\begin{pmatrix}
\phantom{+} 0.12545440 & -0.11198363  \\
-0.11198363  & \phantom{+}  0.10005857
\end{pmatrix},
\end{gather}
which gives a $\chi^2/{\rm d.o.f.}= 1.519/2$. We note that the 
same fit ansatz was used to fit the standard SF results 
of ref.~\cite{Fritzsch:2012wq}. Similarly, for the vector 
current data, $Z_{\rm V,\,sub}^l$, we have,
\begin{gather}
	\nonumber
	Z^l_{\rm V,\,sub} = 1 - 0.129430\,g_0^2 + c_1 g_0^4 + c_2 g_0^6,\\[1.5ex]
	c_{1,2}=
	\begin{pmatrix}
	-0.005952  \\ 
	-0.068180 
	\end{pmatrix}
	\qquad
	{\rm Cov}=
	10^{-3}
	\times
	\begin{pmatrix}
	\phantom{+} 0.14221117 & -0.12684203  \\
    -0.12684203  & \phantom{+}  0.11324469 
	\end{pmatrix},
\end{gather}
which gives a $\chi^2/{\rm d.o.f.}= 1.866/2$. We stress that
although the latter fit functions encode  the expected asymptotic behaviour 
far outside the $\beta$-range covered by the data, it is not recommended to use them
for $\beta$ values much outside this range. For $\beta\in[5.2,5.7]$,
the two sets of fit functions agree within less than $1\sigma$ deviations.

\subsection{$\Nf=3$}

For the case $\Nf=3$, our final $Z_{\rm A,V}$ results are given in 
table~\ref{tab:ZNf3L1} and \ref{tab:ZNf3L2}. Having one additional 
$\beta$-value, it is natural to prefer an interpolation of the 
$L_2$-LCP data of table~\ref{tab:ZNf3L2}. The higher precision of the
data compared to $\Nf=2$, and the availability of a fifth data point 
suggests to use 3-parameter fits in this case. We find that, for
the whole range of $\beta\in[3.4,3.85]$, $Z^l_{\rm A,\,sub}$, is well
described by the quadratic fit:
\begin{equation}
	Z^l_{\rm A,\,sub} = c_1 + c_2 g_0^2 + c_3 g_0^4,
\end{equation}
with coefficients and covariance given by
\begin{gather}
\nonumber
c_{1,2,3}= \begin{pmatrix}
  \phantom{+} 1.35510\\
  -0.501106	\\
  \phantom{+} 0.091656
\end{pmatrix}
\qquad
{\rm Cov}=
10^{-1}
\times
\begin{pmatrix}
\phantom{+}0.229571866  &            -0.278151898  & \phantom{+}0.084105454 \\
         -0.278151898   &  \phantom{+}0.337131945  &           -0.101975449 \\
\phantom{+}0.084105454  &            -0.101975449  &  \phantom{+}0.030856380
\end{pmatrix},
\end{gather}
and $\chi^2/{\rm d.o.f.}= 0.622/2$.
	
For the vector current 
 data, $Z_{\rm V,\,sub}^l$, we use the same fit function,
\begin{equation}
Z^l_{\rm V,\,sub} = c_1 + c_2 g_0^2 + c_3 g_0^4,
\end{equation}
and obtain
\begin{gather}
\nonumber
c_{1,2,3}= \begin{pmatrix}
\phantom{+} 1.32353\\
-0.459016\\
\phantom{+} 0.066995
\end{pmatrix}
\qquad
{\rm Cov}=
10^{-1}
\times
\begin{pmatrix}
\phantom{+}0.247244906  &           -0.299424391  &  \phantom{+}0.090493309 \\
          -0.299424391  &  \phantom{+}0.362743281 &            -0.109668066 \\
\phantom{+}0.090493309  &           -0.109668066  &  \phantom{+}0.033167490 
\end{pmatrix},
\end{gather}
which gives a $\chi^2/{\rm d.o.f.}=1.801/2 $.  

\subsubsection{Matching with perturbation theory}

Also in this case we consider fit functions with the  
correct perturbative 1-loop behaviour for $g_0^2\to0$ (cf.~sect.~\ref{sec:PTImprovement}). 
Applying a 3-parameter polynomial fit of the form
\begin{equation}
Z^l_{\rm A,\,sub} = 1 -0.090488\,g_0^2 + c_1 g_0^4 + c_2 g_0^6 + c_3 g_0^8,\\[1.5ex]
\end{equation}
we obtain
\begin{gather}
\nonumber
c_{1,2,3}= \begin{pmatrix}
\phantom{+}  0.127163\\
 -0.178785\\
\phantom{+}  0.051814
\end{pmatrix}
\qquad
{\rm Cov}=
10^{-2}
\times
\begin{pmatrix}
\phantom{+}0.29841165  &            -0.36050066 & \phantom{+}0.10868891 \\
          -0.36050066  &  \phantom{+}0.43567202 &           -0.13140137 \\
\phantom{+}0.10868891  &            -0.13140137 & \phantom{+}0.03964605 
\end{pmatrix},
\end{gather}
which gives a $\chi^2/{\rm d.o.f.}= 0.403/2$.  We have also tried various Pad{\'e} fits
e.g.~of the type used in~\cite{Bulava:2016ktf}. With these fits we experienced
some technical problems with the bootstrap technique, when trying to determine
the covariance matrix for the fit parameters. We therefore also tried the
the automatic differentiation procedure of ref.~\cite{Ramos:2018vgu}
which completely solved this technical problem. It turns out, however, 
that the best fit function of this type develops a singularity at a $\beta$-value slightly above 4, 
and therefore does not provide a smooth interpolation to the perturbative region. 
Given the good quality of the linear fits we did not pursue any further non-linear options.

Regarding the vector current data, $Z_{\rm V,\,sub}^l$, we have,
\begin{equation}
Z^l_{\rm V,\,sub} = 1 -0.100567\,g_0^2 + c_1 g_0^4 + c_2 g_0^6 + c_3 g_0^8,\\[1.5ex]
\end{equation}
with
\begin{gather}
\nonumber
c_{1,2,3}= \begin{pmatrix}
\phantom{+}  0.130134 \\
 -0.182926\\
\phantom{+}  0.051526
\end{pmatrix}
\qquad
{\rm Cov}=
10^{-2}
\times
\begin{pmatrix}
\phantom{+}0.32342173  &            -0.39055319 &  \phantom{+}0.11769835\\
          -0.39055319  &  \phantom{+}0.47179235 &  -0.14223242          \\
\phantom{+}0.11769835  &            -0.14223242 &  \phantom{+}0.04289459
\end{pmatrix},
\end{gather}
which gives a $\chi^2/{\rm d.o.f.}= 1.443/2$.  

To conclude this appendix, we emphasize again that all given fits to the data
are very good if used as interpolations in the range of
the non-perturbative data. Using them outside this range is 
at the user's own risk, even where perturbative information 
is used as a constraint.

\clearpage

%\section*{References}

%\bibliographystyle{JHEP}
\usebiblio{bibfile.bib}

\end{document}